\documentclass[12pt]{article}
\usepackage{amssymb}
\usepackage{amsmath}
\numberwithin{equation}{section}
\def\dd{{\rm d}}\def\ee{{\rm e}}\def\ii{{\rm i}}
\def\SU{{\rm SU}}\def\vd{\vec{\dd}}
\def\half{{\textstyle{\frac12}}}
\def\fourth{{\textstyle{\frac14}}}
\def\proj{{\rm proj}\,}
\def\beq{\begin{equation}}\def\eeq{\end{equation}}
\def\bea{\begin{eqnarray}}\def\eea{\end{eqnarray}}

\begin{document}
\null\vfill
\thispagestyle{empty}
\begin{center}
\noindent{\bf QUANTUM FIELD THEORY AND GRAVITY IN CAUSAL SETS}

\noindent by

\noindent Roman M. Sverdlov

\vfill

\noindent A dissertation submitted in partial fulfillment\\ of the requirements for the degree of \\ Doctor of Philosophy \\(Physics)\\ in The University of Michigan (2009)

\vfill
\end{center}

Doctoral Committee:

Professor Luca Bombelli, Co-Chair

Emeritus Professor Marc H. Ross, Co-Chair

Professor Igor Dolgochev

Professor Roberto Merlin

Associate Professor James Liu

Assistant Professor Alexei Tkachenko

\begin{center}
\pagenumbering{roman}
\newpage
\tableofcontents
\end{center}
\pagenumbering{arabic}
\bigskip
\newpage
\section{Chapter 1: Introduction}
While both quantum mechanics and gravity have been repeatedly tested and verified, combining the two into a single theory has not been done to this day. Due to the fact that the strength of the gravitational interaction between elementary particles is many magnitudes smaller than that of any other interaction, there are no experiments that detect gravity at the quantum scale. At the same time, once we are dealing with objects heavy enough for their gravity to become detectable, they are too large for any quantum effects to be observed. Thus, there is no scale at which both quantum mechanics and gravity are significant. This leaves no experimental data from which to start in building a quantum theory of gravity. 

Ironically, despite the fact that the gravitational field is very weak, the fact that general relativity identifies gravity with geometry makes the gravitational field more important than other fields, since the latter need geometry to propagate. Thus, even though we do not need to combine quantum field theory with gravity in order to explain outcomes of experiments performed so far, a theory combining the two is important in order to answer a number of conceptual questions. In particular, if gravity did not exist, the ``fuzziness" of quantum mechanics, albeit counterintuitive, would still have been well defined since ``fuzzy" particles would have existed in non-fuzzy spacetime. But due to the quantization of gravity, spacetime becomes fuzzy as well, leaving no non-fuzzy background. 

String theory provides one way of answering this question. The string spectrum generates all the known particles, including gravitons. The latter replace spacetime curvature in defining gravity. However, string theory has made a number of predictions that, due to their very small scale, are not verifiable in the lab to this day. The possibility of both positive and negative outcomes makes it both a prime pro as well as prime con when it comes to pursuing the research in this field.

The main alternative to string theory is spacetime discretization. This approach takes the traditional definitions of quantum field theory and gravity and attempts to combine them directly. Even in regular non-gravitational quantum field theory, in order to make rigorous sense of path integrals we have to introduce a lattice (see Zee's book on quantum field theory \cite{Zee}). Coordinates of a point will then be defined in terms of a position of that point in a lattice, which conceptually means that the bond structure is fundamental, while coordinates are only a bi-product. 

Strictly speaking, coordinates simply designate a label, there is nothing new in saying that they come secondary to actual geometry. However, in practice they almost always correspond to manifold structure. Thus, the idea of making them secondary invites one to think that manifold structure comes second to something more basic, which ultimately invites us to abandon the manifold structure. 

Strictly speaking, a lattice is not a manifold to start with, since the former is discrete and the latter is continuous. But intuitively it can still be viewed as manifold-like. Its manifoldlike appearance is a consequence of the specifics of bond structure. Since one can think of other kinds of bond structures that don't resemble a manifold, one is hard pressed to invent a specific Lagrangian that would ``align" the ``molecules" of spacetime into the manifold. In order to be able to do that, the Lagrangian has to be well defined for all structures, not necessarily manifold-like. This implies that manifold-ness is not fundamental. 

In this framework, the difference between curved and flat spacetime might be simply the difference in that bond structure. Thus, the presence of gravity means that the ``solid" spacetime is allowed to ``bend". However, in the quantum context, this means introducing extra degrees of freedom. These extra degrees of freedom enable spacetime not only to ``bend" but also to ``melt" into ``liquid". In the latter case, instead of being ``curved" it would loose manifold structure altogether. 

This is consistent with the uncertainty principle: at small scales, the uncertainty of the gravitational field should increase to infinity. On the other hand, that same uncertainty principle allows the gravitational field to be less uncertain at larger scales, which would explain why the observed universe is manifold-like. However, it is not clear just how can this happen: how can very non-manifoldlike building blocks line together into a manifold? Engineering a spacetime that meets the above description is the ultimate challenge of quantum gravity. 

In most such approaches to quantum gravity, the spacetime is built according to some rules. The dynamical triangulation model is one example of such approaches, in which one unwanted side effect is the violation of relativity due to the fact that simplices, or other such structures, are not Lorentz symmetric. Causal set theory, on the other hand, avoids all such violations of relativity by avoiding postulating such structures from the start. In other words, while most discrete approaches to quantum gravity replace a ``solid" with a ``liquid", causal set theory replaces it with a ``gas". The goal of causal set theory is to develop a non-manifold generalization of a Poisson distribution of points, since this is the only discrete structure that is relativistically invariant.

While causal set theory still uses links between the points, these links are interpreted as light-cone causal relations as opposed to spatial bonds. In other words, classically, signals can travel only along these links, in the directions specified on each link. These links are referred to as ``causal relations". The fact that causal relations are the only fundamental structure, corresponds to the principle of relativity that the speed of light is invariant. In order for this to resemble a Poisson distribution, the causal relations are determined at random: each point can have an arbitrary number of links, and the presence or absence of causal relations between any of the pairs of points is independent of the presence or absence of causal relations between any other pairs of points.

This approach is motivated by an observation made by Hawking that if we have a Lorentzian manifold, then the metric can be completely determined based on Weyl scaling and light cones (causal relations) alone. Here, ``Weyl scaling" refers to the information about the volumes  of the regions of spacetime, and ``causal relations" refer to an information as to whether or not one can travel from one given point to another without going faster than the speed of light. 

In the discrete case, Weyl scaling is defined by a simple count of points, since each point is assumed to take up exactly the same volume. Thus, causal relations alone are now defining features of a metric. This idea is beautiful quite independently of an issue of quantum gravity. After all, causal relations have a much more basic physical significance than a metric tensor, and in the discrete case the result shows that spacetime geometry can be described in a purely combinatorial way.

However, there is a gap in the above argument. Namely, Hawking's observation only applies to the situations where we already know we are dealing with manifold, we just don't know the metric. According to the proposal of this dissertation, the issue can be addressed by replacing the stochastic process with a Lagrangian-based dynamics as a determining factor of causal relations. This is motivated by a well established fact of general relativity, that geometry and gravitational field are the same thing. In light of the fact that we now identify geometry with causal relations, the natural conclusion is that causal relations are, too, identified with gravitational field. As such, they can be thought of as subject to various Lagrangians, which may mean that applying a pure stochastic process is not the way to go. 

Designing such Lagrangians is the main goal of this dissertation. Its contents are divided into three parts. In the first part, Lagrangians are defined for gravity and various types of matter fields. In the second part, a model of quantum ``collapse" is designed. One of the applications  of that model is a way to ``collapse" the causal relations into some background topology which is needed in order to be able to go from Lagrangian to propagator. Finally, we have to argue that after causal relations are subjected to such ``collapse" they would collapse into manifold-like ones with very high probability. The first step was done in body of the thesis, while the second and third parts, due to their controversial nature, are done in  the Appendix.  

\newpage
\section{Chapter 2: Structure of a Causal Set}

\subsection{Introduction}

The purpose of this chapter is to discuss the basic geometry of a causal set. In Section 2.2 distances and volumes will be defined on a causal set. For time being, it will be assumed that the causal set is given from the start; questions such as where causal set came from and why it has the structure that it does are left for Sections 2.5-2.7. In these latter sections, two alternative models of the origin of causal structure will be presented. One is dynamical, where events are being added one by one and causal relations are decided in the process; this is called ``classical sequential growth dynamics" , and the best-studied model of this type is known as ``transitive percolation" (Ref \cite{percolation}). The other approach is static, where all possible versions of a ``completed" causal set will be considered. That is, it will be assumed that all events and their causal relations are there from the start (perhaps each possibility existing in its own parallel universe) and probabilities are assigned to them. 

In Section 2.5 it will be shown that the transitive percolation model leads to unwanted ``big bangs" (which are referred to as ``posts"). This is blamed on the fact that the transitive percolation model is purely statistical and does not incorporate any Lagrangians. For that reason, in Section 2.7 an alternative will be presented: A Lagrangian-determined structure for a causal set. For that reason, the notion of stochastic process will be abandoned as redundant. 

However, in Section 2.7 it will be assumed assumed that geometry, determined by the causal structure, is a quantum field (namely gravity). The fluctuating geometry collapses into a fixed one. This will require an interpretation of quantum mechanics, specifically a theory of quantum measurement. The proposed model of quantum mechanics will be a theory of ``quantum corridors" (Chapter 5 of Ref \cite{decoherence}). According to that model, a quantum system, including causal relations, is being continuously measured while the measurement is not precise; the measurement error allows for quantum fluctuations and determines a range of path integral. 

Qualitative arguments regarding the compatibility of the above model with decoherence theory are left for the Appendix. While this will no longer pertain to causal sets, this part is vital in terms of justifying the causal set construction of Section 2.7, which is the only proposed model of topology of a causal set as far as this dissertation is concerned.

\subsection{Volumes and Distances on a Causal Set}

Due to the discreteness of causal sets, volumes are straightforward to define. In general, the definitions for a general causal set are motivated by observations of Poisson distribution of points in Lorentzian manifold. In the latter case, the average volume taken up by a single point is some fixed number, $v_0$. The variation of that value is due to the fluctuations of the density of scattering. 

In case of causal set, however, the points are no longer assumed to be embedded in a manifold; thus, it no longer makes sense to speak of variation of their density. Therefore, it is assumed that $v_0$ is an exact value of a volume taken up by every single point, rather than the average. Typically, this is assumed to be a volume based on the Planck scale, although there is no need to make this specific assumption. This implies
\beq
V(T) = v_0\, \sharp T\;,
\eeq
where $T$ is a subset of $S$ and $\sharp T$ stands for number of elements of $T$. Since there is no continuum, there is no well-defined notion of a ``region". Thus, the definition of a volume applies to any $T \subset S$.

A definition of distance, on the other hand, requires the notion of sequences of points that are discrete analogues of geodesics. The number of points on the segments of these curves defines their length. Consider a flat Minkowski space and two timelike-separated points in that space, and rotate the coordinate system in such a way that the two points of interest are lying on $t$-axis, with coordinates $t_1$ and $t_2$. Let $\gamma$ be an arbitrary future-directed curve that connects them. The length of $\gamma$ is given by 
\beq
l(\gamma) = \int_{\gamma} \sqrt{(\dd t)^2-\sum (\dd x_k)^2}
\leq \int_\gamma \vert \dd t \vert = \int_\gamma \dd t = t_2 - t_1 = \tau(p,q)\;.
\eeq
The second equal sign in the above equation is based on the assumption that the curve is future-directed. Thus, while it is not true that the length of every single curve that connects $p$ and $q$ is less than the Lorentzian distance between them, the statement is true for future-directed curves.  This, for example, rules out the scenario involving traveling to the future or back either arbitrary far or arbitrary many times, resulting in a curve of arbitrarily large length.  

The above statement can be easily adjusted to the discrete case of causal sets (see Refs \cite{length1} and \cite{length2}). A future-directed curve can be viewed as a future-directed set of points, $p \prec r_1 \prec . . . \prec r_n \prec q$. Selecting the longest possible curve corresponds to selecting a chain of points of maximal cardinality. This has a side benefit: maximizing the number of points excludes the possibility of removing segments of a curve and/or skipping points at random. Thus, the points are spaced as densely as the discretization allows, making sure that the chain of points approximates a continuous curve. 

Thus, the definition of a distance on a causal set can be summarized as follows:
\beq
\tau (p,q) = \max \{ n \mid \exists\, r_1, . . . , r_n \colon p \prec r_1 \prec . . . \prec r_n \prec q \}\;.
\eeq
Numerical studies (see Refs \cite{length1} and \cite{length2})  confirmed that in the case of a Poisson distribution of points on a Lorentzian manifold there is, in fact, a close correlation between the Lorentzian distance defined in the usual way and the one defined in terms of chains of points as above. The coefficient of proportionality, however, is still unknown. 

There was recent work done in \cite{Wallden} where they tried to define distances between two space-like events. This was done under the assumption that pairs of points in interest are close enough to each other that spacetime in that region looks flat. 

First let us describe an approach that \cite{Wallden} discarded and explain their reason, and then, afterwords,  describe the approach that they later advocated. Suppose $p$ and $q$ are spacelike-separated events separated a distance $\sigma$ apart. Choose the coordinate system so that they both lie on $x$-axis and origin is in the middle. Thus, $p = (-\sigma/2, 0, 0, 0)$ and $q= (\sigma/2, 0, 0, 0)$. Suppose there are points $r$ and $s$ satisfying $r \prec p \prec s$ and $r \prec q \prec s$. 

The fact that $r \prec p$ means that 
\beq
r_t^2- (r_x+\frac{\tau}{2})^2  - r_y^2 - r_z^2 > 0\;,
\eeq
and the fact that $r \prec q$ means that  
\beq
r_t^2- (r_x-\frac{\tau}{2})^2  - r_y^2 - r_z^2 > 0\;.
\eeq
Adding the two together gives 
\beq
0 < 2 r_t^2 - 2r_y^2 - 2r_z^2 -(r_x-\frac{\tau}{2})^2 + r_t^2- (r_x+\frac{\tau}{2})^2 = 2 (r_t^2 - r_x^2 - r_y^2 - r_z^2 ) - \frac{\tau^2}{2}\;.
\eeq
This implies
\beq
\sqrt{r_t^2 - r_x^2 - r_y^2 -r_z^2} \geq \frac{\tau}{2}\;,
\eeq
which means that the distance between $r$ and the origin is greater or equal to $\frac{\tau}{2}$ In a similar way, the distance between point $s$ and the origin is also greater or equal to $\frac{\tau}{2}$. 

But in the discussion of timelike distances it was shown that the distance between two causally related events is equal to the length of the longest possible curve between them. Thus, 
\beq
\tau(r,s) \geq \tau(r, 0) + \tau(0,s) \geq \frac{\tau}{2} + \frac{\tau}{2} = \tau\;.
\eeq
At the same time, there is at least one choice of $(r,s)$ satisfying $\tau (r,s) = \tau$, namely, it is 
\beq
r =( - \frac{\tau}{2}, 0 , 0 , 0) \; ; \; s=(  \frac{\tau}{2}, 0 , 0 , 0)\;.
\eeq
This proves that 
\beq
\tau (p,q) = \inf\{\tau(r,s)\mid r\prec p\prec s\wedge r\prec q\prec s\}\;.
\eeq

However, even though this is true in ideal flat Minkowski space, it is no longer true in a discrete case. As a result of the fact that one can perform a Lorentz boost in the $yz$-plane, there are infinitely many choices of $r$ and $s$, all of which are separated by the distance $\tau$. In the discrete case, as a result of random fluctuations, some are separated slightly further away from each other than $\tau$, and some are slightly closer to each other than $\tau$. 

By definition, infimum selects the ones that are slightly closer. But, in light of the fact that there are infinitely many of them, there will be few that are A LOT closer, much like if one is to throw a dice infinitely many times, one might get $100$ heads in a row at some point, with absolute certainty. Now, even one pair of points that is a lot closer than $\tau$ is sufficient for infimum to be a lot smaller as well and thus not a good estimation.

For that reason, in \cite{Wallden} they proposed a different definition: they consider all possible points $s$ that are linked to $p$ and $q$ by a direct link. Here, when $a \prec b$ are related by ``direct link" it is meant that there is no point $c$ satisfying $a \prec c \prec b$.  Thus, 
\beq
\tau(p,s) = \tau(q,s) = 0\;,
\eeq
which implies that $s$ is either $(\tau/2, 0, 0, 0)$ or a Lorentz boost of that in $xy$ plane. For each such $s$, they find a point $r(s)$ such that $r(s) \prec p$ and $r(s) \prec q$ which minimizes $\tau (r, s)$ with respect to that constraint. Then, they average $\tau (r(s),s)$ over all possible $s$. Since they average as opposed to minimize, they count both the cases when the distance is slightly lower than needed and the cases where the distance is slightly larger than needed, which averages out to a correct distance.

\subsection{Vector Fields on a Causal Set}

Since in the causal set framework the spacetime is no longer viewed as a manifold, a tangent bundle is not well defined. This is also seen from the fact that Lorentzian indices are needed in order to specify value of gauge field.  Therefore, the definition of gauge field needs to be replaced with something else. The latter should satisfy two properties:

1) It is well defined for the case of a manifold, and in the latter case it has a known correspondence to vector fields

2) It continues to be well defined if a given set is not a manifold.  

First, consider a simple case where any pair of points $p \in {\cal M}$ and $q \in {\cal M}$ are connected by a unique geodesic. In this case, there is a one to one correspondence between functions $V \colon {\cal M} \rightarrow T{\cal M}$ and $v \colon {\cal M} \times {\cal M} \rightarrow \mathbb{R}$, as long as both sets of functions are restricted to infinitely differentiable ones:
\beq
V^{\mu}(p) = g^{\mu\nu}\, \frac {\partial v(p,q)}{\partial q^{\nu}}
\eeq
and 
\beq
v(p,q) = \int_{\gamma(p,q)}  g_{\mu \nu} V^{\mu}\, \dd x^{\mu}\;,
\eeq
where $\gamma(p,q)$ stands for the geodesic segment connecting the two points.

Since $V \colon {\cal M} \rightarrow T {\cal M}$ and $v \colon {\cal M} \times {\cal M} \rightarrow \mathbb{R}$ are in one to one correspondence, it is possible that the key player is $v \colon {\cal M} \times {\cal M} \rightarrow \mathbb{R}$ rather than $V \colon {\cal M} \rightarrow T {\cal M}$ and we were fooled by the one to one correspondence thinking that it is $V \colon {\cal M} \rightarrow T {\cal M}$. If such is the case,  $v \colon {\cal M} \times {\cal M} \rightarrow \mathbb{R}$ can be generalized from the manifold to a causal set in a straightforward manner since, being a real valued function, it makes no reference to Lorentzian indices or any other indicator of manifold structure. 

For the case where there is more than one geodesic that connects some pairs of points, things are no longer as simple. Consider, for example, a cylinder $\mathbb{R} \times [0,1]$ where $0$ and $1$ are identified, and distances are defined in Euclidian sense. Suppose a vector field $V$ is defined as follows: 

1) $V_x=0$ for $y < 1/4$ or $ y \geq 3/4$

2) $V_x = x$ for $1/4 \leq y < 3/4$

3) $V_y = 0$ everywhere.

There is more than one way of connecting pairs of points via a geodesic. To remove the ambiguity, establish a rule that for any pair of points $p$ and $q$ the geodesic is selected whose length is the closest to $1$. This means that if $p$ and $q$ are very close to each other, the geodesic will be selected that circles the cylinder once. It is easy to see that in this case 
\beq
v(p, q) \approx \frac{1}{2} (q_x - p_x)\;,
\eeq
implying that 
\beq
\frac{\partial}{\partial q_x} v(p,q) \vert_{p=q} = \frac{1}{2} \neq V_x\;.
\eeq

One way to establish a one-to-one correspondence while avoiding that difficulty is to make the following criteria:

1) Suppose that there exist at least one disc $D$ such that 

a) $p$ and $q$ are both elements of $D$

b) If $r$ and $s$ are any two elements of $D$ then there is only one way they can be connected by a geodesic segment, $\gamma (r, s)$ in such a way that it does not escape the interior of $D$, that is, $r, s \in \gamma (r,s) \subset D$.  

In this case, $\gamma (p,q)$ is used for the definition of $v(p,q)$

2) Suppose $p$ and $q$ are so far away from each other that the above-mentioned $D$ does not exist. In this case, $v (p, q) =0$

Thus, the expression of $v$ as an integral of $V$ is true only locally; if the points are spaced far enough from each other that one has to make a choice of what path to take, none of these paths would give $v$. On the other hand, the expression of $V$ as a derivative of $v$ is always true since the latter, by definition, is local. 

This means that $v$ is more fundamental than $V$ since the latter is expressed in terms of the former and not the other way around. This confirms that the vector field is really $v$ and we were mistaken in thinking it was $V$, which would allow the generalization of vector field on non-manifoldlike situations.

For the case of Minkowski space, the disc can be replaced by an Alexandrov set, which sets the following correspondence:

DEFINITION: Let $\cal U$ be an open subset of a manifold $\cal M$. $\cal U$ is said to be UNIQUELY CONNECTED if for any pair of points $r \in {\cal U}$ and $s \in {\cal U}$ there exist a unique geodesic segment $\gamma (p,q)$ which is contained in $\cal U$. In other words, there exist a unique geodesic segment $\gamma$ satisfying $\{ p, q \} \subset \gamma ([ \gamma^{-1} (p), \gamma^{-1} (s)]) \subset {\cal U} $ . Such $\gamma$ is denoted as $\gamma_U (p,q)$

DEFINITION: Let $r$ and $s$ be two elements of $\cal M$. They are said to be UNIQUELY CONNECTED if the following is true:

1) There exist at least one uniquely connected Alexandrov set $\alpha (p,q)$ which contains both $r$ and $s$

2) If $\{ r, s \} \subset \alpha (p_1, q_1) \cap \alpha (p_2, q_2) $ then $\gamma_{alpha (p_1, q_1 )} (r, s) = \gamma_{\alpha (p_2, q_2)} (r,s)$

The above common value of $\gamma_{\alpha (p,q)} (r, s)$ is denoted by $\gamma_(r, s)$

DEFINITION: Let $\cal M$ be a Lorentzian manifold, and let $V \colon {\cal M} \rightarrow T {\cal M}$ be a vector field. Then $f_v \colon {\cal M} \times {\cal M} \rightarrow \mathbb{R}$ is a function defined as follows

1) If $p$ and $q$ are uniquely connected, then 
\beq
v(p,q) = \int_{\gamma (p,q)} V_{\mu}\, \dd x^{\mu}\;.
\eeq
2) If $p$ and $q$ are NOT uniquely connected, then $v(p,q)=0$. 

As was mentioned earlier, the fact that $V$ is always a derivative of $v$ but $v$ is not always an integral of $V$ implies that $v$ is more fundamental than $V$, which is what is desired since the former does not appeal to manifold structure while the latter does. 

This means that, for example, it is possible to set $v(p,q)=1$ instead of $v(p,q)=0$ for points that are not uniquely connected, and this will amount to a physical field having different values. However, despite the fact that the value of the field will, in fact, be different, the value of Lagrangian density will still be the same. In other words, Lagrangian density is symmetric under the variation of the value of the field for pairs of points that are not uniquely connected. 

In performing the path integral, all possible two point functions $v \colon {\cal M} \times {\cal M} \rightarrow \mathbb{R}$ will be taken into account, not just the ones that can be derived from $V$. After all, in the general case there is no such thing as $V$, so it should not factor into the definiton of path integral. This means that different $v$-s that are related to each other by symmetry will be added to each other, resulting in overcounting.

However, Fadeev-Popov ghosts are NOT introduced to deal with that overcounting. Instead, the infinity is avoided by the fact that the range of integration is limitted to a fixed tunnel (see sec 2.6 and 5.4). This tunnel corresponds to the ``collapse" of the quantum system into a roughtly defined trajectory up to smallest classical scale. Since all fields, including the vector field, are ``collapsed" to the range defined by classical trajectory, overcounting will not produce an infinite result. 

In terms of more rigorous math, a generalized notion of a vector field is defined to be a differentiable function $g \colon {\cal M} \times {\cal M}$. In the case of a differentiable manifold, to every differentiable function $g \colon {\cal M} \times {\cal M} \rightarrow \mathbb{R}$, corresponds a vector field $v_g \colon {\cal M} \times {\cal F} \rightarrow \mathbb{R}$ such that, for every $p \in {\cal M}$ and $f \in {\cal F}$,
\beq
v_g (p, f) = (\partial^{\mu} g\,  \partial_{\mu} f) \vert_p
\eeq
and, in case that any pair of points are connected by unique geodesic, to any vector field $v$ defined in a usual way for a manifold, the corresponding holonomy $g \colon {\cal M} \times {\cal M} \rightarrow \mathbb{R}$ is given by
\beq
g(p,q) = \int_{\gamma (p,q)} v(r, \dd\gamma)\,\dd r\;,
\eeq
where $\dd\gamma$ is a vector defined as 
\beq
\dd\gamma(f) = \frac{\dd(\gamma(f(t))}{\dd t}\;.
\eeq

\subsection{Lack of Symmetry and Definition of Vectors}

In the last section it was stated that due to the lack of manifold structure,  the notion of tangent vector can not be used, which is the ultimate reason why an alternative definition of vector, as a two-point function, is introduced. This raises a question: what about a square lattice? Strictly speaking it is not a manifold, yet one can easilly define Lorentzian indices corresponding to its discrete coordinates.

At first one might say that it is not the issue of discreteness by itself, but rather the issue of lack of coordinates that forced us to abandon the standard definition of a vector. However this is not quite true either. Strictly speaking, it might still be possible to ``label" different elements of a causal set, and adjust these labels in a way that the causal relations correspond to what the labels would tell us -- or, if that proves to be too difficult, adjusting the metric tensor link-wise might be another avenue. On the other hand, in the case of a regular manifold, a lot of differential geometry books define vectors in a coordinate-independent way in order to emphasize their pure geometric nature. Thus, it is important to look at the real reason, beyond the presence or lack of labels, as to why in the case of causal sets a new definition is needed.
 
In the standard theory of manifolds, a vector at a point is defined as a linear operator on a set of differentiable functions. That is, if ${\cal M}$ is a manifold and ${\cal F}$ is a set of differentiable functions ${\cal M} \rightarrow \mathbb{R}$, then a vector field is a function $v \colon {\cal M} \times {\cal F} \rightarrow \mathbb{R}$ such that for any point $p \in {\cal M}$ and any two functions $f$ and $g$, and any real numbers $k_1$ and $k_2$,
\beq
v(p, k_1 f + k_2 g) = k_1 v(p, f) + k_2 v(p, g)\;.
\eeq
and
\beq
v(p,fg)=f(p) v(p,g)+g(p)v(p,f)
\eeq
Intuitively, a vector $v^{\mu} \colon {\cal M} \rightarrow T{\cal M}$ corresponds to the differential operator.
\beq
v(p,f) = v^\mu(p)\,\partial_\mu f\vert_p\;.
\eeq
On the other hand, a coordinate system can be thought of as a set of $d$ functions, $x^{\mu} \colon {\cal M} \rightarrow \mathbb{R}$. Here, $\mu = 0$, ..., $d-1$ are simply the names of the functions and are not to be confused with coordinates. At every point $p \in \mathbb{R}$, $v^{\mu} (p)$ can be defined as
\beq
v^\mu(p) = v(p,x^\mu)\;.
\eeq

The reason that the above definition of vector field, despite being coordinate-independent, does not work in general is that it has too few degrees of freedom. When Lorentz index is used, one, strictly speaking, looks only at the derivatives along the coordinate axes, and nowhere else. Now consider the following example. Suppose a function $f$ is defined as follows: 

1) $f(t,0,0,0) = t$;

2) $f(t,x,y,z) = 0$, if any of $x$, $y$ or $z$ are non-zero;

In this case, $\partial^{\mu} \phi\, \partial_{\mu} \phi = 1$.

On the other hand, suppose that instead of singling out the $t$ axis, we single out the$t'$ axis, which is tilted relatively to the $t$ axis. In other words,

1) $g(t,x,0,0) = t$, if $x = kt$;
2) $g(t,x,y,z) = 0$ if either $x$ is NOT equal to $kt$ OR if either of $y$ or $z$ is non-zero.

In this case, $\partial^{\mu} \phi\, \partial_{\mu} \phi=0$.

This reveals that $\partial^{\mu} \phi\, \partial_{\mu} \phi$ is NOT truly Lorentz covariant. The reason we don't run into this issue is that in case of differentiable functions we HAPPEN to get the same result after performing the rotation. This, of course, is a definition of symmetry. 

Thus, the moral of the story is that the fact that the expression involving Lorentz indices makes sense is a consequence of symmetry. Once the symmetry no longer exists, as in examples presented above, the contraction of Loretnz indices is no longer physical since it begins to depend on our choice of labels.

This applies not only to derivatives but to anything that has a Lorentzian index. When a vector is denoted by $V^{\mu}$ what is really meant is that once FOUR aspects of that vector are specified, namely its projections on the four axes, ALL other aspects are specified as well -- namely the projection on any other axis is a linear combination of the projections on the four given axes. This is ultimately why $V^{\mu}$ refers to only FOUR components. Now, that relation between projections is called symmetry, thus once there is no more symmetry this is no longer true. 

In case of a causal set, the symmetry is lacking due to the fact that different locations in a causal set can not be ``perfectly matched" on each other. Thus, based on what we have just learned, due to lack of symmetry the vector-based expressions are meaningless. 

True, a regular lattice does not obey the rotational symmetry either. In fact, that is the main reason why it is not an acceptable model. However, due to the fact that it does possess translational symmetry, there is some rough correspondence between the two on the scale of large numbers of points, which allows us to recover the notion of vector field at that scale. The definition of derivative in terms of neighboring points is nothing but an extrapolation of what is expected to happen for pairs of points that are close on the scale in which we live in, but are still separated by many points and thus are not true ``neighbors". Its definition on the few-point scale is simply an extrapolation of the latter. 

On the smaller scale, on the other hand, the regular lattice structure is our enemy rather than our friend, since it is not symmetric under rotation or Lorentz boosts. True, random Poisson distribution is not symmetric under these transformations either. But while Poisson distribution violates these symmetries stochastically, the regular lattice structure violates them in a consistant matter, which makes the latter worse. On the language of symmetries, this means that translational and rotational symmetries are not compatible with one another for discrete structures. Poisson distribution, by violating both on a small scale, allows to maintain both on larger scale without putting one above another. On the other hand, regular lattice, by never violating translational symmetry, violates rotational symmetry more consistently.

This raises a question: since there are no longer rotational symmetries, how is Lorentz covariance defined, especially since causal set theory is advertized as the one that is loyal to the latter? The answer is that Lorentz covariance is now defined as a statement that light cone causal relations are the only defining features of geometry. Inside the Lorentzian manifold, this is equivalent to saying that speed of light is covariant which is fundamental tenat of relativity. At the same time, once phrased in a language of partial order, this statement continues to be well defined for a general causal set. 

In the language of symmetries, one can say that while there is no longer active Lorentzian symmetry, there is still a passive one. Passive symmetry is defined in terms of labeling of points. While there is no reason to do that, nor is it done in the theory, one can, if they want, label points with local coordinates as well as choose the values of metric 2-tensor in such a way that causal relations derived from these will match the aforegiven causal structure of the set. Since this labeling is not implimented in the theory, the theory is covariant with respect to the choices of such labelings, which means that it possesses passive symmetry. 

The core of causal set theory is to abandon such labels altogetgher as opposed to claiming the covariance with respect to relabelings. However, this would not have been possible if there were no passive symmetry. Thus, lack of labels might be a new, causal set based, version of the definition of passive symmetry. In this language, it is easy to see that as long as no geometric structures besides causal relations are being used, any and every theory possesses passive symmetry by default. 

That symmetry, however, is neither rotation nor translation; but since its manifold analogue is Lorentz symmetry it is still defined as such. For the latter reason, while the rotational covariance is taken care of passively, the translational one is not, which means that making causal set manifold-like remains to be one of the most serious problems of the theory.

\subsection{Causal Set Growth by Transitive Percolation}

Before moving on to the novel way of doing dymamics on a causal set, let's get some taste of the more standard, stochastic-based dynamics, and prove one of its basic results, that there will be a chain of infinitely many ``big bangs" or ``posts". 

Consider a process of steady growth of the causal set. Initially, the entire causal set consists of only one point, point ``1". Then another point, point ``2", is added. Then a random decision is made between two options: point $2$ is either causally after point $1$ or the two points are unrelated (here, the term ``unrelated" means that neither point is either before or after the other, which is equivalent to a statement that the two points are spacelike-separated). The former option is selected with probability $p$ and the latter with probability $1-p$.   After that yet another point, point ``3", is added. Again, $3 \succ 2$ is set with probability $p$, or $3$ and $2$ are left causally bun-related with probability $1-p$. If it happens that $1 \prec 2$ and $2 \prec 3$ then $1 \prec 3$ is enforced by transitivity. Otherwise, the causal relation between $1$ and $3$ is selected by the same random process: $1 \prec 3$ with probability $p$ and they are un-related with probability $1-p$. Then point $4$ is added and its causal relations to the existing points are determined by the same random procedure, and so on. Whenever point $n$ is added, all points from $i=n-1$ to $i=1$ are viewed in decreasing order. For each  $i<n$, the question is asked whether or not there is any point $j$ for which it is already established that $i \prec j$ and $j \prec n$. If so, $i \prec n$ is enforced. Otherwise, the relation $i \prec n$ is established with probability $p$ and while $i$ and $n$ are left unrelated with probability $1-p$. After repeating that procedure starting from $i=n-1$ and finishing with $i=1$, $n$ is increased by $1$ and the same procedure is repeated. 

A point is said to be a ``post" if it is causally related to every single other point in a causal set. Consider, for example, a situation right after the second point is added, i.e., when entire causal set consists only of two points. If these two points are causally related, then they are both posts. Point $1$ is a post because, by being ``before" point 2 it is formally ``before every single point in the set", and point $2$ is a post because, by being ``after" point $1$ it is formally ``after every single point on a set. On the other hand, if these two points are causally un-related, then neither of them are posts. 

In general, if a causal set is homeomorphic to subset of $\mathbb{N}$, then every single element of that set is a post. On the other hand, if it is homeomorphic to the region of a square lattice, $\mathbb{N}^2$ with causal relations 
\beq
(m_1,n_1) \prec (m_2,n_2) \Leftrightarrow (m_2 - n_2)^2 - (m_1 - n_1)^2 < 0\;,
\eeq
then whether or not a given point is a post depends on what subset it is homeomorphic to. For example, if it is homeomorphic to $\{ (m, n) \} \vert m \vert < \vert n \vert $, then its only post is $(0, 0)$. On the other hand, if it is homeomorphic to $\{ (m, n) \vert 1 < m < M , 1 < n < N \}$ then it has $2M$ posts, and they are points of the form $(m, 1)$ and $(m, N)$. 

In the remainder of this section it will be shown that if a causal set is growing according to the transitive percolation model, then infinitely many posts will be generated with absolute certainty. 

Consider again the situation of points being added one by one, and denote each point by a number. At the moment when causal set consist of exactly $n$ points, let $a_n$ be the number of elements of a causal set that are in the causal past of the last point added (and it is assumed that every point is in a causal past of itself). Thus, $a_n =1$ if and only if $n$-th point is a post at the moment when it was first added. However, this does not imply that $n$-th point will continue to be a post when more points are added; in fact probability of that is rather small. But, the only points of interest are the ones that survive as posts to the very end. In other words, point $n$ is selected if, in addition to the fact that $a_n=1$, all points that are added later will end up being causally after $n$-th point.  

Thus, in order to prove that there are infinitely many posts, two things need to be shown:

1) With absolute certainty, $a_n=1$ holds for infinitely many $n$. Thus, there are infinitely many points that start out as posts at the moment of their creation. 

2) If a given point is a post at the moment of its birth, there is non-zero probability that it will continue to be a post forever. 

Part 2 implies that the number of posts is a finite fraction of the number of points that start out as posts. Furthermore, part 1 implies that the number of points that start out as posts is infinite. Thus, part 2 together with part 1 implies that the number of points that survived as posts is also infinite. 

Since part 2 is easier, it will be done first.

\noindent LEMMA: if $0<c_k<1$ for all $k$ then, for all $n \geq 2$ 
\beq
\prod_{k=1}^n (1-a_k) > 1 - \sum_{k=1}^n a_k\;.
\eeq
PROOF: It is obvious that the above is true for $n=2$:
\beq
(1-a_1)(1-a_2)=1-a_1-a_2+a_1 a_2 > 1 - a_1 - a_2\;.
\eeq
Now, suppose the above is true for some $n$. Then 
\bea
& &\prod_{k=1}^{n+1}(1-a_k)
= (1-a_{n+1}) \prod_{k=1}^n (1-a_k) > (1-a_{n+1}) (1-\sum_{k=1}^n a_k) \nonumber \\
& &\kern60pt= 1- \sum_{k=1}^n a_k -a_{n+1} + a_{n+1} \sum_{k=1}^n a_k = \nonumber \\
& &\kern60pt= 1- \sum_{k=1}^{n+1} a_k + a_{n+1} \sum_{k=1}^{n+1} a_k\;.
\eea 
Since $a_k >0$ the above naively implies that 
\beq
\prod_{k=1}^n (1-a_k) > 1- \sum_{k=1}^{n+1} a_k\;.
\eeq
Thus, by induction, the hypothesis is true for all $n>2$.\hfill QED

\noindent LEMMA: If a given point, $n$, started out being a post, it will survive as a post with non-zero probability.

\noindent PROOF: Let $P_k$ be a probability that point $n$ survived as a post through $k$ steps. In other words, at the point when the causal set has $n+k$ points, point $n$ is still a post. Furthermore, let $p_k$ be a probability that point $n$ will survive as a post after the addition of point number $n+k+1$ provided that it was a post right before that point was added. Thus,
\beq
     P_k = \prod_{i=1}^{k-1} p_i\;.
\eeq
 To compute $p_k$, notice that $1-p_k$ is the probability that point $n$ will stop being a post after the addition of point $n+k+1$. The latter is satisfied if and only if the point $n+p+1$ is NOT causally related to any of the points between $n+1$ and $n+k$, inclusively. The probability of that is $1-p_k = (1-p)^k$, and therefore
\beq
     p_k = 1- (1-p)^k
\eeq
and
\beq
     P_k = \prod_{i=1}^{k-1} (1-(1-p)^i)\;.
\eeq
Let $j$ be some number satisfying $1<j<k$. Then, from the previous lemma, 
\bea
& &P_k = \prod_{i=1}^{j-1} (1-(1-p)^i) \prod_{i=j}^{k-1} (1-(1-p)^i) > \prod_{i=1}^{j-1} (1-(1-p)^i) (1-\sum_{i=j}^{k-1} (1-p)^i ) \nonumber \\
& &\kern18pt= \Big( \prod_{i=1}^{j-1} (1-(1-p)^i) \Big) \Big(1-\frac{(1-p)^j-(1-p)^k}{p} \Big)\;.
\eea

Now, select 
\beq
j = 1+\Big[ \log_{1-p} \frac{p}{2} \Big]\;.
\eeq
This implies 
\beq
     1-\frac{(1-p)^j - (1-p)^k}{p} \geq \frac12\;.
\eeq
Therefore, 
\beq
     P_k > \prod_{i=1}^{j-1} (1-(1-p)^i)\;,
\eeq
or, substituting the above-given value of $j$, 
\beq
     P_k > \prod_{i=1}^{[\log_{1-p} \frac{p}{2}]} (1-(1-p)^i)\;.
\eeq
Thus, the probability that point $n$ will survive as a post indefinitely satisfies the inequality
\beq
     P_{\infty} \geq \prod_{i=1}^{[\log_{1-p} \frac{p}{2}]} (1-(1-p)^i)\;.
\eeq
In other words, it is bounded below by a non-zero constant, as desired.\hfill QED

\noindent LEMMA: If infinitely many points started out as posts, then infinite subset of them will indefinitely continue to be posts.

\noindent PROOF: Let the candidates for post be $a_1, a_2, . . . $. For any given $k$, the probability that NONE of the points $a_j$, $j>k$ will survive as posts is 
\beq
\lim_{n \rightarrow \infty} (1-P_{\infty})^n = 0\;,
\eeq
since $P_{\infty}>0$. 

Thus, for any given $k$ there exists, with absolute certainty, at least one $j$ for which $a_j$ survives as a post; call it $j_1$. By setting $k=j_1$ one finds that there exists, with absolute certainty, a number $j_2$ such that $a_{j_2}$ is a post. Likewise, by setting $k=j_2$ there exists with absolute certainty a $j_3$ such that $a_{j_3}$ is also a post, etc. This implies that infinitely many points survive as posts in the case $n \rightarrow \infty$.\hfill QED

Since it was just shown that if infinitely many points start out as posts then an infinitely large subset of them will survive as posts, the only task right now is to prove that the former is, indeed, the case. 

For any set $S$, let its future boundary surface, $\partial_{\,\rm future}(S)$ be the set of all elements of $S$ that are NOT before any other elements of $S$; however, there might be elements of $S$ to which they are spacelike separated: 
\beq
\partial_{\,\rm future}(S)= \{ i \in S \mid \forall j \in S\ \neg (i \prec j) \}\;.
\eeq
Furthermore, $S_n$ is defined to be
\beq
S_n = \{1, . . . , n \}\;.
\eeq
Finally, $a_n$ is defined to be the number of elements of the future boundary of $S_n$:
\beq
a_n = \sharp \partial_{\,\rm future} S_n\;.
\eeq
The statement that infinitely many points start out as posts as they appear is equivalent to the statement that $a_n=1$ for infinitely many $n$. 

It is easy to see that if the point $n+1$ is causally unrelated to any of the elements of $S_n$, then $a_{n+1} = a_n +1$. If there is exactly one element of $S_n$ to which the point number $n+1$ is causally related, then $a_{n+1}=a_n$. Therefore, the probabilities corresponding to scenario $a_{n+1}=a_n+1$, $a_{n+1}=a_n$ and $a_{n+1}<a_n$ are given by
\bea
& & {\rm prob}(a_{n+1}=a_n+1)=(1-p)^{a_n} \nonumber \\
& & {\rm prob}(a_{n+1}=a_n)= a_np\,(1-p)^{a_n-1} \nonumber \\
& & {\rm prob}(a_{n+1}<a_n)= 1-(1+(a_n-1)p)(1-p)^{a_n-1}\;.
\eea
It is easy to see that, as long as $p>0$, ${\rm prob}(a_{n+1}<a_n)$ approaches $1$ as $a_n$ approaches infinity. This means that there exist a fixed number $A(p)$ such that ${\rm prob}(a_{n+1}<a_n)>2/3$ whenever $a_n>A(p)$. 

\noindent LEMMA: if $A(p)$ is defined as above, and $a_n>A(p)$ for some $n$ then, with absolute certainty, there will be an $m>n$ such that $a_m \leq A(p)$. 

\noindent PROOF: Suppose, along with random assignments of $a_i$, there are also random assignments of $b_i$, and each basic step of that process is defined as follows:

1) Set $i=n$ and $b_n = a_n$;

\noindent Then keep repeating 2, 3 and 4 over and over:

2) Use the prescription above to randomly select $a_{i+1}$ based on the information about $a_i$

3) Define $b_{i+1}$ based on the following rules:

a) If $b_i \leq A(p)$ then $b_{i+1} = b_i$

b) If $b_i >A(p)$ and $a_{i+1}=a_i +1$ then $b_{i+1} = b_i +1$

c) If $b_i >A(p)$ and $a_{i+1}=a_i$ then $b_{i+1} = b_i$

d) If $b_i >A(p)$ and $a_{i+1}<a_i$ then $b_{i+1}$ can be either $b_i-1$ or $b_i$ or $b_i+1$ with the following probabilities:
\bea
& &{\rm prob}(b_{i+1}=b_i +1)\vert_{a_i, a_{i+1}}  =  \frac{1/3 - (1-p)^{a_n}}{1-(1+(a_n-1)p)(1-p)^{a_n-1}} \nonumber \\
& &{\rm prob}(b_{i+1}=b_i) \vert_{a_i, a_{i+1}} = \frac{1/3 - a_np(1-p)^{a_n-1}}{1-(1+(a_n-1)p)(1-p)^{a_n-1}} \nonumber \\
& &{\rm prob}(b_{i+1}=b_i -1) \vert_{a_i, a_{i+1}} = \frac{5/3 - (1+ (a_n -1)p)(1-p)^{a_n-1}}{1-(1+(a_n-1)p)(1-p)^{a_n-1}}\;;
\eea
4) Increase $i$ by $1$ and go back to step $2$.

It is easy to see that if no information is available about $a_i$, then $b_i$ obeys the following rules of random walk:

1) $b_n = a_n$;

2) $b_i \leq A(p) \Rightarrow b_{i+1} = b_i$;

3) $b_i > A(p) \Rightarrow p(b_{i+1} = b_i+1)
= p(b_{i+1}=b_i) = p(b_{i+1} = b_i-1)=1/3$.

Furthermore, by induction it can be shown that if $a_j>A$ for all $j$ satisfying $n \leq j \leq i$, then $b_i \geq a_i$. This would trivially imply that $b_i >A$. By contradiction, this implies that if $b_i =A$ for some $i$, this means that $a_j \leq A$ for at least one $j$ satisfying $n \leq j \leq i$. Thus, it has to be shown that, with absolute certainty, there is, indeed, $i >n$ for which $b_i =A$. 

For any given $b$, let $f(b)$ be the probability that an above-described random process started at $b_i = b$ will generate $b_j = A$ for some $j>i$. The process of going from $b$ to $A$ can be separated into two parts: first part consists of a very first step, which has only 3 possible outcomes: $b+1$, $b-1$ or $b$, and the second part is a remainder of a journey to $A$. From this construction, it is easy to see that probabilities obey the following relation:
\beq
f(A(p)) = 1 \; ;\ \forall b > A\;,\ (f(b) = \frac13\, (f(b+1)+f(b)+f(b-1)))\;.
\eeq
The rearranging of the above equation is
\beq
f(A(p)) = 1 \; ;\ \forall b > A(p)\;,\ (f(b)-f(b-1) = f(b+1)-f(b))\;.
\eeq
This implies that, for some $k$, 
\beq
\forall b > A(p)\;,\ (f(b) = f(A(p)) + k(b-A(p)))\;.
\eeq
However, since $f(b)$ represents probability, which means that it has to be in the range between $0$ and $1$. This implies that
\beq
k = 0 \; ;\ \forall b > A (f(b)=f(A(p)))\;.
\eeq
Thus, the fact that $f(A(p))=1$ implies that  
\beq
\forall b \geq A(p)\;,\ (f(b)=1)\;.
\eeq
This means that, with absolute certainty, there will be $i>n$ such that $b_i =A$. Now, if $a_j>A$ is to hold for all $j$ satisfying $n<j<i$, then, by induction, $b_j \geq a_j$ would hold for all $j \leq i$. In particular, this would imply that $b_i \geq a_i >A$, which would contradict $b_i =A$. So, by contradiction, $a_j =A$ for at least one $j$ satisfying $n<j \leq i$, which proves the statement of the lemma. \hfill QED

\noindent LEMMA: In the above-described random process starting from any arbitrary value of $a_1$, $a_i=1$ holds for infinitely many $i$-s, with absolute certainty. 

\noindent PROOF: The first step is to show that $a_i \leq A(p)$ for infinitely many values of $i$. This can be seen as follows: if there exists $n$ such that $a_i \leq A(p)$ for every single $i>n$, the answer is self-obvious. So assume otherwise: for every single $n$ there exists at least one $i>n$ for which $a_i > A(p)$. In the previous lemma it was shown that whenever $a_i > A(p)$, then, with absolute certainty, there exists a $j>i$ such that $a_j \leq A(p)$. Since $i>n$, the fact that $j>i$ implies that $j>n$. This shows that, with absolute certainty, for every single $n$ there exists $j>n$ such that $a_j \leq A(p)$. This means that there are infinitely many such $j$.

The fact that there are infinitely many such $j$ means that they form a sequence $j_1$, $j_2$, .... I claim that $a_{j_k + 1}=1$ for at least one $k$. First assume that such is not the case. In this case infinitely many statements have to simultaneously be true, where statement $S_k$ implies that $a_{j_k+1}>1$.  Each of these events occurs with probability $1-p^{a_{j_k}} \leq 1-p^{A(p)}$. Since $1-p^{A(p)}$ is a fixed number, the probability of infinitely many of such events occurring is $0$. Thus, with absolute certainty, this is bound to fail for at least one $k$. At the point of ``failure", by definition, $a_{j_k +1} =1$.

The above statement can be weakened: the statement ``$a_{j_k+1}=1$ for at least one $k$" can be replaced with ``$a_j=1$ for at least one $j$". Now, if that process is started with $j+1$ it will similarly show that $a_k=1$ for at least one $k>j$. Then, again, the imaginary starting point can be moved to $k+1$ which will show that $a_l=1$ for at least one $l>k$, and so forth. This will generate an infinite sequence $k_1, k_2 , . . .$ for which $a_{k_i}=1$.\hfill QED

\noindent THEOREM. Transitive percolation generates infinitely many $n$ with absolute certainty. 

\noindent PROOF. Every single $k$ for which $a_k=1$ corresponds to a post. Thus, the fact that there are infinitely many such $k$ shows that infinitely many points are starting out as posts. Furthermore, it was previously shown that if infinitely many points start out as posts, then their infinitely large subset will survive as posts. Thus, infinitely many posts will be generated. \hfill QED

\subsection{Generating Causal Sets: Using Statistics or Lagrangians?}

In the last section, it was shown that if causal set was generated according to the transitive percolation model, this would imply infinitely many ``big bangs" in the form of posts. While the particular model presented is certainly not the only one that has been tried for causal sets (albeit the most common and studied one) other models that have been tried still lead to similar conclusions.

It is very likely that this is an obstacle that prevent causal set from being manifold-like, especially since much of numerical work predict posts to be spaced close to each other. It is possible to hope that by setting the probability of causal relation to be extremely low, posts will not be so close, and perhaps the manifold structure will appear in the regions between them. 

Furthermore, if gravity can indeed be generated statistically, perhaps there is connection between posts predicted on statistical bases and the successive big bangs due to universe collapse. However, while the above possibilities can not be completely ruled out, extensive numerical work shows that such does not happen. Again, it is possible that there might be some other stochastic processes that do generate desired manifold, but so far such was not done. On the contrary, the results obtained up till now indicate that as the size of causal set grows to infinity, the probability of it being manifold-like goes down to $0$.  

I believe that the reason for this problem is that typically physics starts from experimental data and then the theory in one way or the other adjusts to that data, even though it pretends that it doesn't. This did not happen for sequential growth model. In order for the stochastic process to have experimental support, one has to first find all of its results and then verify that these results match the observations. While, strictly speaking, nothing rules out the possibility that such might happen at some point in future, so far it hasn't, and it would be very risky to try to stick to a model on a mere basis of the hope that one would get lucky at some point.  

However, there were examples where, as risky as that might have been, the theory ended up being successful despite the fact that its starting point was not motivated by experiments. One of the prime examples of this is the classical thermodynamics, where $PV=nRT$ based on statistics and momentum conservation alone. The spirit of causal set theory is that the spacetime itself is broken into atoms, much like gas is.  Thus, it is hoped to ``simulate" gravitation stochastically so that one can say that gravitation to stochastic process is the same as $PV=nRT$ to conservation of momentum.

Of course, if one can actually invent a stochastic process that meets the above criteria, it would be a much better and more beautiful theory than the one proposed in this dissertation. The problem, of course, is that so far no one was able to invent such a process. For this reason, the approach proposed in this dissertation directly utilizes Lagrangians and abandons the statistical method altogether. Of course, this might look ugly, since it is similar to a statement that  gas molecules are ``choosing" their velocities in just the right way so as to ``force" $PV=nRT$ to be a statistical average. But, as ugly as it might be, it is a safer option since the Lagrangians used do, in fact, go back to experimental observations, albeit on scales much larger than the ones of a causal set. 

There is, however, a philosophical argument to justify this. Consider, for example, real numbers. At first glance, they appear to be very simple. However, their rigorous definition is not as simple: they are defined as Dedekind cuts of the rationals (here, a cut is defined as a set $S \subset \mathbb{Q}$ such that for any two rational numbers $a$ and $b$, if $a \in S$ and $b<a$ then $b \in S$). While the definition is complicated on the surface, it is no longer questioned once its correspondence to real numbers is understood. This is because we are believers in real numbers and not believers in the cuts. So, as long as cuts ``serve our agenda", it makes perfect sense why they were ``invented".

However, if one hadn't experienced real numbers in everyday life, one would have been forced to be a true believer in cuts, which would lead to a similar question: how come the Dedekind cuts happened to be defined in this particular way rather than some other way? Perhaps, one would have decided to make some random generation of a cut and then show that the specific structure that corresponds to real number is the most probable one. This would lead to disappointment, since such is not the case.   

What the above example teaches is that existence of real numbers is, too, an experimental observation, since the specific definition of Dedekind cut is not the most probable one stochastically. Yet, due to the fact that we have to ``observe" that over and over within seconds, we became so used to that observation that it appears to us as ``math". Similarly, it can be claimed that a causal set, too, is generated by physics rather than by math. But due to the fact that we have lived inside of the causal set for such a long time, we got so used to it that it appears to us as math. 

Since we have only experienced the causal set on a very large scale, the only part that appears to us as math is its large-scale geometry. If, however, we were to live on a microscopic scale for just as long, then what happens on the level of, say, 10 points, would also appear very natural to us and it won't even occur to us to ask whether or not it is the most probable stochastically; we would regard it as ``simple geometry" even though it would have nothing in common with the geometry we are used to.

\subsection{Causal Set as a Static Geometry}

In Section 2.5 it was shown that the transitive percolation model for generating causal sets leads to the unpleasant fact of the presence of too many posts and lack of manifold structure. Then in Section 2.6 it was suggested that perhaps the answer to the puzzle is to replace statistics by a Lagrangian. In this section, a Lagrangian-based model of causal set structure will be proposed that will be used for the dissertation. A necessary component of this model is the ``collapse" of fluctuating causal relations into a well-defined structure. The meat and justification for the specific way of doing that will be extensively discussed in the Appendix. For now, it will suffice to simply outline the setup.

According to the proposed model, both fields and causal relations are undergoing quantum fluctuations within a certain tunnel around their ``central" configurations. The size of that tunnel is roughly at the transition point between quantum and classical scales. For any specified behavior of fields throughout space time, ${\cal F}_0 (s)$, there is a ``neighborhood" $n({\cal F}_0)$ that consists of all possible variations of these behaviors, ${\cal F} = {\cal F}_0 + \delta {\cal F}$, that are regarded as ``similar" to ${\cal F}_0$ by some specified criteria.  

For example, for a scalar field $\phi$, the ``$\delta$-neighborhood" around $\phi (x) = \phi_0 (x)$ is given by 
\beq
n_{\delta}(\phi_0)= \big\{ \phi \colon S \rightarrow \mathbb{R} \mid \forall p \in S\;, \ \vert \phi (p) - \phi_0 (p) \vert < \delta \big\}\;.
\eeq
The definitions for neighborhoods of other fields will be discussed in section 2 of the Appendix.

One important aspect of the theory is that the notion of neighborhoods is also defined for the set of all possible causal relations, i.e., for gravitational fields. That is, for any causal relation $\prec_0$ there is a neighborhood $n(\prec_0)$ consisting of all causal relations, $\prec$, that are ``similar to" $\prec_0$. Again, the actual definition of that neighborhood is left for the Appendix. 

To every $({\cal F}_0, \prec_0)$ corresponds a probability amplitude given by
\beq
Z(\prec_0, {\cal F}_0) = \sum_{\prec \in n(\prec_0)}
\int_{{\cal F} \in n({\cal F}_0)} [{\cal D} {\cal F}]\, \ee^{\ii\,S(\prec, {\cal F})}\;.
\eeq
For the sake of mathematical rigor, the values of ${\cal F}$ can be discretized at every point and the integration over ${\cal F}$ can be replaced with a sum. For example, for the simple case of the universe consisting of scalar field and gravity by itself (in other words, ${\cal F}=\phi$), 
\beq
Z(\prec_0, \phi_0) = (\epsilon)^{\sharp S} \sum_{\prec \in n_{\delta}(\prec_0)}
\sum_{\phi \in n (\phi_0) \; ; \; \forall p (\phi(p)=k\epsilon)} \ee^{\ii\,S(\prec, \phi)}\;,
\eeq
where
\beq
n_{\delta}(\phi_0)= \big\{ \phi \colon S \rightarrow \mathbb{R} \mid \forall p \in S\;, \ \vert \phi (p) - \phi_0 (p) \vert < \delta \big\}\;.
\eeq
The constraint that the relation $\prec$ belongs to $n(\prec_0)$ will provide the background geometry that the theory used to lack. It is important that $\cal F$ is constrained to $n({\cal F}_0)$ as well, since the behavior of $\cal F$ plays a role in Einstein's equation. 

This defines a probability amplitude determined entirely based on a Lagrangian and not on a stochastic process. This by itself by no means implies a manifold-like structure, unless the Lagrangian is designed in a way that it would. However, by restricting fluctuations of gravitational field to a spedified range, the argument against the manifold structure is removed as well, which is a first step along the way.

\newpage
\section{Chapter 3: Type-1 Bosonic Fields}

\subsection{Fields on a Causal Set} 

In order to define Lagrangians it is important to first define fields on a causal set. Since scalar field is defined without reference to manifold structure, it's definition can be borrowed from a regular manifold case, replacing manifold ${\cal M}$ with causal set $S$:

DEFINITION: A real scalar field is a function $\phi \colon S \rightarrow \mathbb{R}$ and complex scalar field is a function $\phi \colon S \rightarrow \mathbb{C}$.

A vector field is trickier to define, and its definition is discussed extensively in section 2.3. Here, it will simply be borrowed:
 
DEFINITION: A vector field on a causal set $S$ is a function $v \colon S \times S \rightarrow \mathbb{R}$
 
Putting these two definitions together, a general definition of a bosonic field in a manifold combines both scalar field, $\phi \colon S \rightarrow \mathbb{C}$ as well as vector field $v \colon S \times S \rightarrow \mathbb{R}$ is as follows:

\noindent DEFINITION: Let $S$ be a causal set. A field on $S$ is either $f \colon S^n \rightarrow \mathbb{R}$ or $f \colon S^n \rightarrow \mathbb{C}$,  where $A \subset \mathcal{P} S$.

The key here is that these fields are all real valued functions, thus are well defined for arbitrary causal set.

\subsection{Type-1 Lagrangian Generators} 

Our next goal is to define Lagrangians for causal sets. In the case of a lattice field theory, the Lagrangian density at any given point is defined in terms of the field at that point as well as at the neighbors of that point, where ``neighbor" is one of the $2d$ points that are directly linked to the point of interest by the square lattice. In the case of a causal set, however, the notion of a neighbor is no longer well defined. 

Even in the special case of a manifold, due to lightcone singularity, the region whose Lorentzian distance to any given point is bounded by $\epsilon$ has infinitely large volume. Since being connected by a ``direct link" is equivalent to having Lorentzian distance equal to Plank scale, this means that any given point is connected by direct links to infinitely many other points, which qualifies all of them as possible candidates for ``neighbors". 

For these reasons the Lagrangian cannot be defined in terms of a linear superposition of some function of all possible neighbors. However, it is still possible to define a ``Lagrangian generator" which will be a function of all possible choices of a few-point collections of neighbors. Since these collections contain more than one point, there is no more need to define any further ``neighbors" of points that are being considered, thus the above difficulties are avoided.  

Furthermore, a standard procedure will be introduced that dictates the definition of a Lagrangian based on the Lagrangian generator. Roughly speaking, that procedure is designed to select ``representative" collections of points that would correspond to the extremum of the Lagrangian generator with respect to some well defined constraints. It is shown by the examples of well known fields that the value of Lagrangian generator at that select set of points is, in fact, proportional to what is usually defined as Lagrangian density.  

Consider the example of electrodynamics. In order to be able to define the Lagrangian, one has to define electric and magnetic field in terms of vector potential. Due to the lack of a coordinate system, it is not possible to write down the definition of electromagnetic field directly. However, some information about electric or magnetic fields, albeit very insufficient, can be extracted from any triple or quadruple of points: an integral of vector potential around the loop defined by these points. In principle, a value of electromagnetic field at a point can be deduced from the values of the field at all the possible loops containing that point. 

Lagrangian generator is defined a function on set of loops rather than set of points, and it will be defined based on circulation of electric and magnetic fields around these loops. After that, a standard procedure will be introduced to go from Lagrangian generator on set of loops to Lagrangian density on set of points. 

The set of such loops is identified with $S^4$. In generalizing from specific example of electromagnetic field to a general case, $S^4$ is generalized to $S^i$. This leads to the following definition: 

\noindent DEFINITION: A Lagrangian generator is a function ${\cal K} \colon D \times {\cal F}_{S^k \rightarrow \mathbb{C}} \rightarrow \mathbb{R}$. Here, $D$ is a subset of $S^i$ for some $i$ and ${\cal F}_{X \rightarrow Y}$ denotes the set of all functions $f \colon X \rightarrow Y$.

It can be shown that as long as constant distances are used as opposed to constant volumes, the set has sufficiently many elements:

THEOREM: Suppose $p \prec q$ and $\tau (p,q)=A$. Then, for any $B<A$ there exist $r \succ p$ and $s \prec q$ such that $\tau (p,r) = \tau (s,q) = B$

PROOF: By definition, there exist a sequence of points $p \prec r_1 \prec . . . \prec r_{A-1} \prec q$ and this is the longest possible future directed chain of points connecting $p$ and $q$. Thus, $p \prec r_1 \prec . . . \prec r_B$ is a chain of points connecting $p$ and $r_B$, of length $B$. Suppose there is another chain of points connecting $p$ and $r_B$, $p \prec a_1 \prec . . . \prec a_C \prec r_B$. Then this gives another chain of points connecting $p$ and $q$, $p \prec a_1 \prec . . . \prec a_C \prec r_B \prec r_{B+1} \prec . . . \prec r_{A-1} \prec q$ The length of this chain is $A+C-B$. Its length has to be smaller or equal to the length of original chain: $A+C-B \leq A$. This implies $C \leq B$. Since this argument works for arbitrary  $p \prec a_1 \prec . . . \prec a_C \prec r_B$, this implies that $\tau (p, r_B) \leq B$. At the same time, $p \prec r_1 \prec . . . \prec r_B$ implies $\tau (p, r_B) \geq B$, which means $\tau (p, r_B) = B$

Identical argument, replacing $\prec$ with $\succ$, $p$ with $q$ and $r_k$ with $r_{A-k}$ implies that $\tau (r_{A-B}, q) = B$.\hfill QED

In case of gauge field,  ${\cal F}_{S^2 \rightarrow \mathbb{R}}$ is a set of all possible gauge field holonomies. Thus, ${\cal K} \colon S^4 \times {\cal F}_{A \rightarrow \mathbb{R}}$ defines a rule by which one can start from gauge field holonomies and get a real number corresponding to each square loop:
\beq
{\cal K}(a; p_1, p_2, p_3, p_4)
= (a(p_1, p_2) + a(p_2, p_3) + a(p_3, p_4) + a(p_4, p_1))^2\;.
\eeq
That real number is interpreted as the flux of the gauge field through that loop. 

The procedure of going from Lagrangian generator to Lagrangian will be puzzling at first, but at subsequent sections it will become clear why it is defined in a way that it is:

\noindent DEFINITION: Let ${\cal K} \colon D \times \mathbb{R}^j \times {\cal F}_{A \rightarrow \mathbb{C}} \rightarrow \mathbb{R}$ be Lagrangian generator; furthermore, suppose that $p \prec q$. Then  
\beq {\cal K}_{\rm max} (f; p,q) = \max \{{\cal K}(T, f) \vert T \in D \cup \alpha(p,q))^i \} \nonumber \eeq
\beq {\cal K}_{\rm min} (f; p,q) = \min \{{\cal K}(T, f) \vert T \in D \cup (\alpha(p,q))^i \} \eeq
\beq\Delta {\cal K} (f; p,q)
= {\cal K}_{\rm max}(f; p,q) -{\cal K}_{\rm min}(f;p,q) \nonumber \eeq
The only Alexandrov sets $\alpha (p,q)$ that are relevant for the Lagrangian density at $x$ are the ones that minimize $\Delta {\cal K}$ with constraints that $x \subset$ and $\tau (p,q) = \tau_0$. The set of these Alexandrov sets $\alpha (p,q)$ corresponds to a set of pairs of points $p \prec q$, and is formally defined as follows: 

\noindent DEFINITION: Let  ${\cal K} \colon D \times \mathbb{R}^j \times {\cal F}_{A \rightarrow \mathbb{C}} \rightarrow \mathbb{R}$ be Lagrangian generator and let $x$ be any element of $S$, then
\beq
 \alpha_{\tau}(x) = \{ p, q \in S \mid p \prec x \prec q \; ; \; \alpha (p,q) = \tau \; ; \; \nonumber \eeq
\beq  \forall r, s (r \prec x \prec s \wedge \alpha (r,s) = \tau \Rightarrow \Delta {\cal K} (f;r, s)
\geq \Delta {\cal K} (f;p,q) ) \}\;.
\eeq
In most cases $\alpha (x)$ is a one-element set, but it is defined as a set in order to accommodate a few cases where it isn't.

Finally, the Lagrangian density based on ${\cal K}$ is given by
\beq
{\cal L}_{\cal K; \tau} (f; x) = \frac{\sum_{(p,q) \in \alpha_{\tau} (x)} {\cal K}_{\rm max} (f; p, q)}{\sharp Q_{\tau} (p)}\;,
\eeq
and the total Lagrangian is given by
\beq
S_{\cal K; \tau} (f)= v_0 \sum_{x \in S} {\cal L}_{\cal K; \tau} (f;x)\;,
\eeq
where $v_0$ is a volume taken up by one point, which is often assumed to be Plank volume.

While the above is done for arbitrary $\tau$, it is assumed that in reality $\tau$ is equal to some constant $\tau_0$. Its value, in principle, can be measured although is currently unavailable.  The Lagrangian is given by 
\beq
S_{\cal K} = S_{\cal K; \tau_0} \;.
\eeq

It is important to stress that in the same way as in usual quantum field theory writing down a theory of a particular field is equivalent to writing down its Lagrangian, in the same way in causal set case, writing down a theory is equivalent to writing down Lagrangian generator. However, in light of the fact that the transition from Lagrangian generator to Lagrangian density is non-linear, it is not correct to simply add all the Lagrangian generators into one single generator. Instead, they have to be listed as sets. Thus, 
\beq
{\cal K} = \{ {\cal K}_1 , . . . , {\cal K}_n \} \;,\quad
{\cal L}_{\cal K} = \sum_{i=1}^n {\cal L}_{{\cal K}_i}\;.
\eeq
For example, in case of charged spin-0 particle interacting with electromagnetic field, the Lagrangian generator is given by ${\cal K}= ({\cal K}_1, {\cal K}_2, {\cal K}_3, {\cal K}_4, {\cal K}_5)$ where
\bea
& &{\cal K}_1 (p_1,q_1; \phi, a)= c_1\vert a(p_1,q_1) \phi(q_1) - \phi (p_1)\vert^2
\nonumber \\
& &{\cal K}_2 (p_2,q_2; \phi, a)= -c_1\vert a(p_2,q_2) \phi (q_2) - \phi (p_2)\vert^2
\nonumber \\
& &{\cal K}_3 (p_3,q_3,r_3,s_3) = c_2(a(p_3,q_3)+a(q_3,r_3)+a(r_3,s_3)+a(s_3,p_3))^2
\nonumber \\
& &{\cal K}_4 (p_4,r_4,q_4,s_4) = -c_2(a(p_4,q_4)+a(q_4,r_4)+a(r_4,s_4)+a(s_4,p_4))^2
\nonumber \\
& &{\cal K}_5 (r, s) = m^2 (r \prec^? s) (\phi(r) + \phi(s))^2 
\eea
(here $r \prec^? s=1$ if $r \prec s$ and  $r \prec^? s=0$ otherwise)

under the following constraints:

1) $p_1 \prec q_1$ ; 

2) $p_2$ and $q_2$ are space-like separated from each other;

3) $p_3$, $q_3$ , $r_3$, and $s_3$ are spacelike separated from each other;

4) $p_4 \prec r_4 \prec q_4$ and $p_4 \prec s_4 \prec q_4$.

\noindent CLAIM: Let $S$ be a Poisson distribution of points in a Lorentzian manifold $\cal M$. Furthermore, suppose there are scalar and gauge fields on $\cal M$, which are differentiable and vary slowly enough. $S$ is viewed as a causal set with respect to causal structure inherited from $\cal M$. Scalar and gauge fields are ``inherited" into that causal set in the form of a scalar field and a holonomy.  If that Poisson distribution is dense enough, then Lagrangian densities of the scalar and vector fields on $\cal M$ as defined in regular quantum field theory approximate the Lagrangian density defined based on Lagrangian generators on $S$. 

The proof of the validity of this claim is the subject of the next two chapters.

\subsection{Type-1 Scalar Field}

Assuming that there is no interaction with electromagnetic field (thus, setting $a(r,s)=1$) Lagrangian generators for the scalar field, corresponding to timelike (t) and spacelike (s) pairs of points are
\bea
& &{\cal K}_t (r_t,s_t; \phi)= c_t (\phi(s_t) - \phi (r_t))^2
\nonumber \\
& &{\cal K}_s (r_s,s_s; \phi)= -c_s (\phi (r_s) - \phi (s_s))^2 
\nonumber \\
& &{\cal K}_m (r, s, \phi)=  \frac{1}{4} m^2 (r \prec^? s) (\phi (r) + \phi(s))^2 
\eea
under the following constraints:

1) $r_t \prec s_t$ ; 

2) $r_s$ and $s_s$ are space-like separated from each other;

Lets first start with mass term, ${\cal K}_m$. It is easy to see that if $\phi$ is roughly the same inside an Alexnadrov set, then the variations of $\frac{1}{4}\,m^2\, (r \prec^? s) (\phi (r) + \phi(s))^2$ are mostly due to $r \prec^? s$ changing from $0$ to $1$. Thus, the desired variation is roughly equal to $\frac{1}{2} m^2 \phi^2$, which corresponds to mass term of Lagrangian. It is also easy to see that if $\phi$ were to vary, the variation would be larger rather than smaller. Thus, indeed, the desired mass term corresponds to minimized variation.

Now lets move on to ${\cal K}_t$ and ${\cal K}_s$. The prescription to go from ${\cal K}$ to ${\cal L}$ requires the minimization of the maximum variation of $\phi$ between two points inside it:
\beq
{\cal L} = \min_{\tau(p,q) = \tau_0}\, \max_{r,s \in \alpha(p,q)}(\phi(r)-\phi(s))^2
\eeq
This means that the problem splits into four parts:
 
1) ${\cal K} = {\cal K}_t$ ; The gradient of $\phi$ is timelike
 
2) ${\cal K} = {\cal K}_t$; The gradient of $\phi$ is spacelike
 
3) ${\cal K} = {\cal K}_s$; The gradient of $\phi$ is timelike

4) ${\cal K} = {\cal K}_s$; The gradient of $\phi$ is spacelike

\noindent CASE 1: ${\cal K} = {\cal K}_t$ ; The gradient of $\phi$ is timelike.

Let's start from ${\cal K}_1 \rightarrow {\cal L}_1$. In other words, the pairs of points of interest are causally related. 

Choose a coordinate system in which the gradient of $\phi$ points in the $t$ direction, while the remaining coordinates $x^k$ vary along spacelike directions normal to $\partial_\mu\phi$. That is, $\partial_k \phi =0$. If $r$ and $s$ are elements of the Alexandrov set $\alpha (p,q)$ then $|r^0 - s^0| \leq q^0 - p^0$, thus $|\phi(r)- \phi(s)| \leq |\phi(q)-\phi(p)|$. Thus, the maximal variation of the scalar field inside the Alexandrov set $\alpha(p,q)$ is given by $|\phi(q) - \phi(p)|$. But due to the assumption that the gradient of $\phi$ points in the $t$ direction, the latter is proportional to $q^0 - p^0$. Thus, the variation of $\phi$ inside the Alexandrov set $\alpha(p,q)$ is minimized whenever $q^0 - p^0$ is. If the constraint $\tau (p,q) = \tau_0$ is imposed, the above minimization implies $p^k - q^k =0$, or in other words the direction of the line passing through $p$ and $q$ should coincide with the direction of the gradient of $\phi$. Thus, 
\beq
{\cal L}_1 =c \min_{\tau(p,q) = \tau_0}\, \max_{r,s \in \alpha(p,q)}(\phi(r)-\phi(s))^2 \approx (\delta t)^2 (\partial_0 \phi)^2 = (\delta t)^2 \partial^{\mu} \phi \partial_{\mu} \phi\;.
\eeq
Even though $(\delta t)^2$ is a finite version of infinitesimal, due to the fact that it is constant, it is absorbed in an overall coefficient, which might be very large and thus produce finite result. Thus, the Lagrangian takes the expected form, 
\beq
{\cal L}_1 \approx d_t\, \partial^\mu \phi\, \partial_\mu \phi\;,
\eeq
where $d_t= c_t (\delta t)^2$.

It would have been simpler if $p$ and $q$ were used instead of $r$ and $s$. After all one can predict ahead of time that the maximization criterion will select $r=p$ and $s=q$. The reason the above expression is left in $r$-$s$ form is that the $p$-$q$ simplification would not work in the case of a spacelike gradient of $\phi$ and for the purposes of consistency of the theory it is preferred that the expressions for timelike and spacelike cases be the same.

It can be shown (see \cite{paper1}) that in dimension $d$ the volume of an Alexandrov set is
\beq
V(\alpha(p,q)) = k_d\, \tau^d(p,q)\;,\qquad
k_d:= \frac{\pi^{(d-1)/2}}{d\,(d-1)\,2^{d-2}\, \Gamma((d-1)/2)}\;.
\eeq
Thus, the above expression can be rewritten as
\beq
|\partial_0\phi| = \bigg(\frac{k_d}{V_0} \bigg)^{\!1/d} \min_{\tau (p,q) = \tau_0}
\,\max_{r,s \in \alpha (p,q)} |\phi(r) - \phi (s)|\;,
\eeq
and its covarient generalization is
\beq
\partial^\mu\phi\, \partial_\mu\phi = \bigg(\frac{k_d}{V_0}\bigg)^{\!2/d}
\min_{V(\alpha(p,q)) = \tau_0}\, \max_{r,s \in \alpha (p,q)}
(\phi(r) - \phi(s))^2\;.
\eeq
\noindent CASE 2: ${\cal K} = {\cal K}_t$; The gradient of $\phi$ is spacelike.

Choose coordinate system in such a way that gradient of $\phi$ points in $x$ direction. It will be shown that, again, the Alexandrov set that minimizes fluctuations is the one whose axis points in the $t$ direction (although there is a lot of freedom of choosing $t$ axis in a way that $x$ axis still coincides with the direction of gradient of $\phi$). 

This can be done in two steps: (a) Show that if the axis of Alexandrov set coincides with $t$ axis then the sought-after fluctuation is equal to $\frac{1}{4} \tau^2 \vert\partial \phi \vert^2$, and (b) Show that in all other cases fluctuation is greater or equal to $\tau^2$.

(a) Assume that the axis of Alexandrov set is parallel to $t$ axis, and suppose $p \prec r \prec s \prec q$. The goal is to show that $\vert r_x - s_x \vert \leq \tau/2 $. Assume otherwise. That is, assume $\vert r_x - s_x \vert > \tau/2 $. Applying that to right hand side of $\tau/2 = \tau - \tau/2$ gives
\beq
\frac{\tau}{2} = \tau - \frac{\tau}{2} > \tau - (s_t - r_t) = \Big(\frac{\tau}{2}-s_t)+ (\frac{\tau}{2}+r_t \Big) = (q_t - s_t) + (r_t - p_t)\;.
\eeq 
The causal relations $p \prec r$ and $s \prec q$ respectively imply $q_t - s_t \geq \vert s_x - q_x \vert$ and $r_t - p_t \geq \vert p_x - r_x \vert$ which gives
\beq \frac{\tau}{2} > \vert s_x - q_x \vert + \vert p_x - r_x \vert \eeq
Implementing the fact that $p_x = q_x =0$ , the above inequality becomes 
\beq \frac{\tau}{2}> \vert s_x \vert + \vert r_x \vert = \vert s_x \vert + \vert -r_x \vert \eeq
By the triangle inequality, this implies
\beq
\frac{\tau}{2} > |s_x - r_x|\;,
\eeq
as desired.

(b) Now suppose that the axis of Alexandrov set is tilted with respect to $t$ axis. If it is tilted in $yz$ plane, then the coordinate system can be rotated so that in new coordinates it is again parallel to $t$ axis leading to previous result. Thus, the only situation of interest is when the axis of Alexandrov set has $x$ component in which case it can not be gotten rid of since $x$ axis is set to be parallel to gradient of $\phi$. In this case, again rotate coordinate system so as to get rid of $y$ and $z$ components of Alexandrov set, thus its axis lies on $xt$ plane. 

Consider two lightlike lines, $l_1$ passing through $p$ in $x+t$ direction and $l_2$ passing through $q$ in $x-t$ direction. They are given as
\bea
& &l_1 = \{ (a_1+p_t, a_1+p_x, 0, 0) \mid a_1 \in \mathbb{R} \} \nonumber\\
& &l_2 = \{ (q_t-a_2, a_2+q_x, 0, 0) \mid a_2 \in \mathbb{R} \}\;.
\eea
Let $s$ be the unique elements of $l_1 \cap l_2$. The cases of $r=p$ and $r=2$ imply that
\bea
& &\sup \{ r_t - s_t \mid r, s \in \alpha (p,q) \}
\geq \max \{ |s_t - p_t|, |s_t - q_t| \} \nonumber\\
& &\kern140pt\geq\ \max \{ |a_1|, |a_2| \}\;.
\eea
The values of $a_1$ and $a_2$ can be derived by treating the respective equalities of non-zero components of $l_1$ and $l_2$ as a system of two equations and two unknowns:
\beq
a_1+p_t =  q_t-a_2 \; ; \; a_1+p_x = a_2 + q_x\;.
\eeq
This can be rewritten as
\beq
a_1+ a_2 =  q_t-p_t \; ; \quad a_1-a_2 = q_x-p_x\;,
\eeq
which gives
\beq
a_1 = \frac12\, (q_t-p_t+q_x-p_x) \; ; \quad a_2 = \frac12\, (q_t-p_t+p_x-q_x)\;.
\eeq
Thus,
\bea
& &\kern-40pt\sup \{ r_t - s_t \mid r, s \in \alpha (p,q) \} \geq \max \{ |s_t - p_t|, |s_t - q_t| \} \nonumber\\
& &\kern100pt\geq \max \{ |a_1|, |a_2| \} = \frac12\, (q_t - p_t + |q_x - p_x| ) \nonumber\\
& &\kern100pt> \frac12\, (q_t - p_t) > \sqrt{(q_t - p_t)^2 - (q_x - p_x)^2} \nonumber\\
& &\kern100pt= \frac12\, \tau(p,q)\;.
\eea
But in part (a) it was shown that for the case $p_x = q_x$ the above supremum is exactly equal to $\tau /2$. This, combined with the fact that gradient of $\phi$ is parallel to $x$ axis, implies that Alexandrov sets satisfying $p_x = q_x$ minimize the fluctuations of interest. 

This means that the $p_x = q_x$ Alexandrov set will be selected as the representative one in defining the Lagrangian, which implies that
\beq
{\cal L}_1 = \frac{c_t \tau^2}{4}\, (\partial_1 \phi)^2
= -\frac{c_t \tau^2}{4}\, \partial^\mu \phi\, \partial_\mu \phi
= -\frac{d_t}{4}\, \partial^\mu \phi\, \partial_\mu \phi\;.
\eeq
\noindent CASE 3: ${\cal K} = {\cal K}_s$; The gradient of $\phi$ is timelike.
In this case, just like for the case 1, the minimizing Alexandrov set is the one whose axis is parallel to the gradient of $\phi$. This will be shown by similar steps as in case 2: in step a the fluctuation will be computed for that specific Alexandrov set, and in step b it will be shown that all other Alexandrov sets have larger fluctuations. 

a) Let $\proj v$ denote the projection of $v$ on the $xyz$ hyperplane. The goal is to show that whenever $s_0 - r_0 > \tau/2$, $r$ and $s$ are timelike related. This automatically implies that for any spacelike related $r$ and $s$, $\vert s_0 - r_0 \vert \leq \tau/2$. 

Assume $s_0 - r_0 > \tau/2$. This implies 
\beq
- \tau/2 \leq r_0 <0 < s_0 < \tau/2\;.
\eeq
Therefore, 
\beq
|\proj r| \leq \tau/2 + r_0 \leq \tau/2 + s_0 - \tau/2
= \tau/2 - (\tau/2 - s_0) \leq \tau/2 - |\proj s|\;.
\eeq
By the triangle inequality, this implies
\beq
\vert \proj (r-s) \vert = |\proj s - \proj r| \leq |\proj r| + |\proj s|
\leq \tau/2\;.
\eeq
But the fact that $\vert \proj (r-s) \vert \leq \tau/2$ combined with the fact that $s_t - r_t > \tau/2$ implies that $r \prec s$. Therefore, if $r$ and $s$ were spacelike separated then, by contradiction, the original assumption that $s_t - r_t > \tau/2$ has to be wrong, i.e. $\vert s_t - r_t \vert \leq \tau/2$.

b) In part b of case 2 it was shown that if Alexandrov set is ``tilted" then one can produce a point $s$ that is lightlike separated from both $p$ and $q$ such that its coordinate difference with one of the two points is greater than $\tau/2$. The identical argument carries through to the situation at hand. The only modification is that instead of using $p$ and $q$, one has to use $p+ \epsilon (q-p)$ and $q - \epsilon (q-p)$ in order to ensure that points of interest are spacelike separated. The $\epsilon \rightarrow 0$ limit will imply that the fluctuations are greater than $\tau/2$. 

The Lagrangian predicted from this generator is 
\beq {\cal L}_s = \frac{c_s \tau^2}{4} (\partial_0 \phi)^2 = \frac{c_s \tau^2 }{4} \partial^{\mu} \phi \partial_{\mu} \phi = \frac{d_s}{4} \partial^{\mu} \phi \partial_{\mu} \phi \eeq 
\noindent CASE 4: ${\cal K} = {\cal K}_s$; gradient of $\phi$ is spacelike

In this case, start by selecting a frame in which the gradient of $\phi$ points along the $x$ axis. That is, for any Alexandrov set $\alpha(p,q)$ select a frame in which $p$ and $q$ lie in the $t$-$x$ plane. By considering the cases $p_x < 0 < q_x$ and $q_x < 0 < p_x$, in both of which $p^\mu = -q^\mu$, it is easy to see that the two intersections of the Alexandrov set with the $x$ axis lie on $-q_t - |q_x|$ and $q_t + |q_x|$. Since the gradient of $\phi$ points in the $x$ direction, the variation of $\phi$ is minimized whenever $q_t + |q_x|$ is. But $q_t$ is bounded below since the distance between the end points of Alexandrov set is fixed. This implies minimization of $|q_x|$. This requires $q_x = 0$. Since it is assumed that $q$ lies in the $t$-$x$ plane, this means $q = (q^0,0,0,0)$ and $p^\mu = -q^\mu$ implies $p = (-q^0,0,0,0)$. The intersections of the Alexandrov set with the $x$ axis are at $x = -\half\, \tau(p,q)$ and $x = \half\, \tau(p,q)$. Thus,
\beq
{\cal L}_s = c_s (\partial_1\phi)^2 = -c_s \tau^2(p,q)\, (\partial^{\mu} \phi \partial_{\mu} \phi = - d_s \partial^{\mu} \phi \partial_{\mu} \phi\;.
\eeq
By combining all four cases, the total Lagrangian based on $( {\cal K}_t, {\cal K}_s )$ is given by 
\beq
{\cal L} = \begin{cases}
(d_t + \frac{d_s}{4})\, \partial^{\mu} \phi\, \partial_{\mu} \phi
& \hbox{if}\quad \partial^{\mu} \phi\, \partial_{\mu} \phi > 0\\
-(d_s + \frac{d_t}{4})\, \partial^{\mu} \phi\, \partial_{\mu} \phi
& \hbox{if}\quad \partial^{\mu} \phi\, \partial_{\mu} \phi <0\;.
\end{cases}
\eeq
Since whether or not $\partial^{\mu} \phi \partial_{\mu} \phi$ greater than 0 is relativistically covariant, the above definition of Lagrangian is relativistically covariant as well. Thus, in principle, there is nothing wrong with the above Lagrangian, except for the fact that, of course, such was not observed in the lab. However, the observed Lagrangian is a special case of above for $c_s = - c_t$ or, equivalently, $d_s = -d_t$. Thus, such will be assumed from now on.

\subsection{Type-1 Gauge Field}

We would like to define Lagrangian generator for electrodynamics as a flux through the loop defined by a set of points $r_1, . . . , r_n$ : 
\beq {\cal K}_n (a ; r_1, . . . , r_n) = a(r_1, r_2) + a(r_2, r_3) + . . . + a(r_{n-1}, r_n) + a(r_n, r_1) \eeq
 Naively, it is tempting to consider flux through the triangular loop, $a(r_1, r_2) +  a(r_2, r_3) + a(r_3, r_1)$.  However, there is a problem: the equator of an Alexandrov set is spherically shaped, while its intersection with, say, the $x$-$t$ plane is not. Thus, the maximal area of the triangles lying in the intersection of the Alexandrov set with the $x$-$y$ plane is not the same as the maximal area of its intersection with the $x$-$t$ plane.
 
On the other hand, if fluxes are taken over rectangles, $a(r_1, r_2)+ a(r_2, r_3)+ a(r_3, r_4)+ a(r_4, r_1)$, it will be shown that in both cases the area is maximized by a square. In the spacelike case, the square will lie on the equator while in timelike case only two of its points will lie on equator while the other two points will lie at the poles. These two kinds of squares look identical to each other except for the 90 degree rotation, which means that they have the same areas as well.  For this reason, Lagrangian generator is given by
\beq {\cal K} = {\cal K}_4 \eeq
One might first ask the following question: if a causal set is inherently relativistic, doesn't it mean that the answer is covariant no matter what shape of the contour is chosen, as long as the procedure is covariant? If so, why wouldn't that apply to ${\cal K} = {\cal K}_3$? The answer to the covariance question is yes, but we have to be more careful as to what is meant by relativistically invariant. 

Strictly speaking, relativistically invariant means it doesn't change under rotations and boosts. Now, there is no rotation or boost that would take a spacelike vector into a timelike vector, or vice versa. Thus, we have two different answers for the spacelike and timelike contours, without violating relativistic invariance. However, despite the fact that this would be an invariant answer, it is not an answer that agrees with experiments. So, in order to get an answer that does, we chose to use rectangles instead of triangles.

While it might be interesting to investigate more general domains of the Lagrangian generator in the future work, for the purposes of this thesis we would select the domains that would save our time the most: 
\beq D= \{ (r_1, r_2, s_1, s_2, s_3, s_4) \vert \tau(r_1, r_2) = \tau_0 ; \nonumber \eeq
\beq r_1 \prec s_k \prec r_2 \; ; \; \tau(r_1, s_k) + \tau(s_k, r_2) < \epsilon \} \eeq
where $\epsilon$ is some small number.  The constraint $\tau(r_1, r_2)= \tau_0$ implies that at the step of going from Lagrangian generator to Lagrangian, the points $r_1$ and $r_2$ will be forced to coincide with $p$ and $q$ respectively. It is also easy to see that the constraint $\tau(r_1, s_k) + \tau(s_k, r_2) < \epsilon$ is equivalent to saying that points $s_1$ through $s_4$ lie very close to the surface of equator of Alexandrov set. 

We will have two Lagrangian generators, the timelike and spacelike one:
\beq {\cal L}_t (r_1,r_2, s_1, s_2, s_3, s_4) = (a(r_1, s_1) + a(s_1, r_2) + a(r_2, s_2) + a(s_2, r_1))^2 \eeq
\beq {\cal L}_s (r_1,r_2, s_1, s_2, s_3, s_4) = (a(s_1, s_2) + a(s_2, s_3) + a(s_3, s_4) + a(s_4, s_1))^2 \eeq
Since $ {\cal K}_t$ is independent of $s_3$ and $s_4$, we will simply write it as 
\beq {\cal K}_t (r_1,r_2, s_1, s_2) = (a(r_1, s_1) + a(s_1, r_2) + a(r_2, s_2) + a(s_2, r_1))^2 \eeq
From the fact that $r_1$ and $r_2$ coincide with poles of the Alexandrov set while $s_k$ lie on the equator, it is easy to see that in the reference frame defined by that Alexandrov set ${\cal L}_t$ measures the electric field while ${\cal L}_s$ measures the magnetic one. 

It is easy to see that 
\beq {\cal K}_{t} (p,q, r,s) = (\vec{E} \cdot (\vec{r} - \vec{s}))^2 \eeq
which means that it varies from $0$ to $\vert \vec{E} \vert^2 \tau^2$. 

Now lets talk about ${\cal L}_s$. Choose a reference frame in which $B$ points in $z$ direction. The flux through the loop is proportional to the area of a rectangle which is formed by the projections of points $r_1$, $r_2$, $r_3$ and $r_4$ on $xy$-plane. 

 That area can be computed as follows: we connect each of these 4 points to the origin. This would break the picture into 4 triangles. The angles of the two adjacent lines of the triangle that meet at the origin are $\theta_1$, $\theta_2$, $\theta_3$ and $\theta_4$. Each triangle can be further broken into two triangles by drawing the perpendicular line from the origin to the line connecting the opposite side of that triangle. The area of each of the two pieces is
\beq
\half\, (\half\,\tau\, \cos \half\,\theta_i)\,
(\half\,\tau\, \sin \half\,\theta_i) = \fourth\,\tau^2\, \sin \theta_i\;.
\eeq
Thus, the area of the whole thing is
\beq
2\, \sum_{i=1}^4 \fourth\,\tau^2 \sin\theta_i = \half\,\tau^2\sum_{i=1}^4\theta_i\;.
\eeq
Thus we would like to maximize $\sum_{\i=1}^4 \sin n\theta_i$, with the constraint that
$\sum_{i=1}^4 \theta_i = 2 \pi$.
 
This means that the gradient of $\sum_{\i=1}^4 \sin \theta_i$ should be parallel to the gradient of $\sum_{i=1}^4 \theta_i = 2 \pi$. In other words, there is a constant $c$ such that
\beq 
\frac{\partial}{\partial \theta_k}Ê\sum_{\i=1}^4 \sin \theta_i
= c\, \frac{\partial}{\partial\theta_k} \sum_{i=1}^4 \theta_i\;.
\eeq
This implies that
\beq
\cos\theta_k = c\;.
\eeq
In other words, all angles are equal. Since their sum is $2 \pi$ this means that they are all equal to $\frac{\pi}{2}$. So this can be accomplished by putting $r_1$ and $r_3$ at the intersections of the $x$ axis with the boundaries of the Alexandrov set, and $r_2$ and $r_4$ at the intersections of the $y$ axis with the boundaries of the Alexandrov set. This tells us that ${\cal K}_s$ varies between  $0$ and $\vert \vec{B} \vert^2 \tau^2$.

Thus, the variations of ${\cal K}_t$ and ${\cal K}_s$ are minimized in reference frame that minimizes $\vert \vec{E} \vert$ and $\vert \vec{B} \vert$ respectively. 

I claim that that frame coincides with the one in which $\vec{E}$ and $\vec{B}$ are parallel. In order to show that, we have to first show that such frame exists, and then, while working in that frame, we have to show that magnitudes of $\vec{E}$ and $\vec{B}$ will be larger in any other frame. 

Let's start with finding a reference frame in which $\vec{E}$ and $\vec{B}$ are parallel to each other. We start from the reference frame where they are not, and then find the kind of Lorentz boost that would bring us into the frame where they are. In the original frame, rotate coordinate system in such a way that 
\beq E_x = B_x =0 \eeq
Our guess is that the boost in $z$ direction will bring us into the reference frame where $\vec{E}$ and $\vec{B}$ are parallel. In the boosted frame, the electric and magnetic fields written in a form of two separate space-only vectors are
\beq \vec{E}' = (0, \gamma (E_y - vB_z), \gamma (E_z + vB_y)) \eeq
\beq \vec{B}' = (0, \gamma (B_y + vB_z), \gamma (B_z - vB_y)) \eeq
Thus, in order for them to be proportional to each other the following equation needs to be satisfied:
\beq \frac{E_y - vB_z}{B_y + vE_z} = \frac{E_z+vB_y}{B_z - vE_y} \eeq
This can be re-written as a quadratic equation in $v$: 
\beq v^2 (B_y E_z - B_z E_y) + v(E_y^2 + E_z^2 + B_y^2 + B_z^2) + E_z B_y - E_y B_z \eeq
Its solutions are
\beq v_1 = \frac{-E_y^2 - E_z^2 - B_y^2 - B_z^2 - \sqrt{(E_z^2 + E_y^2 - B_z^2 - B_y^2)^2 + 4 (E_y B_y + E_z B_z)^2}}{2(E_z B_y - E_y B_z)} \eeq
\beq v_2 = \frac{-E_y^2 - E_z^2 - B_y^2 - B_z^2 + \sqrt{(E_z^2 + E_y^2 - B_z^2 - B_y^2)^2 + 4 (E_y B_y + E_z B_z)^2}}{2(E_z B_y - E_y B_z)} \eeq
which, in light of the fact that $E_x=B_x=0$, can be rewritten as
\beq v_1 = \frac{- \vert \vec{E}  \vert^2 - \vert \vec{B} \vert^2 - \sqrt{(\vert \vec{E} \vert^2 - \vert \vec{B} \vert^2)^2 + 4 (\vec{E} \cdot \vec{B})^2}}{2 \vert \vec{B} \times \vec{E} \vert} \eeq
\beq v_2 = \frac{- \vert \vec{E}  \vert^2 - \vert \vec{B} \vert^2 + \sqrt{(\vert \vec{E} \vert^2 - \vert \vec{B} \vert^2)^2 + 4 (\vec{E} \cdot \vec{B})^2}}{2 \vert \vec{B} \times \vec{E} \vert} \eeq
However, in order to either of these solutions to be physically valid, they have to be between $-1$ and $+1$, which is what I am about to check. 

Let's start with $v_1$. We notice that 
\beq 0 \leq (\vert \vec{E} \vert - \vert \vec{B} \vert)^2 = \vert \vec{E} \vert^2 + \vert \vec{B} \vert^2 - 2 \vert \vec{E} \vert \vert \vec{B} \vert \eeq
This implies that 
\beq \vert \vec{E} \vert \vert \vec{B} \vert \leq \frac{1}{2} ( \vert \vec{E} \vert^2 + \vert \vec{B} \vert^2) \eeq
We also know that 
\beq \vert \vec{B} \times \vec{E} \vert \leq \vert \vec{E} \vert \vert \vec{B} \vert \eeq
which implies that 
\beq \vert \vec{E} \times \vec{B} \vert \leq \frac{1}{2} ( \vert \vec{E} \vert^2 + \vert \vec{B} \vert^2) \eeq
Therefore 
\beq v_1 \leq - \frac{\vert \vec{E} \vert^2 + \vert \vec{B} \vert^2}{2 \vert \vec{B} \times \vec{E} \vert} \leq -1 \eeq
which means that $v_1$ is not physically valid. 

Now let's look at $v_2$.  We notice that 
\beq (\vert \vec{E} \vert^2 - \vert \vec{B} \vert^2)^2 + 4 (\vec{E} \cdot \vec{B})^2 \leq (\vert \vec{E} \vert^2 - \vert \vec{B} \vert^2)^2 + 4 \vert \vec{E} \vert^2 \vert \vec{B} \vert^2 = (\vert \vec{E} \vert^2 + \vert \vec{B} \vert^2)^2 \eeq
We also konw that 
\beq  (\vert \vec{E} \vert^2 - \vert \vec{B} \vert^2)^2 + 4 (\vec{E} \cdot \vec{B})^2 \geq 0 \eeq
Together, these imply that 
\beq 0 \leq \sqrt{(\vert \vec{E} \vert^2 - \vert \vec{B} \vert^2)^2 + 4 (\vec{E} \cdot \vec{B})^2} \leq \vert \vec{E} \vert^2 + \vert \vec{B} \vert^2 \eeq
which implies that 
\beq v_2 \leq 0 \eeq
So $v_2$ is physically valid if and only if it is greater than $-1$. We will denote the inequalities that need to be satisfied for that to happen by a question mark: $<^?$ and $>^?$. Thus, by noticing that denominator is positive, we have 
\beq \sqrt{(\vert \vec{E} \vert^2 - \vert \vec{B} \vert^2)^2 + 4 (\vec{E} \cdot \vec{B})^2} - \vert \vec{E} \vert^2 - \vert \vec{B} \vert^2  >^? \eeq
\beq >^? - 2 \vert \vec{E} \times \vec{B} \vert \nonumber \eeq
which is equivalent to 
\beq \sqrt{(\vert \vec{E} \vert^2 - \vert \vec{B} \vert^2)^2 + 4 (\vec{E} \cdot \vec{B})^2} >^? \vert \vec{E} \vert^2 + \vert \vec{B} \vert^2  - 2 \vert \vec{E} \times \vec{B} \vert \eeq
Squaring both sides implies that
\beq (\vert \vec{E} \vert^2 + \vert \vec{B} \vert^2)^2 + 4 (\vec{E} \cdot \vec{B})^2 >^? \eeq
\beq >^?(\vert \vec{E} \vert^2 + \vert \vec{B} \vert^2 )^2 + 4 \vert \vec{B} \times \vec{E} \vert^2 - 4 \vert \vec{B} \times \vec{E} \vert ( \vert \vec{E} \vert^2 + \vert \vec{B} \vert^2 ) \nonumber \eeq
After expanding squares on both sides this becomes 
\beq \vert \vec{E} \vert^4 + \vert \vec{B} \vert^4 - 2 \vert \vec{E} \vert ^2 \vert \vec{B} \vert^2 + 4 (\vec{E} \cdot \vec{B})^2 >^? \nonumber \eeq
\beq >^? \vert \vec{E} \vert^4 + \vert \vec{B} \vert^4 + 2 \vert \vec{E} \vert^2 \vert \vec{B} \vert^2 + 4 \vert \vec{B} \times \vec{E} \vert^2 - 4 \vert \vec{B} \times \vec{E} \vert ( \vert \vec{E} \vert^2 + \vert \vec{B} \vert^2 ) \eeq
After cancelling $\vert \vec{E} \vert^4$ and $\vert \vec{B} \vert^4$ terms and moving things around between left and right sides, this becomes
\beq 4 (\vec{E} \cdot \vec{B})^2 - 4 \vert \vec{B} \times \vec{E} \vert^2 + 4 \vert \vec{B} \times \vec{E} \vert ( \vert \vec{E} \vert^2 + \vert \vec{B} \vert^2 )  >^? 4 \vert \vec{E} \vert^2 \vert \vec{B} \vert^2 \eeq
which, factoring out $4$, is
\beq (\vec{E} \cdot \vec{B})^2 -  \vert \vec{B} \times \vec{E} \vert^2 +  \vert \vec{B} \times \vec{E} \vert ( \vert \vec{E} \vert^2 + \vert \vec{B} \vert^2 ) >^? \vert \vec{E} \vert^2 \vert \vec{B} \vert^2  \eeq
Let's now express things in terms of the angle $\theta$ between vectors $\vec{E}$ and $\vec{B}$:
\beq \vert \vec{B} \times \vec{E} \vert = \vert \vec{B} \vert \vert \vec{E} \vert \vert \sin \theta \vert \eeq
and
\beq (\vec{E} \cdot \vec{B})^2 - \vert \vec{B} \times \vec{E} \vert^2 = \vert \vec{E} \vert^2 \vert \vec{B} \vert^2 (\cos^2 \theta - \sin^2 \theta) = \vert \vec{E} \vert^2 \vert \vec{B} \vert^2 \cos (2 \theta) \eeq
Substituting these we obtain
\beq \vert \vec{E} \vert^2 \vert \vec{B} \vert^2 \cos (2 \theta) + \vert \vec{E} \vert \vert \vec{B} \vert (\vert \vec{E} \vert^2 + \vert \vec{B} \vert^2 ) \vert \sin \theta \vert >^? \vert \vec{E} \vert^2 \vert \vec{B} \vert^2 \eeq
which, after factoring out $\vert \vec{E} \vert^2 \vert \vec{B} \vert^2$ becomes
\beq \cos (2 \theta) + \Big( \frac{ \vert \vec{E} \vert}{\vert \vec{B} \vert} + \frac{ \vert \vec{B} \vert}{\vert \vec{E} \vert} \vert \sin \theta \vert \Big) >^? 1 \eeq
Let's denote the ratio of magnitudes of electric and magnetic field by $\lambda$: 
\beq \lambda = \frac{ \vert \vec{E} \vert}{\vert \vec{B} \vert} \eeq
Thus, our inequality becomes
\beq \cos (2 \theta) + (\lambda + \lambda^{-1}) \vert \sin \theta \vert >^? 1 \eeq
We notice that 
\beq \frac{d}{d \lambda} (\lambda + \lambda^{-1}) = 1 - \lambda^{-2} \eeq
which means that it is decreasing when $\lambda<1$, increasing when $\lambda>1$ and reaches minimum when $\lambda=1$. That minimum is $2$. Thus
\beq \lambda + \lambda^{-1} \geq 2 \eeq
This means that if $ \cos (2 \theta) + 2  \vert \sin \theta \vert > 1$, it would automatically imply that $ \cos (2 \theta) + (\lambda + \lambda^{-1}) \vert \sin \theta \vert > 1$ for all $\lambda$. On the other hand, if $ \cos (2 \theta) + 2  \vert \sin \theta \vert < 1$, then the vicinity of $\lambda=1$ will serve as counter-examples for the above statement. Thus, instead of checking the original inequality, we can simply check the following: 
\beq \cos (2 \theta) + 2 \vert \sin \theta \vert >^? 1 \eeq
Let 
\beq f(\theta) = \cos (2 \theta) + 2 \sin \theta \eeq
Thus, what we need to check is 
\beq 0 < \theta < \pi \Rightarrow^? f (\theta) >1 \eeq
It can be easilly seen that derivative of $f$ is given by 
\beq f' (\theta) = 2 (\cos \theta - \sin (2 \theta)) = 2 \cos \theta (1 - 2 \sin \theta) \eeq
Thus, the maxima or minima of $f$ occur where 
\beq \cos \theta = 0 \; {\rm or} \; \sin \theta = \frac{1}{2} \eeq
This means that they occur at 
\beq \theta_1 = \frac{\pi}{6} \; , \; \theta_2 = \frac{\pi}{2} \; , \; \theta_3 = \frac{5}{6} \pi \; , \; \theta_4 = \frac{3}{2} \pi \eeq
The values of $f$ at these points are
\beq f(\theta_1) = f(\theta_3) = \frac{3}{2} \; , \; f ( \theta_2) = 1 \; , \; f( \theta_4) = -3 \eeq
The fact that $f(\theta_4) < 1$ does not bother us because $\theta_4 >  \pi$ and therefore is outside of the range of the angles we are interested in. 

The above values tell us that local maxima occur at $\theta_1$ and $\theta_3$, while local minima occur at $\theta_2$ and $\theta_4$. However, since we are restricting our range to the interval between $0$ and $\pi$, these two values also serve as potential minima. This means that 
\beq 0 \leq \theta \leq \pi \Rightarrow f(\theta) \geq min (f(0), f(\theta_1), f( \pi)) \eeq
Now, $f(0)$ and $f(\pi)$ are given by
\beq f(0) = f(\pi) =1 \eeq
This implies that
\beq 0 \leq \theta \leq \pi \Rightarrow f(\theta) \geq 1 \eeq
as desired.

Using the earlier established fact that 
\beq \lambda^{-1} + \lambda \geq 2 \eeq
this implies that 
\beq \cos (2 \theta) + (\lambda + \lambda^{-1}) \vert \sin \theta \vert >1 \eeq
which, as shown earlier, is equivalent to 
\beq v_2 > -1 \eeq
Since, as shown earlier, 
\beq v_2<0 \eeq
this implies that $v_2$ is a physical, thus confirming the existence of frame where $\vec{E}$ and $\vec{B}$ are parallel. 

Now, in the above work it was assumed that the Lorentz boost is performed in $x$ direction. This means that, by itself, this does not prove the uniqueness of the frame where electric and magnetic fields are parallel. In fact, if we start out from the frame in which they are parallel, then performing a Lorentz boost in the direction in which they both point, their values will stay unchanged, which means that they will remain parallel.

However, they will not stay parallel if Lorentz boost is pefromed in any other direction. Suppose the mutual direction of $\vec{E}$ and $\vec{B}$ is $z$ axis. By rotational symmetry, we can say that the direction of Lorentz boost lies in $xz$-plane. This boost can be done by first performing a boost in the $z$ direction and then in the $x$ direction. The boost in the $z$ direction will leave both fields unchanged, while the boost in the $x$ direction will produce 
\beq
\vec{E}' = (0, - \gamma v B, \gamma E)
\eeq
and 
\beq
\vec{B}' = (0, \gamma v E, \gamma B)\;.
\eeq
If the above vectors were parallel, then $y$ components would imply $E=-B$ while $z$ components would imply $E=B$, which would imply $E=B=0$ in which case the whole notion of them being or not being parallel is silly. 

Thus, in the frame where they are parallel, any other such frame is produced by boosts in their mutual direction, and in these other frames their values are identical. This means that in the case where everything is flat, smooth and linear, the answer to the Lagrangian generators given earlier is independent of which of the frames in the above class is chosen. 

In the real situation, the frame will be chosen in a way that would minimize the random fluctuations due to discreteness and non-linearity. By throwing away these effects, Lagrangian generators will return something very close to the values of electric and magnetic fields in the case of sufficiently dense Poisson distribution on a manifold. 

By going back to the original Lagrangian generators, we see that we have two separate ones: one for the electric field, and the other for the magnetic field. On the first glance, this might appear bothersome since neither is Lorentz covariant. However, by realizing that the values these fields are taken from specific reference frame (i.e. where they are parallel to each other) implies that they are replaced by covariant expressions in an arbitrary frame. This is similar to $\partial^0 \phi$ taken from the reference frame where $t$-axis is parallel to the gradient of $\phi$ being replaced with $\sqrt{\partial^{\mu} \phi \partial_{\mu} \phi}$ in an arbitrary frame. 

Let's now show explicitly what the covariant expressions are. In the reference frame in which $E$ is parallel to $B$, we can treat them as scalars and have the following system of equations
\bea
& &E^2 - B^2 = F^{\mu\nu}\, F_{\mu\nu} \\
& &EB = \frac{1}{k}\, \epsilon_{\alpha\beta\gamma\delta}\,
F^{\alpha\beta}\, F^{\gamma \delta}\;.
\eea
This amounts to a system of two equations and two unknowns, which means that its solutions for  $E^2$ and $B^2$ are covariant expressions for each of these terms separately. This should not surprise us. After all, when $E$ and $B$ were treated as scalars, it was done in a special frame where $E$ and $B$ are parallel. Thus, the procedure of {\em first\/} finding such frame and {\em then\/} evaluating $E$ and $B$ in that frame {\em is\/} covariant, despite the fact that $E$ and $B$ in an arbitrary frame are {\em not}.
 
Now, to satisfy ourselves, let us solve that system of equations to get an expression for $E$ and $B$. The second equation implies that $B = \epsilon_{\alpha\beta\gamma\delta}\, F^{\alpha\beta}\, F^{\gamma\delta}/kE$. Substituting this into the first equation, we obtain
\beq
F^{\mu\nu}\, F_{\mu\nu} = E^2 - \frac{(\epsilon_{\alpha\beta\gamma\delta}\,
F^{\alpha\beta}\, F^{\gamma\delta})^2}{k^2\, E^2}\;.
\eeq
Multiplying it by $E^2$, and moving all terms to the left-hand side, we get the following equation
\beq
E^4 - E^2\, F^{\mu\nu}\, F_{\mu\nu} - \frac{(\epsilon_{\alpha\beta\gamma\delta}\,
F^{\alpha\beta}\, F^{\gamma\delta})^2}{k^2\, E^2} = 0\;.
\eeq
This solves to
\beq
E^2 = \frac12\,\bigg[F^{\mu\nu}\, F_{\mu\nu} + \sqrt{(F^{\mu\nu}\, F_{\mu\nu})^2 + \frac{4}{k^2}\, (\epsilon_{\alpha\beta\gamma\delta}\, F^{\alpha\beta}\, F^{\gamma\delta})^2} \bigg]\;.
\eeq 
Thus, first equation of the system of two equations implies
\beq
B^2 = \frac12\,\bigg[ \sqrt{(F^{\mu\nu}\, F_{\mu\nu})^2 + \frac{4}{k^2}\, (\epsilon_{\alpha\beta\gamma\delta}\, F^{\alpha\beta}\, F^{\gamma\delta})^2}  - F^{\mu\nu}\,F_{\mu\nu}\bigg]\;.
\eeq
The Lagrangian generators for these are ${\cal L}_t$ and ${\cal L}_s$ respectively, defined earlier. Thus,   we see that indeed they are both Lorentz covariant; they simply include a Lorentz contraction that does not exist in practice. This can be easily fixed by setting 
\beq {\cal L} = ({\cal L}_t, - {\cal L}_s) \eeq

\subsection{Getting Rid of Unwanted Fluctuations}

There is a side-benefit of the Lagrangian generator approach that is worth mentioning: it automatically adresses the issue of unwanted fluctuations. 

In the last two sections we have shown that if we have a Poisson distribution of points on a Lorentzian manifold, then the Lagrangian density derived from Lagrangian generator on that scattering will approximate continuum-based Lagrangian density defined in a conventional way. However, we have implicitly made the following assumptions:

1) The Alexandrov set that is being selected in a prescribed manner is not too small. In particular, it has enough points to make stochastic fluctuations nearly $0$, which would make my arguments reliable.  

2) The above Alexandrov set should not be too large either. In particular, in the interior of that Alexandrov set both the curvature of spacetime as well as all fields are assumed to be linear.

The first condition is taken care of explicitly when the constraint $\tau (p,q) = \tau_0$ is imposed in the ${\cal L}_k = \min \{ {\cal K}_{\rm max} \}$ part of my definition, as long as $\tau$ is assumed to be large enough for most of such Alexandrov sets to contain sufficiently many points to be statistically reliable. 

The second constraint can not be imposed as naively as the first constraint can. In a near light cone region the Lorentzian distance between two points can be arbitrary small, while the coordinate differences can be arbitrary large. Due to large coordinate differences, we can not assume linearity in the interior of Alexandrov set, despite the small Lorentzian distance.

As a result of Lorentzian covariance, the above statement means that smallness of Lorentzian distance does not imply linearity, period. This statement is true regardless of whether coordinate difference small or not. This can be illustrated by using a specific example of small Lorentzian distance and large coordinate one. For example, we can consider an electron flying from lab A to lab B with near lightlike velocity. The Lorentzian distance between events of emission and absorbtion of electron is small, while coordinate one is large.

We now make a Lorentz transformation to the reference frame of the electron. In this frame, the coordinate difference between the two events is also small. What the electron will see is labs A and B flying very close to each other with near lightlike velocity. Both labs fit inside of a very small region since both underwent Lorentzian contraction. As a result of this, fields change very fast in space. And due to the fact that this spatial picture moves with near lightlike velocity, they change in time as well at any fixed space location, in particular in the fixed location occupied by an electron.

Thus, what we see is that, on the one hand, it is not true that fields are locally linear in an arbitrary chosen reference frame. On the other hand, however, they are in fact locally linear in the reference frame in which we live in. This seems to suggest that there is ether. This, however, would contradict letter and spirit of causal set theory, since one of its goals is to maintain manifest Lorentz invariance. So, instead of saying that there is an ether, we would like to say that the way in which fields interact with each other makes them ``slow down" with respect to each other to the extent that their ``center of mass" would have a well defined reference frame (which would not have happened otherwise due to the non-comactness of the Lorentz group). 

This situation is something we already are used to in standard physics. For example, if we consider earth physics, there will also be a ``preferred frame", namely the frame of the Earth. But this does not raise a concern of violation of Lorentz covariance because the process of Earth formation was  caused by interactions that are defined in Lorentz covariant terms. Thus, in order to ``pick a preferred frame without violating relativity" we have to define a relativistically covariant physics that would do the work for us. 

This is exactly what the theory of Lagrangian generators is doing. On the one hand, the criteria of selecting of Alexandrov set with minimal $\Delta {\cal L}$ is relativistically covariant. On the other hand, however, the Alexandrov set selected by that criteria will imply a ``preferred frame" (namely, the one in which the two end points of Alexandrov set have the same space coordinates). This ``preferred frame" is determined in light of the behavior of the fields as opposed to being imposed from outside. Thus, this is analogous to the situation with the earth rather than the situation with the ether.

\newpage
\section{Chapter 4: Type-2 Bosonic Fields}

\subsection{Failure to View Gravity as a Type-1 Field}

Before proceeding to introduce type-2 bosonic fields, it is vital to understand the prime example of a bosonic field that can not be viewed as type 1: a gravitational field. In particular, it is important to show that if type-1 gravity were to exist, then yes we would still be able to write down its Lagrangian in covariant form; the only problem will be the presence of unwanted terms such as $R^{\mu}_{\nu} R^{\nu}_{\mu}$. 

This is crucial because if the failure to produce Lorentz covariant contraction was the problem it would imply a hole in a theory, as its major premise is that, by making sure that only causal relations are used to define geometry, Lorentz covariance is maintained.

 On the other hand, by producing covariant terms that are not observed in nature I make a point that the logic of the type-1 theory is perfectly self consistent and relativistically covariant; it is simply that experiments (in particular the ones involving gravity) have shown us that there is something more. That ``something more" is type-2 fields, and its presence does not negate the possibility of other fields still being type 1.  
 
Consider Lagrangian generator defined as follows:
\beq
{\cal K} (\prec; p, q) = \begin{cases}
V( \alpha (p,q))
& \hbox{if} \; p \prec q \hbox{ and } \tau (p,q) = \tau_0 \\
 0 
& \hbox{otherwise}\;.
\end{cases}
\eeq 
Here, $\tau_0$ is the same number that is used as a constraint on the selection of Alexandrov set in the procedure of going from Lagrangian generator to the Lagrangian. This means that, for any Alexandrov set that is allowed to be selected, exactly one pair of points will have non-zero value of gravitational Lagrangian generator, namely the end points of that Alexandrov set. This means that variation of Lagrangian generator is given by 
\beq \Delta {\cal K} = Vol ( \alpha (p,q)) \eeq
As is shown in Ref \cite{GibSol}, the volume of Alexandrov set defined by arbitrary $r \prec s$ is given by 
\beq Vol (\alpha (a,b)) = \tau^d (a,b) (k_d + (ARg_{\mu \nu} + BR_{\mu \nu})(b^{\mu}-a^{\mu})(b^{\nu} - a^{\nu})) + 0 (\tau^{d+3}) \eeq
By substituting $p$ and $q$ into the above, 
\beq \Delta {\cal K} = \vert k_d \tau_0^d +AR \tau_0^{d+2} + B \tau_0^d R_{\mu \nu}(b^{\mu}-a^{\mu})(b^{\nu} - a^{\nu}) + 0 (\tau_0^{d+3}) \vert \eeq
In light of the fact that the Ricci tensor has 10 degrees of freedom, while Lorentz group has 6 (3 rotations and 3 boosts), 10-6=4 implies that it is possible to choose a coordinate system in such a way that Ricci tensor is diagonal:
\beq
R_{\mu \nu} = \rho_{\mu} \delta^{\mu}_{\nu}\;.
\eeq
In this coordinate system, the variation of the Lagrangian generator becomes
\beq
\Delta {\cal K} = \vert k_d \tau_0^d +AR \tau_0^{d+2} + B \tau_0^d \sum \rho_{\mu}(b^{\mu}-a^{\mu})^2  + 0 (\tau_0^{d+3}) \vert\;.
\eeq
We will first evaluate the Lagrangian density in that particular coordinate system, and then generalize it to other ones. We will do that by cases. 

For arbitrary $p$ and $q$ satisfying $\tau (p,q) = \tau_0$, 
\beq \sum \rho_{\mu}(b^{\mu}-a^{\mu})^2 = \rho_0 (b^0 -a^0)^2 + \sum_k \rho_k (b^k - a^k)^2 =  \eeq
\beq= \rho_0 (\tau_0^2 - \sum_k (b^k - a^k)^2) + \sum_k \rho_k (b^k - a^k)^2 = \rho_0 \tau_0^2 + \sum_k (\rho_k - \rho_0 )( b^k - a^k)^2  \nonumber \eeq
If for all $k$, $\rho_k > \rho_0$, then the Alexandrov set that minimizes $\Delta {\cal K}$ is the one whose axis parallel to $t$ axis, $b^{\mu}-a^{\mu} = \tau \delta^{\mu}_{\nu}$. This means that the latter Alexandrov set will be used to define the Lagrangian density, which gives
\beq \forall k (\rho_k > \rho_0 ) \Rightarrow {\cal L} = \rho_0 \tau_0^2 \eeq
Now suppose there is at least one $i$ for which $\rho_i < \rho_0$. Then, regardless of the values of $\rho_j$ for $j \neq i$,  if we select
\beq q_0 = p_0 + \sqrt{\tau_0^2 + \frac{\rho_0}{\rho_0 - \rho_i}}\; , \; q_i = p_i + \sqrt{\frac{\lambda_0}{\lambda_0 - \lambda_3}} \; , \; q_j = p_j \; , \; j \neq i \eeq
we would get $\delta {\cal K}=0$. This means that either the above Alexandrov set, or some other one with $\delta {\cal K}=0$ is used to define the Lagrangian. In either case, this implies ${\cal L}=0$. Thus,
\beq \exists i (\lambda_i < \lambda_0) \Rightarrow {\cal L} = 0 \eeq
This can be summarized as
\beq
{\cal L} (\prec; p, q) = \begin{cases}
\rho_0 \tau_0^2
& \hbox{if} \; \forall k (\rho_k > \rho_0) \\
 0 
& \hbox{otherwise}\;.
\end{cases}
\eeq 
In the above expressions, the components of the Ricci tensor selected in the special frame where it is diagonal were wisely replaced with $\rho_i$. The latter are simply eigenvalues of $R$. Since the notion of eigenvalue is covariant, the above expression is covariant as well. This means that we can remember the above result, while dropping the assumption of a specifically chosen reference frame. 

Lets satisfy ourselves and actually find out the covariant expressions for eigenvalues. The four eigenvalues of Ricci tensor are the solutions of the equation
\beq
\det (\lambda g_{\mu \nu} - R_{\mu \nu}) = 0\;.
\eeq
Here, we are using $\lambda$-s instead of $\rho$-s, because we are not sure in what ``order" to write down the eigenvalues once we found them. Thus, $\lambda$-s will be some permutation of $\rho$-s. 

The above expression can be rewritten as 
\bea
& &\kern-10pt0 = \det(\lambda\, g_{\mu \nu} - R_{\mu \nu}) \nonumber \\
& &= \epsilon^{\alpha \beta \gamma \delta} \epsilon ^{\mu \nu \rho \sigma} (\lambda g_{\alpha \mu} - R_{\alpha \mu})(\lambda g_{\beta \nu} - R_{\beta \nu})(\lambda g_{\gamma \rho} - R_{\gamma \rho})(\lambda g_{\delta \sigma} - R_{\delta \sigma})\nonumber \\
& &= \lambda^4 \epsilon^{\alpha \beta \gamma \delta} \epsilon^{\mu}{}_{\beta \gamma \delta}\, R_{\alpha\mu} + 6 \lambda^2 \epsilon^{\alpha\beta\gamma\delta}
\epsilon^{\mu\nu}{}_{\gamma\delta}\, R_{\alpha \mu} R_{\beta \nu} \nonumber \\
& &\kern20pt-\ 4\, \lambda \epsilon^{\alpha\beta\gamma\delta}
\epsilon^{\mu\nu\rho}{}_{\delta}\, R_{\alpha\mu}\, R_{\beta\nu}\, R_{\gamma \rho} + \epsilon^{\alpha \beta \gamma \delta} \epsilon^{\mu \nu \rho \sigma} R_{\alpha \mu} R_{\beta \nu} R_{\gamma \rho} R_{\delta \sigma}\;.
\eea
We would now like to compute the contractions of the $\epsilon$ tensor.  First, note that 
\bea
& &\kern-10pt
-\epsilon^{\alpha \beta \gamma \delta} \epsilon_{\mu \nu \rho \sigma} \\
& &= \delta^{\alpha}_{\mu}\, \delta^{\beta}_{\nu}\, \delta^{\gamma}_{\rho}\, 
\delta^{\delta}_{\sigma} - \delta^{\alpha}_{\mu}\, \delta^{\beta}_{\nu}\, 
\delta^{\gamma}_{\sigma}\, \delta^{\delta}_{\rho} - \delta^{\alpha}_{\mu}\, 
\delta^{\beta}_{\rho}\, \delta^{\gamma}_{\nu}\, \delta^{\delta}_{\sigma} +
\delta^{\alpha}_{\mu}\, \delta^{\beta}_{\rho}\, \delta^{\gamma}_{\sigma}\, 
\delta^{\delta}_{\nu} + \delta^{\alpha}_{\mu}\, \delta^{\beta}_{\sigma}\, 
\delta^{\gamma}_{\nu}\, \delta^{\delta}_{\rho} - \delta^{\alpha}_{\mu}\, 
\delta^{\beta}_{\sigma}\, \delta^{\gamma}_{\rho}\, \delta^{\delta}_{\nu} \nonumber \\
& &
-\ \delta^{\alpha}_{\nu}\, \delta^{\beta}_{\mu}\, \delta^{\gamma}_{\rho}\, 
\delta^{\delta}_{\sigma} + \delta^{\alpha}_{\nu}\, \delta^{\beta}_{\mu}\, 
\delta^{\gamma}_{\sigma}\, \delta^{\delta}_{\rho} + \delta^{\alpha}_{\nu}\, 
\delta^{\beta}_{\rho}\, \delta^{\gamma}_{\mu}\, \delta^{\delta}_{\sigma} -
\delta^{\alpha}_{\nu}\, \delta^{\beta}_{\rho}\, \delta^{\gamma}_{\sigma}\, 
\delta^{\delta}_{\mu} - \delta^{\alpha}_{\nu}\, \delta^{\alpha}_{\sigma}\,
\delta^{\gamma}_{\mu}\, \delta^{\alpha}_{\rho} + \delta^{\alpha}_{\nu}\, 
\delta^{\beta}_{\sigma}\, \delta^{\gamma}_{\rho}\, \delta^{\delta}_{\mu} \nonumber \\
& &
+\ \delta^{\alpha}_{\rho}\, \delta^{\beta}_{\mu}\, \delta^{\gamma}_{\nu}\, 
\delta^{\delta}_{\sigma} - \delta^{\alpha}_{\rho}\, \delta^{\beta}_{\mu}\, 
\delta^{\gamma}_{\sigma}\, \delta^{\delta}_{\nu}-\delta^{\alpha}_{\rho}\, 
\delta^{\beta}_{\nu}\, \delta^{\gamma}_{\mu}\, \delta^{\delta}_{\sigma} +
\delta^{\alpha}_{\rho}\, \delta^{\beta}_{\nu}\, \delta^{\gamma}_{\sigma}\, 
\delta^{\delta}_{\mu} + \delta^{\alpha}_{\rho}\, \delta^{\beta}_{\sigma}\, 
\delta^{\gamma}_{\mu}\, \delta^{\delta}_{\nu} - \delta^{\alpha}_{\rho}\, 
\delta^{\beta}_{\sigma}\, \delta^{\gamma}_{\nu}\, \delta^{\delta}_{\mu} \nonumber \\
& &
-\ \delta^{\alpha}_{\sigma}\, \delta^{\beta}_{\mu}\, \delta^{\gamma}_{\nu}\, 
\delta^{\delta}_{\rho} + \delta^{\alpha}_{\sigma}\, \delta^{\beta}_{\mu}\, 
\delta^{\gamma}_{\rho}\, \delta^{\delta}_{\nu} + \delta^{\alpha}_{\sigma}\, 
\delta^{\beta}_{\nu}\, \delta^{\gamma}_{\mu}\, \delta^{\delta}_{\rho} -
\delta^{\alpha}_{\sigma}\, \delta^{\beta}_{\nu}\, \delta^{\gamma}_{\rho}\, 
\delta^{\delta}_{\mu} - \delta^{\alpha}_{\sigma}\, \delta^{\beta}_{\rho}\, 
\delta^{\gamma}_{\mu}\, \delta^{\delta}_{\nu} + \delta^{\alpha}_{\sigma}\, 
\delta^{\beta}_{\rho}\, \delta^{\gamma}_{\nu}\, \delta^{\delta}_{\mu}\;. \nonumber
\eea
Setting $\delta = \sigma$ implies
\bea
& &\kern-20pt - \epsilon^{\alpha \beta \gamma \chi} \epsilon_{\mu \nu \rho \delta}
= \sum_{\delta \neq \alpha, \beta, \gamma} (\delta^\alpha_\mu \delta^\beta_\nu \delta^{\gamma}_{\rho} - \delta^{\alpha}_{\mu} \delta^{\beta}_{\rho} \delta^{\gamma}_{\nu} - \delta^{\alpha}_{\nu} \delta^{\beta}_{\mu} \delta^{\gamma}_{\rho} + \delta^{\alpha}_{\nu} \delta^{\beta}_{\rho} \delta^{\gamma}_{\mu} + \delta^{\alpha}_{\rho} \delta^{\beta}_{\mu} \delta^{\gamma}_{\nu} - \delta^{\alpha}_{\rho} \delta^{\beta}_{\nu} \delta^{\gamma}_{\mu} ) \nonumber\\
& &\kern43pt =\delta^{\alpha}_{\mu} \delta^{\beta}_{\nu} \delta^{\gamma}_{\rho} - \delta^{\alpha}_{\mu} \delta^{\beta}_{\rho} \delta^{\gamma}_{\nu} - \delta^{\alpha}_{\nu} \delta^{\beta}_{\mu} \delta^{\gamma}_{\rho} + \delta^{\alpha}_{\nu} \delta^{\beta}_{\rho} \delta^{\gamma}_{\mu} + \delta^{\alpha}_{\rho} \delta^{\beta}_{\mu} \delta^{\gamma}_{\nu} - \delta^{\alpha}_{\rho} \delta^{\beta}_{\nu} \delta^{\gamma}_{\mu}\;.
\eea
Now setting $\gamma = \rho$ implies
\beq
- \epsilon^{\alpha \beta \gamma \delta}\epsilon_{\mu \nu \gamma \delta} = \sum_{\gamma \neq \alpha, \beta} (\delta^{\alpha}_{\mu} \delta^{\beta}_{\nu} - \delta^{\alpha}_{nu} \delta^{\beta}_{\mu})= 2(\delta^{\alpha}_{\mu} \delta^{\beta}_{\nu} - \delta^{\alpha}_{\nu} \delta^{\beta}_{\mu})\;.
\eeq
Setting $\beta=\nu=\lambda$ implies
\beq
- \epsilon^{\alpha \beta \gamma \delta} \epsilon_{\mu \beta \gamma \delta} = 2 \sum_{\beta \neq \alpha} \delta^{\alpha}_{\mu} = 6 \delta^{\alpha}_{\mu}\;.
\eeq
Finally, setting $\alpha= \mu = \eta$ implies
\beq
- \epsilon^{\alpha \beta \gamma \delta} \epsilon_{\alpha \beta \gamma \delta} = 6 \sum_{all \; \mu} \delta^{\mu}_{\mu} = 24\;.
\eeq
Substitution these into the expression for the determinant implies 
\bea
& &0 = -24 \lambda^4 + 24 \lambda^3 R + 12 \lambda^2 (R^{\alpha}_{\beta} R^{\beta}_{\alpha} -R^2 ) + 4 \lambda (3RR^{\alpha}_{\beta}R^{\beta}_{\alpha} - 2R^{\alpha}_{\beta}R^{\beta}_{\gamma}R^{\gamma}_{\alpha} - R^3 ) \nonumber\\
& &\kern22pt-R^4 +6R^2 R^{\alpha}_{\beta}R^{\beta}_{\alpha} + 6 R^{\alpha}_{\beta}R^{\beta}_{\gamma}R^{\gamma}_{\delta}R^{\delta}_{\alpha} - 8RR^{\alpha}_{\beta} R^{\beta}_{\gamma} R^{\gamma}_{\alpha} - 2(R^{\alpha}_{\beta} R^{\beta}_{\alpha} )^2\;.
\eea
Thus, the four solutions to this equation are 
\beq \lambda_{1,2,3,4} = f_{1,2,3,4}(-24,24, 12(R^{\alpha}_{\beta} R^{\beta}_{\alpha} -R^2 ), 4 (3RR^{\alpha}_{\beta}R^{\beta}_{\alpha} - 2R^{\alpha}_{\beta}R^{\beta}_{\gamma}R^{\gamma}_{\alpha} - R^3 ) , \nonumber \eeq
\beq -R^4 +6R^2 R^{\alpha}_{\beta}R^{\beta}_{\alpha} + 6 R^{\alpha}_{\beta}R^{\beta}_{\gamma}R^{\gamma}_{\delta}R^{\delta}_{\alpha} -  8RR^{\alpha}_{\beta} R^{\beta}_{\gamma} R^{\gamma}_{\alpha} - 2(R^{\alpha}_{\beta} R^{\beta}_{\alpha} )^2)\eeq
where $f_{1,2,3,4}(a,b,c,d,e)$ represent the four solutions to the equation
$$
a\,x^4 + b\,x^3 + c\,x^2 + d\,x + e = 0\;.
$$
We would now like to be able to single out an eigenvalue corresponding to the time-like eigenvector. Noticing that 
\beq
R_{00}=\lambda_i \Leftrightarrow R = 2\lambda_i - \sum_{j=1}^d \lambda_j\;.
\eeq
$R_{00}$ can be produced in the following covariant form:
\beq
R_{00} = \lim_{n \rightarrow \infty} \sum_{i=1}^d \lambda_i
\,\ee^{-n (2\lambda_i - \sum_{j=1}^d \lambda_j  - R)^2}\;.
\eeq
In the case of a Poisson distribution of points on a manifold, the above equation is not true since it relies on the exact equality to $1$, and once things are discrete nothing is exact. Thus, instead the limit can be replaced with the same expression where $n$ is assumed to be large, but finite, constant:
\beq
R_{00} \approx \sum_{i=1}^d \lambda_i
\,\ee^{-n (2\lambda_i - \sum_{j=1}^d \lambda_j  - R)^2}\;.
\eeq
Substituting this into the earlier defined expression for the Lagrangian in terms of eigenvalues,
\beq
{\cal L} (\prec; p, q) = \begin{cases}
\tau_0^2 \sum \lambda_i \ee^{-n (2\lambda_i - \sum_{j=1}^d \lambda_j  - R)^2} 
& \hbox{ if } \sharp \{ k \vert \lambda_k > \epsilon + \ee^{-n (2\lambda_i - \sum_{j=1}^d \lambda_j  - R)^2} \} = d-1 \\
 0 
& \hbox{ if } \sharp \{ k \vert \lambda_k < - \epsilon + \ee^{-n (2\lambda_i - \sum_{j=1}^d \lambda_j  - R)^2} \} > 0\;.
\end{cases}
\eeq 
In the above expression, 
\beq \sharp \{ k \vert \lambda_k > \epsilon + \ee^{-n (2\lambda_i - \sum_{j=1}^d \lambda_j  - R)^2} \} = d-1 \eeq means that the number of points that are ``visibly" (i.e., by more than $\epsilon$) greater than $\rho_0$ is $d-1$. If $n$ and $\epsilon$ are adjusted in such a way that the expression for estimation of $\rho_0$ can not be visibly larger than $\rho_0$ itself, it implies that every single $\rho_k$ is visibly larger than $\rho_0$.

On the other hand,  
\beq \sharp \{ k \vert \lambda_k < - \epsilon + \ee^{-n (2\lambda_i - \sum_{j=1}^d \lambda_j  - R)^2} \} > 0 \eeq
implies that there is at least one $\lambda$ that is visibly smaller than the expression for $\rho_0$. Again, assuming that $\epsilon$ and $n$ are adjusted in such a way that $\rho_0$ can not be visibly smaller than the equation that approximates it, this implies that there is at least one $\rho_i$ that is smaller than $\rho_0$. 

The above makes it obvious why the ${\cal L}$ was expressed in the form that it was. Now, remembering that, once the quartic equation is solved, $\lambda_i$ themselves are expressed in Lorentz covariant form, this implies that the expression for Lagrangian density is covariant as well. Its only problem is the presence of unwanted contractions of the Ricci tensor with itself. But these contractions are still covariant; they were just never observed in nature. 

\subsection{Difficulties with Non-Gravitational Type-1 Fields}

In order to motivate the introduction of type-2 version of non-gravitational fields, let's briefly outline other difficulties that type-1 theory encounters. One such difficulty was mentioned to me by my advisor, Luca Bombelli. It is related to introducing SU($n$) fields. While it might still be possible to do better in the future work, the current type-1 definition of Lagrangian generator relies on a choice of coordinate system in which electric and magnetic fields are parallel. Existence of that coordinate system was guaranteed by the comparison of 6 degrees of freedom of $F^{\mu \nu}$ and 6 degrees of freedom of Lorentz group, which would not work for SU($n$) which has far more than 6 degrees of freedom.

There is a way around the issue, if one is willing to sacrifice the fundamental SU($n$) covariance. In particular, one can view $A^1$, $A^2$, etc. as completely different fields which, independently from each other, have identical type-1 Lagrangian generators. As a result, they have identical Lagrangians, and their sum just happens to APPEAR to possess SU($n$) symmetry. But this is a result, rather than a cause, of specific Lagrangians. In other words, the symmetry is entirely coincidental. Furthermore, as a consequence of non-linearity, the symmetry is no longer exact. After all, the Alexandrov sets that used to minimize variation of each individual $A^a$ might be completely different from the ones used to minimize their linear combinations. 

Up to the order of some approximation, however, this would still reproduce predictions of quantum field theory, since the Lagrangians match in linear case, whether we call it symmetry or not. So this can be viewed as an ideological sacrifice as we are no longer calling it electroweak theory, but rather it is separate electric and weak fields. This sacrifice might still be worth it if type-1 theory proves to be the only one producing real results in light of type-2 theory being more complicated. But still one has to be aware of what is being sacrificed. 

There is also another difficulty, which has to do with type-1 fields in general, not only SU($n$) ones. This has to do with the stochastic nature of assigning numerical values of fields to points or pairs of points of a causal set. For example, suppose the values of fields are distributted on a bell curve of width $1/1000$ around $1$. Despite the fact that the width of bell curve is extremely small, statistics tell us that, as long as there are enough points (or enough pairs of points), there is $99$ percent chance that some values of fields will be far from $1$, say, they might they might be greater than $10$ or less than $-10$. This would have been okay if all values of fields were averaged out. But in light of the maximization procedure, only these ``obscure" field values would ever get a chance to be looked at. 

There is a way of adressing this. Namely, if we live in a world full of type-1 fields, we would never have a chance to learn that any of the ``more common" values occur. So, no matter how far apart the ``obscure" values are spaced, according to us they are very close to each other, since we never observe any variation that is smaller than that. Furthermore, even if in reality they are very rare, to us they would be the most common ones, if not the only ones that ever exist.

However, whether such an answer is satisfactory or not is ultimately up to the numeric tests since, in light of the fact that Alexandrov sets overlap, it is possible that less obscure values of fields will figure out somewhere, allowing us to ``catch ourselves". So, until such numeric work is done, it is best to hope for the best and, at the same time, have ammunition for the worst. In other words, it is best to retain type-1 theory in case it works, and also introduce type-2 theory in case it doesn't.

\subsection{Type-2 Lagrangian Generators}

In light of the above difficulties we will introduce an alternative model of Lagrangian generators, type-2 one.  The key idea to the model of type-2 fields is that, as will be shown in the next three sections, in the linear case,  
\beq k_{scalar} \partial^{\mu} \phi \partial_{\mu} \phi
= \tau^{-2} (p,q) (\phi (q) - \phi (p))^2 + \nonumber \eeq
\beq + E_{scalar} \tau^{-2d-2} (p,q)
\int_{\alpha (p,q)} \dd^d r\, \dd^d s\,(\phi (r) - \phi (s))^2 \eeq
\beq k_{EM} F^{\mu \nu} F_{\mu \nu} = \tau^{-d-2}
\int \dd^d r\, (a(p,r) + a(r, q) + a(q, p)) + \nonumber \eeq
\beq + E_{gauge} \tau^{-3d-2} \int \dd^d r\, \dd^d s\, \dd^d t\,
(a(r, s) + a (s, t) + a (t, r)) \eeq
\beq k_{grav} R= V(\alpha (p,q)) + E_{grav} \Big( \int_{V(\alpha (p,r)) \leq V (\alpha (p,q))} V(\alpha (p,r)) + \nonumber \eeq
\beq +  \int_{V(\alpha (p,r)) > V (\alpha (p,q))} V(\alpha (r,q)) \Big)\;,
\eeq
for appropriately adjusted $k$ and $E$. 

This means that it has the following advantages and disadvantages:

ADVANTAGES:

1) Lagrangian can be estimated based on arbitrary Alexandrov set rather than specifically selected one

2) Gravitational Lagrangian is reproduced

DISADVANTAGE:

1) The coefficients $E$ have to be appropriately adjusted. 

Adressing the above-mentioned disadvantage is an important issue that needs to be adressed, is adjusting the coefficients. In light of the fact that the theory is aimed at arbitrary causal set, not just manifold-like, if $E$ is viewed as a constant, this would imply that it was an extremely lucky coincidence that it happened to lead to exact cancelation in a special case of a four-dimensional manifold. This, of course, is not satisfactory.

For that reason, it is viewed as variable, subject to some physical laws that point-wise adjust it to the values that it should have at any given point. In other words, if field of interest is $F \in {\cal F}$, then 
\beq E= E(F, x) \eeq
is a function of both point $x$ as well as the value of fields $F \in {\cal F}$ in its surrounding. Since $E$ is not a separate field, the path integral does not include integration over $E$. Rather, it is given as
\beq
\int {\cal D}F \exp \Big(i \int \dd^d x\, {\cal L} (F, x, E(F, x)) \Big)\;.
\eeq

Definition of $E$ for a general causal set requires a notion of the degree of relativistic non-covariance. $E$ is selected in such a way that minimizes that non-covariance. If reference frame is identified with the axis of Alexandrov set, the degree of non-covariance is how much the answer changes depending on the choice of Alexandrov set. 

In case of linear fields in flat Minkowski space, non-covariance is zero as long as $E$ is appropriately adjusted. In case of discritized curved space, as a consequence of discretization it is not possible to select infinitesimal region, which means that some small non-covariance will appear as a result of specific behavior of a curvature, but as long as the fields are well behaved differentiable functions it should be very small. 

However, in case of general causal set, most of the assumptions that are made for the manifold can no longer be trusted, which means that it is possible that degree of non-covariance is large no matter what $E$ is selected to be. However, it is still possible to FORMALLY select $E$ in such a way that it would minimize the degree of non-covariance, even though the minimum would still be very large.  

In light of the fact that in the near lightcone region of point $p$ fields vary uncontrollably, the value of $E$ might have nothing to do with the one it would have had in linear case if all possible Alexandrov sets ($\alpha (p,q)$, $\tau(p,q)=\tau_0$) were considered. For that reason, the varying Alexandrov set is constrainted to vary within the boundaries of some other, larger, Alexandrov set $\alpha (P, Q)$. These smaller and larger Alexandrov sets satisfy the constraint
\beq \tau_1 = \tau (p,q) < \tau (P, Q) = \tau_2 \eeq
The variation, then, is defined
\beq Var_{\tau_1}({\cal L}, F, E, \alpha (P, Q)) = \eeq
\beq = max \{ {\cal L} (E, \alpha (p,q)) \vert P \prec p \prec q \prec Q \wedge \tau (p,q) = \tau_1 \} \nonumber \eeq
For any point $x$ in a causal set, $\alpha_{\tau 1, \tau_2} (x)$ is defined a set of triples $(E, P, Q)$ for which $Var_{\tau_1}({\cal L}, F, E, \alpha (P, Q))$ is minimized with a constraint $\tau (P,Q)= \tau_2$:
\beq \alpha_{\tau_1, \tau_2, F} (x) = \{ (E, P, Q) \vert \tau (P,Q) = \tau_2 \wedge \forall (E', P', Q') (\tau (P', Q') = \tau_2 \Rightarrow \nonumber \eeq
\beq \Rightarrow Var_{\tau_1}({\cal L}, F, E, \alpha (P', Q')) \geq Var_{\tau_1}({\cal L}, F, E, \alpha (P, Q))) \} \eeq
Typically, $\alpha_{\tau_1, \tau_2, F} (x)$ is a one-element set, and Lagrangian at $x$ is simply given by ${\cal L} (F, E, P,Q) $ where $(E,P,Q)$ is a unique element of that set. But in order to formally accommodate extremely rare cases where $\alpha_{\tau_1, \tau_2, F} (x)$ has more than one element, the Lagrangian at $x$ is formally defined as an averaging of the above over all elements of  $\alpha_{\tau_1, \tau_2, F} (x)$ :
\beq
{\cal L}_{\tau_1, \tau_2, F} (x) = \frac{1}{\sharp \alpha_{\tau_1, \tau_2, F} (x)} \sum_{(E,P,Q) \in \alpha (x)} {\cal L} (F, E, P,Q)\;.
\eeq
Now let's switch gears and go back to the expression for ${\cal L} (E, \alpha (p,q))$, which, from now on, will be referred to as pre-Lagrangian. As the examples in the beginning of this section illustrate, the expressions for ${\cal L} (E, \alpha (p,q))$ look similar for scalar, gauge, and gravitational fields in a sense that they all look like a linear combination of a function of $p$ and $q$ and some form of integral of the same function over the interior of $\alpha (p,q)$. So the natural question arises: why do Lagrangians take this particular form and not any other?

That question is answered by introducing a concept of type-2 Lagrangian generator and formally defining a procedure of going from type-2 Lagrangian generator to pre-Lagrangian in such a way that the above mentioned linear combination arises in a natural way if one formally follows the steps of the procedure. 

Suppose ${\cal J} \colon S^2 \rightarrow \mathbb{R}$ is some real valued function, and suppose $f,g \colon S^3 \rightarrow S^2$ are defined as follows:
\beq f(a,b,c)=(a,b) \; , \; g(a,b,c)= (a,c) \eeq
Then
\beq
\int_{\alpha (p,q)} \dd^d r\, {\cal J} (F, f(p,q,r))
= \int_{\alpha (p,q)} \dd^d r\, {\cal J} (F, p, q) = \nonumber \eeq
\beq = {\cal J} (F, p, q) \int_{\alpha (p,q)} \dd^d r
= V(\alpha (p,q)) {\cal J} (F, p, q) \eeq
and
\beq \int_{\alpha (p,q)} \dd^d r\, {\cal J} (F, g(p,q,r))
= \int_{\alpha (p,q)} \dd^d r\, {\cal J} (F, p, r)\;.
\eeq
Thus, the pre-Lagrangian is expressed as 
\beq {\cal L}_{\cal J} (F, E, p, q)  = \frac{{\cal J} (p,q)}{\tau^2 (p,q)}
+ \frac{E}{\tau^{d+2} (p,q)} \int \dd^d r\, {\cal J} (p, r) \eeq
can be rewritten as 
\beq
{\cal L} = \frac{1}{\tau^{d+2} (p,q)} \int_{\alpha (p,q)} \dd^d r\, ({\cal J}
(f(p,q,r)) + E {\cal J} (g(p,q,r)))\;.
\eeq
In order to preserve time-reversal symmetry, we would like to replace ${\cal J} (f (p,q, r)$ and ${\cal J} (g  (p,q, r)$ with ${\cal J} (f ( p, q, r)) + {\cal J} (f(q,p,r))$ and ${\cal J} (g( p, q, r)) + {\cal J} (g(q,p,r))$ respecitively. Thus, pre-Lagrangian is given by
\beq
{\cal L}_{\cal J} (F, E, p, q) = \frac{1}{\tau^{d+2} (p,q)} \int_{\alpha (p,q)}
\dd^d r \Big( {\cal J} (F, f (p,q,r)) + {\cal J} (F , f(q,p,r)) + \nonumber
\eeq
\beq + E \big( {\cal J} (F, g(p,q,r)) + {\cal J} (F, g(q,p,r)) \big) \Big)
\eeq
Since $\tau (p,q)$ is constant, it can be thrown away. Furthermore, in order to allow for the integrals such as 
\beq
\int \dd^d r\, \dd^d s\, \dd^d t\, (a (r, s) + a (s, t) + a (t, r))
\eeq
in electrodynamics, $f(p, q, r)$ and $g(p,q,r)$ should be generalized to $f(p,q, r_1, . . . , r_n)$ and $g(p,q, r_1, . . . , r_n)$. 
 
This leads to the final definition of type-2 Lagrangian generator and corresponding pre-Lagrangian. For the sake of completeness, and convenience to the reader, the following definition will formally include both the transition from Lagrangian generator ${\cal J}$ to pre-Lagrangian $ {\cal L} (F, E, p, q)$ just discussed as well as the transition from the above pre-Lagrangian to actual point-wise Lagrangian $ {\cal L} (F, x)$ discussed earlier in this section: 

DEFINITION: Let {\cal F} be a set of the possible distributions of a field of interest. A type-2 Lagrangian generator is a triple $({\cal J} \colon {\cal F} \times S^n \rightarrow \mathbb{R}, f \colon S^m \rightarrow S^n, g \colon S^m \rightarrow S^n)$. The pre-Lagrangian corresponding to the above type-2 Lagrangian generator is  ${\cal L}_{{\cal J}} \colon {\cal F} \times \mathbb{R} \times S^n \rightarrow \mathbb{R}$, given by 
\beq  {\cal L}_{\cal J} ( F, E, p, q) =  \sum_{p \prec r_i \prec q} \Big( {\cal J} (F, f (p,q,r_1, . . . , r_n)) + {\cal J} ( F , f(q,p,r_1, . . . ,r_n)) + \nonumber \eeq
\beq + E \big( {\cal J} (F, g(p,q,r_1, . . . ,r_n)) + {\cal J} (F, g(q,p,r_1, . . . , r_n)) \big) \Big) \eeq
The ``variation" of this pre-Lagrangian is a function $Var_{{\cal J}, \tau_1} \colon  {\cal F} \times \mathbb{R} \times \{(P,Q) \vert P \prec Q\}  \rightarrow \mathbb{R}$ given by
\beq Var_{\tau_1}({\cal J}, F, E, \alpha (P, Q)) = \eeq
\beq = max \{ {\cal L} (E, \alpha (p,q)) \vert P \prec p \prec q \prec Q \wedge \tau (p,q) = \tau_1 \} \nonumber \eeq
For any $x \in S$, the $(F, \tau_1, \tau_2)$-based neighborhood of $x$ is given by 
\beq \alpha_{\tau_1, \tau_2, F} (x) = \{ (E, P, Q) \vert \tau (P,Q) = \tau_2 \wedge \forall (E', P', Q') (\tau (P', Q') = \tau_2 \Rightarrow \nonumber \eeq
\beq \Rightarrow Var_{\tau_1}({\cal L}, F, E, \alpha (P', Q')) \geq Var_{\tau_1}({\cal L}, F, E, \alpha (P, Q))) \} \eeq
Finally, the pointwise Lagrangian density corresponding to $\cal J$ is given by 
\beq {\cal L}_{{\cal J}, \tau_1, \tau_2} (F, x) = \frac{1}{\sharp \alpha_{\tau_1, \tau_2, F} (x)} \sum_{(E,P,Q) \in \alpha (x)} {\cal L} (F, E, P,Q) \eeq

\subsection{Type-2 Scalar Fields} 

In the rest of the chapter, the machinery of type-2 Lagrangian generators will be used to predict Lagrangians for particular fields. While it was shown that gravity if viewed as type-1 field leads to erroneous results, the converse is not true: the fields that were successfully viewed as type-1 fields, such as scalar field, can be viewed as type-2 fields with equal success. 

Of course, in reality scalar field, just like other non-gravitational fields, are either type 1 or type 2; they can't be both at the same time. However, there are only two kinds of experimental tests between these theories:

1) Higher order terms created by curvature are predicted to differ between type-1 and type-2 theories.

2) A the type-2 theory predicts that masses, charges and gravitational constant differ from dimension to dimension, while a type-1 theory does not predict that. 

Since neither kind of experiment can be carried out in the near future, both kinds of theories should be studied for non-gravitational fields. In this section we will study type-2 scalar field. Then, in the following two sections we will do type-2 gauge and gravitational fields. 

The Lagrangian generator for scalar field is given by $({\cal J}, f, g)$ where 
\beq {\cal J} (\phi, r, s) =  (\phi (r) - \phi(s))^2 - \frac{1}{2}(m^2 \phi^2 (r))\eeq
\beq f(r_1, r_2, r_3, r_4) = (r_1, r_2) \; , \; g(r_1, r_2, r_3, r_4) = (r_3, r_4) \eeq
For the reasons that will soon become apparent, $m$ is not the actual mass, although it is related to it. In fact, $m$ is assumed to be very small, of the order of $\tau$. 
The above expression implies that the pre-Lagrangian for scalar field is given by 
\beq {\cal L} (\phi, E, p, q) = \int d^d r d^d s ({\cal J}(f (p, q, r, s)) + E {\cal J}(g(p, q, r, s))) = \eeq
\beq = \int d^d r d^d s ({\cal J} (\phi, p,q) + E {\cal J} (\phi r, s)) = {\cal J} (\phi, p, q) V^2 (\alpha (p,q)) + E \int d^d r d^d s {\cal J} (\phi, r, s) \nonumber \eeq
Here, the factors of $2$ appear as a result of permutting $p$ and $q$. By remembering that the volume of $n$ dimensional ball is 
\beq V( \rm Ball) = \frac{2 \pi^{n/2}}{n \Gamma (n/2)} r^n\eeq
we obtain
\beq V(\alpha (p,q)) =  \frac{2 \pi^{(d-1)/2}}{(d-1) \Gamma ((d-1)/2)}  \int_{- \tau/2}^{\tau/2} \Big( \frac{\tau}{2} - \vert t \vert \Big)^{d-1} dt = k_d \tau^d \eeq
where
\beq k_d = \frac{2 \pi^{(d-1)/2}}{(d-1) \Gamma (2^{d-2}d(d-1))} \eeq
for some constant $k_d$ that will be computted later. Substituting the above expression for volume, along the expression for $\cal J$ in the above equation for $\cal L$ we obtain
\beq {\cal L} (\phi, E, p, q) = k_d^2 \tau^{2d} (p,q) \Big(  (\phi (q) - \phi (p))^2 - \frac{m^2}{2} (\phi^2 (p) + \phi^2 (q))  \Big)  \nonumber \eeq
\beq + E \int d^d r d^d s (  (\phi (r) - \phi (s))^2 - m^2 \phi^2 (r)) \eeq
We would now like to compute Lagrangian density, if the space time is assumed to be flat Minkowskian and $\phi$ is assumed to be linear. 

Lets start from the mass term. It is given as
\beq {\cal L}_m (\phi, E, p, q) = \frac{m^2 k_d^2}{2} \tau^{2d} (p,q) (\phi^2 (p) + \phi^2 (q)) + Em^2 \int d^d r d^d s \phi^2 (r) \eeq
The above expression tells us that the leading order of the mass term is $m \tau^{2d}$. When we will get to the kinetic term, it will be shown that the order of magnitude that we are interested in is $\tau^{2d+2}$. Thus, if $m$ is assumed to be of the order of $\tau$, then the leading order is $\tau^{2d+2}$ which coincides with a leading order for the kinetic term. 

This means that, as far as mass term is concerned, we can throw away all the higher order terms, which can be done by using an approximation $\phi \approx \phi_0$ which tells us 
\beq {\cal L}_m  (\phi, E, p) \approx m^2 k_d^2 \tau^{2d} (p,q) \phi_0^2 + Em^2 \phi_0^2 V^2 (\alpha (p,q)) = \nonumber \eeq
\beq = (1+E) m^2 k_d^2 \phi_0^2 \tau^{2d} (p,q) \eeq
Now let's look at the kinetic term. The linearity assumptions imply that
\beq (\phi (r) - \phi (s))^2 = (r^{\mu} - s^{\mu})(r^{\nu} - s^{\nu}) \partial_{\mu} \phi \partial_{\nu} \phi \eeq
Consider a coordinate system in which $t$ axis passes through $p$ and $q$, while origin lies in the midle between these points. Denoting $\tau (p,q)$ by $\tau$, 
\beq p= (-\tau/2, 0,0,0) \; , \; q= (\tau/2, 0, 0 ,0) \eeq
In this coordinate system, 
\beq \int_{\alpha (p,q)} d^d r r^{\mu} = 0 \eeq
since the above integrand is antisymmetric with respect to the center of Alexandrov set. By slicing Alexandrov set on balls $t=const$, we get 
\beq \int_{\alpha (p,q)} (x^0)^2 d^d r = \frac{2 \pi^{(d-1)/2}}{(d-1) \Gamma ((d-1)/2)}  \int_{-\tau/2}^{\tau/2}t^2 \Big( \frac{\tau}{2} - \vert t \vert \Big)^{d-1} d  t = I_{d0} \tau^{d+2}\eeq
where
\beq I_{d0} = \frac{2 \pi^{(d-1)/2}}{(d-1) \Gamma (2^{d-1}(d-1)d(d+1)(d+2))} \eeq
Furthermore, it can be shown that 
\beq \int_{\alpha (p,q)} (x^k)^2 d^d r = I_{d1} \tau^{d+2} \eeq
where, by cylindrical symmetry, the coefficient is the same for each $k$ and is given as 
\beq I_{d1}=  \frac{2 \pi^{\frac{d}{2}-1}}{(d-2) \Gamma \Big(\frac{d}{2}-1 \Big)} \eeq
Substituting these expressions into the integral we obtain
\beq \int d^d r d^d s (\phi (r) - \phi (s))^2 = \Big( \int d^d s \Big) \sum_{\mu =0}^{d-1} d^d r (r^{\mu})^2 + \Big( \int d^d r \Big) \sum_{\mu =0}^{d-1} d^d s (s^{\mu})^2 = \nonumber \eeq
\beq = 2k_d \tau^{2d+2} ((I_{d0} + I_{d1} (d-1)) (\partial_0 \phi)^2 - I_{d1} (d-1) \partial^{\mu} \phi \partial_{mu} \phi ) \eeq
Thus, the kinetic term of the pre-Lagrangian is
\beq {\cal L}_{kin} (\phi, E, p, q) = \eeq
\beq= k_d \tau^{2d+2} ((\partial_0 \phi)^2 (k_d + 2E_d (I_{d0} + I_{d1} (d-1))) - 2 E_d I_{d1} (d-1) \partial^{\mu} \phi \partial_{\mu} \phi ) \nonumber \eeq
Switching from the coordinate system in which $t$ axis passes through $p$ and $q$ to the arbitrary one, the result becomes 
\beq {\cal L} (\phi, E, p, q) = k_d \tau^{2d} (p,q) (q^{\mu} - p^{\mu})(q^{\nu} -p^{\nu})\partial_{\mu} \phi \partial_{\nu} \phi (k_d + 2E_d (I_{d0} + I_{d1} (d-1))) - \nonumber \eeq
\beq - 2 k_d E_d I_{d1}  \tau^{2d+2}(d-1) \partial^{\mu} \phi \partial_{\mu} \phi  \eeq
If the choice of points $p$ and $q$ varies with constraints that both the mid-point $0$ between $p$ and $q$ as well as the Lorentzian distance between the two points are fixed, the ${\cal L} (\phi, E, p, q)$ undergoes the variation of the order of $\tau^{2d+2}$ due to $(q^{\mu} - p^{\mu})(q^{\nu} -p^{\nu})\partial_{\mu} phi \partial_{\nu} \phi$ term. Mass term, on the other hand, only gives variations to higher orders. Thus, if the variations of the orders higher than $\tau^{2d+2}$ are neglected, then the variation can be ``minimized", or in this particular case, set to $0$, if
\beq
k_d + 2E_d (I_{d0} + I_{d1} (d-1)) = 0\;,
\eeq
which determines the value of $E_d$:
\beq
E_d = - \frac{k_d}{2(I_{d0} + I_{d1} (d-1))}\;.
\eeq
Substituting this into the expression for Lagrangian gives
\beq
{\cal L} = \frac{I_{d1}k_d^2 (d-1)}{I_{d0} + I_{d1} (d-1)} \tau^{2d+2} \partial^{\mu} \phi \partial_{\mu} \phi - \eeq
\beq - m^2 \phi^2 k_d^2 \tau^{2d} \Big(1-\frac{k_d}{2(I_{d0} + I_{d1} (d-1))} \Big)\;.
\nonumber \eeq
This Lagrangian can be rewritten as 
\beq
{\cal L} =  \hbar_d ^2 \frac{v_0}{2} \partial^{\mu} \phi_d \partial_{\mu} \phi_d - \frac{v_0 m_d^2}{2} \phi_d^2\;,
\eeq
where $v_0$ is volume taken up by one point and 
\beq \hbar_d \phi_d = \frac{k_d}{v_0} \tau^{d+1} \sqrt{ \frac{I_{d1}(d-1)}{I_{d0} + I_{d1} (d-1)} } \eeq
\beq m_d = \frac{m}{v_0} k_d \tau^d \sqrt{1-\frac{k_d}{2(I_{d0} + I_{d1} (d-1))}} \eeq
In future sections other fields will be similarly scaled, but the coefficients will be different from field to field. At first this might seem wrong since the kinetic terms of all Lagrangians have the same coefficient $1$ in standard quantum field theory. But it can be easilly shown that this difference does not amount to anything but the change of overall factor: 
\beq \int [{\cal D} \phi_1 . . . {\cal D} \phi_n] \exp (iS (\phi_1)  + . . . + iS(\phi_n)) = \eeq
\beq = \rho_{1d} . . . \rho_{nd} \int [{\cal D} \phi_{1d} . . . {\cal D} \phi_{nd}] \exp (iS(\phi_{1d}/ \rho) + . . . + iS(\phi_{nd}/ \rho)) \nonumber \eeq
where
\beq \phi_{kd} = \rho_{kd} \phi_k \eeq

\subsection{Charged Type-2 Scalar Field}

\noindent Consider a charged spin-0 particle, described by a set of complex scalar fields $\phi = (\phi_1, ..., \phi_n)$ coupled to a SU($n$) gauge field. (Notice that in this approach to the dynamics of matter fields in causal set theory, although it will be assumed that spacetime is discretized, the internal degrees of freedom will still have a continuous invariance group.) The dynamics of such a field can be described in the continuum starting with the matter Lagrangian density
\beq
{\cal L}_{\rm m}(g_{\mu\nu},\phi,A_\mu;x) = \half\, |g|^{1/2}\,\big[g^{\mu\nu}\,
(D_\mu\phi)^{\dagger}\,(D_\nu\phi) - m^2\,\phi^{\dagger}\phi\big]\;, \label{Lm}
\eeq
where the gauge covariant derivative is defined as usual by
$D_\mu\phi^a:= \partial_\mu\phi^a + \ii\,e\, A_\mu{}^a{}_b\,\phi^b$,
with $A_\mu = A_\mu{}^k\,T^k$ the Lie-algebra-valued connection form representing the gauge field on a differentiable manifold. (Here, Latin indices $a$, $b$, ..., are Lie-algebra tensor indices, while $k$, $l$, ..., label elements of the basis $T^k$ of the Lie algebra.) In the causal set context, the scalar field will be simply replaced by a corresponding field defined at each causal set element, but to write down the action, it is important to specify what variables will replace $A_\mu$.

As done in the type-1 approach, gauge field is defined in terms of holonomies, where by holonomy refers to the group transformation corresponding to the parallel transport of a Lie-algebra-valued field such as $\phi$ between two points $p$ and $q$. In a differentiable Lorentzian manifold $M$ (of dimension $d$), holonomy is defined as the function $a : M\times M \rightarrow {\rm TSU}(n)$, where T stands for tangent bundle, which means that TSU($n$), being a tangent bundle to $\SU(n)$, consists of all $n \times n$ tensors with trace $0$. This map assigns to any two elements $p,\,q\in M$ the holonomy of $A_\mu$ along the geodesic segment $\gamma(p,q)$ connecting $p$ and $q$ in $M$, given by
\beq
a(p,q) = \int_{\gamma(p,q)} A_\mu{}^k\,T^k\,\dd x^\mu\;, \label{holo}
\eeq
in terms of which the expression $D_\mu\phi(x)$ appearing in the scalar field Lagrangian arises from the leading-order term in the expansion of the expression $(1+a(x,y))\,(\phi(y) - \phi(x))$ .

This means that the causal set version of the charged scalar field Lagrangian can be obtained by making some simple substitutions in the one obtained in the previous section for the Klein-Gordon field. Thus,  if gauge field $a$ is assumed to be fixed and not subject to any Lagrangians, then the type-2 Lagrangian generator for a matter field $\phi \in \mathbb{R}^n$ interacting with $a$ is given by $({\cal J}, f, g)$ where 
\beq {\cal J}_{scal} (\phi, a, r, s) = \vert \phi (r) - a(r, s) \phi (s) \vert ^2 + \frac{m^2}{8} (\phi^* (r) + \phi^*(s))(\phi(r)+ \phi(s))  \eeq
and
\beq f_s(r_1, r_2, r_3) = (r_1, r_3) \; , \; g_s(r_1, r_2, r_3) = (r_1, r_2) \eeq
However, as discussed in the next section, $a$, itself, is subject to type-2 Lagrangian generator given by 
\beq {\cal J}_{YM} (a, r_1, r_2, r_3) = {\rm tr}[(a(r_1, r_2)+a(r_2, r_3)+a(r_3,r_1))^2] \nonumber \eeq
\beq  f_{YM}(r_1,r_2,r_3,r_4,r_5)=(r_3,r_4,r_5) \; , \; g_{YM}(r_1,r_2,r_3,r_4,r_5) = (r_1,r_2,r_3)\;. \eeq
In order for the theory to possess SU($n$) symmetry in the setup, these two Lagrangian generators are combined into one as $({\cal J}_{tot}, f, g)$, 
\beq {\cal J}_{tot} (a, r_1, r_2, r_3) = {\cal J}_{scal} (\phi, a, r_1, r_2) + {\cal J}_{YM} (a, r_1, r_2, r_3) \nonumber \eeq
\beq f_{tot}(r_1,r_2,r_3,r_4,r_5)= f_{YM}(r_1,r_2,r_3,r_4,r_5) \eeq 
\beq g_{tot}(r_1,r_2,r_3,r_4,r_5)= g_{YM}(r_1,r_2,r_3,r_4,r_5) \nonumber \eeq

\subsection{Type-2 Yang-Mills Field}

In this section, the main goal is to express the Yang-Mills Lagrangian density,
\beq
{\cal L}_{\rm YM}(g_{\mu\nu},A_\mu;x)
= \half\,|g|^{1/2}\,{\rm tr}(F_{\mu\nu}F^{\mu\nu})\;,
\eeq
in terms of the holonomy variables for the gauge field introduced in the previous section, as well as variables describing the geometry that are meaningful in the causal set context, namely causal relations, and either volumes or timelike lengths. Once this is done, the Lagrangian density can be easily rewritten in the discrete setting.

In the causal set context, the gauge field is defined as $a \colon S^2 \rightarrow {\rm TSU}(n)$ where TSU($n$) denotes a tangent bundle to SU($n$), thus it consists of $n \times n$ tensors of trace $0$. The type-2 Lagrangian generator for gauge field is given as $({\cal J}, f, g)$, where
\beq
{\cal J} (a, r_1, r_2, r_3) = {\rm tr}[(a(r_1, r_2)+a(r_2, r_3)+a(r_3,r_1))^2]
\eeq
and 
\beq
f(r_1,r_2,r_3,r_4,r_5)=(r_1,r_2,r_3) \; , \; g((r_1,r_2,r_3,r_4,r_5) = (r_3,r_4,r_5)\;. \eeq
This means that in the case of flat Minkowski space, the pre-Lagrangian is given by
\beq {\cal L} (a,E,p, q) = \int_{\tau (p,q)} d^d r d^d s d^d t ({\cal J}(f(p,q, r, s, t)) + {\cal J}(f(q,p, r, s, t)) +  \nonumber \eeq
\beq + E({\cal J}(g(p,q, r, s, t)) + {\cal J}(g(q,p, r, s, t)))) = \nonumber \eeq
\beq = \int_{\tau (p,q)} d^d r d^d s d^d t ({\cal J}(p, q, r) + {\cal J}(p, q, r) + E({\cal J}(r, s, t) + {\cal J}(r, s, t))) = \nonumber \eeq
\beq =  V^2 (\alpha (p,q)) \int_{\tau (p,q)} d^d r ({\cal J}(p,q, r) + \eeq
\beq + {\cal J}(q,p, r)) + 2 E \int_{\tau (p,q)} d^d r d^d s d^d t {\cal J}(r, s, t)  \nonumber \eeq
It can be easily seen that 
\beq {\cal J} (p,q,r) = {\cal J}(q,p,r) \eeq
which implies
\beq {\cal L} = 2 \Big( V^2 (\alpha (p,q)) \int_{\tau (p,q)} d^d r {\cal J}(p,q, r) + E \int_{\tau (p,q)} d^d r d^d s d^d t {\cal J}(r, s, t) \Big) \eeq
Assume that the gauge field is differentiable and reasonably well behaved. In particular, it is well-behaved-enough for $F_{\mu\nu}{}^k$ to be approximately constant in $\alpha(P,Q)$ whenever $(P,Q,E) \in \alpha_{a, \tau_1, \tau_2} (u)$ for some $u \in S$.  Assume for definiteness that the three points are spacelike related. Choose a coordinate system so that $r$ coincides with the origin, the $x$ axis points from $r$ to $s$, and the $y$ axis is perpendicular to the $x$ axis in the $rst$ plane. Then in this coordinate system $a = (0,0,0,...)$, $b = (0,b^1,0,...)$, $c = (0,c^1,c^2,...)$. 

The flux of $F_{\mu\nu}{}^k$ through the interior of that triangle is expressed by the relationship
\beq
a(r,s)+ a(s,t)+ a(t,r) =  \half\, s^1\, t^2\, F_{12}{}^k\,T^k + ...
\eeq
This result generalizes to points at arbitrary locations, and can be written covariantly as
\beq
a(r,s)+a(s,t)+ a(t,r) =  \half\, F_{\mu\nu}{}^k\,T^k
(s^\mu - r^\mu)(t^\nu - r^\nu) + ...
\eeq
Recalling that, for SU($n$), tr$(T^kT^l) = C_2\, \delta_{kl}$, to leading order in the separation between points,
\beq
{\rm tr}[(a(r,s)+a(s,t)+a(t,r))^2] = \label{trabc}\eeq
\beq = \frac{C_2}{4} F_{\mu\nu}{}^k (s^\mu - r^\mu)(t^\nu - r^\nu)F_{\rho\sigma}{}^k\, (b^\rho - a^\rho)(c^\sigma - a^\sigma)\nonumber
\eeq
Let's start from
\beq \int_{\tau (p,q)} d^d r d^d s d^d t {\cal J}(r, s, t) \eeq
Expand the right-hand side of Eq (\ref{trabc}), and integrate term by term. Clearly any term with an odd number of powers of any variable will integrate to 0. Thus, the only terms that may potentially survive the integration are those of the form $r^{\mu}\, r^{\nu}\, r^{\rho}\, r^{\sigma}$ or quadratic terms in two of the three points. Simple counting of terms gives
\bea
& &\int_{p\prec r,s,t \prec q} \dd^dr\,\dd^ds\,\dd^dt\; {\rm tr}\big[(a(r,s)+a(s,t)+a(t,r))^2\big] \nonumber \\
& &= \frac{C_2}{4}\, F_{\mu\nu}{}^k\, F_{\rho\sigma}{}^k \int_{p\prec a,b,c\prec q} \dd^da\, \dd^db\,\dd^dc\; (s^\mu - r^\mu)( t^\nu - r^\nu)( s^\rho - r^\rho)( t^\sigma - r^\sigma) \nonumber \\
& &= \frac{C_2}{4}\; \bigg[ 3\,V\sum_{k,\mu,\nu} (F_{\mu\nu}{}^k)^2
\bigg(\int_{\alpha(p,q)} \dd^da\, (r^\mu)^2 \bigg) \bigg(\int_{\alpha(p,q)} \dd^db\, (s^\nu)^2 \bigg) \nonumber \\
& & \kern60pt -\ V^2 F_{\mu\nu}{}^k\, F_{\rho\sigma}{}^k\int_{\alpha(p,q)}
\dd^da\, r^\mu\, r^\nu\, r^\rho\, r^\sigma \bigg]\;, \label{intabc}
\eea
where $V$ is the volume of the Alexandrov set $\alpha(p,q)$.

The only terms of $F_{\mu\nu}{}^k\, F_{\rho\sigma}{}^k\, r^{\mu}\, r^{\nu} r^{\rho} r^{\sigma}$ that survive integration are the ones whose indices are pairwise equal. But if either $\mu = \nu$  or $\rho = \sigma$  then $F_{\mu\nu}{}^k = 0$ or $F_{\rho\sigma}{}^k = 0$, respectively, which would set the whole thing to 0. Thus, the only options are $\mu = \rho$, $\nu = \sigma$ and $\nu = \rho$, $\mu = \sigma$.  The antisymmetry of $F_{\mu\nu}{}^k$ then implies that these two cases are opposites of each other, which in turn implies that $F_{\mu\nu}{}^k\, F_{\rho\sigma}{}^k\,r^\mu\, r^\nu\, r^\rho\, r^\sigma = 0$. Thus, Eq (\ref{intabc}) becomes
\bea
& &\int_{p\prec r,s,t\prec q} \dd^dr\,\dd^ds\,\dd^dt\;
{\rm tr} \big[ (a(r,s)+a(s,t)+a(t,r))^2 \big]
\nonumber\\
& &= \frac{3\,VC_2}{4} \sum_{k,\mu,\nu} (F_{\mu\nu}{}^k)^2 \Big(\int_{\alpha(p,q)}
\dd^dr\, (r^\mu)^2\Big) \Big(\int_{\alpha(p,q)} \dd^ds\, (s^\nu)^2\Big) \nonumber\\
& &= \frac{3\, k_d C_2 \tau^{3d+4}}{2}\, \sum_k \Big(J^0 J^1 \sum_{i=1}^{d-1}
(F_{i0}{}^k)^2 + (J^1)^2 \sum_{i<j} (F_{ij}{}^k)^2 \Big)\;, \label{intabc2}
\eea
where $J^{\mu}=\tau^{-d-2}\int_{\alpha(p,q)} \dd^dx\, (x^\mu)^2$, or in other words
\beq
J^0 = 
= \frac{2 \pi^{(d-1)/2} }{2^d\, (d-1) \, d\,(d+1)\,(d+2)} \, \Gamma ((d-1)/2) \eeq
\beq J^1 = ... = J^{d-1} = 
=  \frac{2 \pi^{\frac{d}{2}-1} }{2^{d+1} (d-2)d(d+2)} \Gamma((d-2)/2) \int_{- \pi/2}^{\pi/2)} (cos \theta)^d d \theta 
\eeq
Now let's move to the second integral, $\int_{\alpha(p,q)} \dd^dx\;{\rm tr}[(a(p,x)+a(x,q)+a(q,p))^2]$, where $p \prec q$ are the endpoints of the Alexandrov set. Rewriting Eq (\ref{trabc}) in terms of the points $p$, $x$, and $q$ gives
\bea
& &{\rm tr}\big[ (a(p,x)+a(x,q)+f(q,p))^2 \big] \\ \noalign{\medskip}
& &=\ \fourth\,C_2\,F_{\mu\nu}{}^k\,(p^\mu-x^\mu)\,(q^\nu-x^\nu)\,F_{\rho\sigma}{}^k\, (p^\rho-x^\rho)\,(q^\sigma-x^\sigma)\;. \nonumber
\label{trApB}
\eea
Again this can be expanded and integrated term by term. There are several conditions each term has to meet, in order for its integral not to vanish. First of all, it needs to contain an even number of factors of $x$. Secondly, as was shown before, for symmetry reasons
\beq
F_{\mu\nu}{}^k\, F_{\rho\sigma}{}^k \int_{\alpha(p,q)}\, \dd^dx\,
\, x^\mu\, x^\nu\, x^\rho\, x^\sigma = 0\;.
\eeq
Finally, $F_{\mu\nu}\, p^\mu p^\nu = F_{\mu\nu}\, q^\mu q^\nu = 0$
and identities $p = (-\frac{\tau}{2},0,0,0)$ and $q = (\frac{\tau}{2},0,0,0)$ imply $F_{\mu\nu}\, p^\mu q^\nu = -F_{\mu\nu}\, p^\mu p^\nu = 0$. The only terms in Eq (\ref{trApB}) that do {\em not\/} vanish for any of the above reasons are 
\bea
& &F_{\mu\nu}{}^k\, F_{\rho\sigma}{}^k\, p^\mu\, x^\nu\, p^\rho\, x^\sigma\,,\qquad
F_{\mu\nu}{}^k\, F_{\rho\sigma}{}^k\, p^\mu\, x^\nu\, x^\rho\, q^\sigma\,,
\nonumber\\ \noalign{\medskip}
& &F_{\mu\nu}{}^k\, F_{\rho\sigma}{}^k\, x^\mu\, q^\nu\, p^\rho\, x^\sigma\,,\qquad
F_{\mu\nu}{}^k\, F_{\rho\sigma}{}^k\, x^\mu\, q^\nu\, x^\rho\, q^\sigma\,.
\nonumber
\eea
Plugging in the coordinate values of $p$ and $q$, each
of the above four expressions evaluates to 
$\fourth\,\tau^2\, F_{\mu0}{}^k\, F_{\rho0}{}^k\, x^\mu\, x^\rho$.
In order for this not to be an odd function, $\mu = \rho$ has to hold, and
in order for $F_{\mu 0}$ to be non-zero $\mu \not= 0$ has to hold. Thus,
this becomes $\fourth\,\tau^2\, (F_{i0}{}^k)^2\, (x^i)^2$ and, since
there are four such terms, the integral becomes
\bea
& & \int_{\alpha(p,q)} \dd^dx\;{\rm tr}\big[ (a(p,x)+a(x,q)+a(q,p))^2 \big]
\nonumber \\
& &= \fourth\,C_2\, \tau^2 \sum_{i=1}^{d-1} \int_{\alpha(p,q)} \dd^dx\,(F_{i0}{}^k)^2\, (x^i)^2 = \fourth\,C_2\, \tau^{d+4} J^1 \sum_{i=1}^{d-1} (F_{i0}{}^k)^2\;,
\eea
where, based on rotational symmetry, $J^1 = ... = J^{d-1}$ was used. 

Substituting this into original expression for pre-Lagrangian and doing some basic algebra, the latter becomes
\beq
{\cal L} (a,E,p,q) = \tau^{3d+4} C_2 k_d J^1 \Big( (3E J^0 + \frac{k_d}{4}) \sum(F^k_{i0})^2 + 3E J^1 \sum_{i<j} (F^k_{ij})^2 \Big)\;.
\eeq
In order to get rid of the variation that results from different choices of Alexandrov sets, we would like the above expression to be relativistically invariant; in other words, we would like this to be proportional to $F^{\mu\nu} F_{\mu\nu}$. This means
\beq
3E_d j^0 + \frac{k_d}{2} = -3E_d J^1\;,
\eeq
which implies
\beq
E_d = - \frac{k_d}{6(J^0 +J^1)}\;.
\eeq
Substituting this expression for $E_d$ gives a Lagrangian
\beq
{\cal L} (a, p, q) = - \frac{k_d^2 (J^1)^2 C_2}{4(J^0+J^1)} \tau_2^{3d+4} F^{\mu \nu} F_{\mu \nu}\;.
\eeq
This can be expressed as
\beq {\cal L} (a, p, q) = - \frac{v_0}{4} F_d^{\mu \nu} F_{d \mu \nu} \eeq
where $v_0$ is a volume taken up by a single point if $F_d^{\mu \nu}$ is defined as
\beq F_d^{\mu \nu} = k_d J^1 F^{\mu \nu} \sqrt{ \frac{C_2 \tau^{3d+4}}{v_0 (J^0 + J^1)}} \eeq
This means 
\beq A_d^{\mu} = k_d J^1 A^{\mu} \sqrt{ \frac{C_2 \tau^{3d+4}}{v_0 (J^0 + J^1)}} \eeq
From this we can determine how the charge changes from dimension to dimension. On the first glance, since we haven't yet defined Lagrangian for fermions, we are only ready to talk about the charge of bosonic fields. However, a simple symmetry consideration allows us to overcome this barrier and include fermionic charges in a discussion.  Whether a field is bosonic or fermionic, we would like to be able to say 
\beq \partial_{\mu} \rightarrow \partial_{\mu} + eA_{\mu} \eeq
We would also like to be able to say 
\beq \partial_{\mu} \rightarrow \partial_{\mu} + e_d A_{d\mu} \eeq
This means that, regardless whether the field in question is bosonic or fermionic, and regardless of any other properties of the field (such as mass) it should satisfy
\beq eA^{\mu} = e_d A^{\mu}_d \eeq
This immediately implies that, both for bosons and fermions,
\beq e_d = \frac{e}{k_d J^1} \sqrt{\frac{v_0 (J^0 + J^1)}{C_2 \tau^{3d+4}}}\;. \eeq

\subsection{Type-2 Gravity}

Now let us move to the number one reason that type-2 fields were invented: gravitational field.  The Lagrangian generator for gravity is $({\cal J}, f, g)$ where
\beq {\cal J} (r, s, t) =  \begin{cases}
\frac{1}{8 \pi G} \;
& \hbox{if} \; r \prec t \prec s \\
0 \;
& \hbox{otherwise }
\end{cases}
\eeq
The function $f$ and $g$ are defined as
\beq f(r,s,t,u)=(r,s,t) \; , \; g(r,s,t,u) = (r, t, u)\eeq
This means that pre-Lagrangian is given by
\beq {\cal L} = \int d^d r d^d s \sqrt{(-1)^{d-1} \det g(r)} \sqrt{(-1)^{d-1} \det g(s)} \times \eeq
\beq \times ({\cal J}(f(p,q,r,s)) +E {\cal J} (g(p,q,r,s))) \nonumber \eeq
Since ${\cal J}$ is a constant inside a certain domain and is $0$ outside, the above integrals amount to restricting $r$ and $s$ to a certain domain. The first term is ${\cal J} (p, q, r)$ thus its restriction is $p \prec r \prec q$. That statement, of course, is trivial which means that the first term has no restriction at all. On the other hand, the second term is ${\cal J} (p, r, s)$, thus it has a restriction $p \prec r \prec s \prec q$. Thus, the pre-Lagrangian is 
\beq {\cal L} =  \frac{1}{8 \pi G} \Big(\int d^d r d^d s \sqrt{(-1)^{d-1} \det g(r)}\sqrt{(-1)^{d-1} \det g(s)} + \nonumber \eeq
\beq + E \int_{p \prec r \prec s \prec q} d^d r d^d s \sqrt{(-1)^{d-1} \det g(r)}\sqrt{(-1)^{d-1} \det g(s)} \Big) = \eeq
\beq = (V(\alpha (p,q)))^2 + E \int d^d s \sqrt{(-1)^{d-1} \det g(s)}V(\alpha (p,s)) \nonumber \eeq 
One can parametrize the interior of that Alexandrov set with normal geodesic coordinates around $p$. Suppose index $\mu$ stand for arbitrary coordinates, satisfying
\beq g_{\mu \nu} (p) = \eta_{\mu \nu} \eeq
 while not necessary geodesic. Then one can define the normal geodesic coordinates derived from the above in the following way: 
\beq r^{\overline{\mu}} =  \eta^{\mu \nu} \partial_{\nu} \vert_p \gamma_{pr} \sqrt {\int_{\gamma_{pr}} g_{\mu \nu} dx^{\mu} dx^{\nu}}\eeq
where $\gamma_{ab}$ denotes geodesic segment connecting $a$ and $b$. 
From this, it is straightforward to see that the following equation is satisfied exactly:
\beq \tau (p,r) = \eta_{\overline{\mu} \overline{\nu}} r^{\overline{\mu}} r^{\overline{\nu}} \eeq
It should be noticed, however, that the above is true only if one of the two points is $p$:
\beq \tau (r,s) \neq \eta_{\overline{\mu} \overline{\nu}} (s^{\overline{\mu}}-r^{\overline{\mu}})(s^{\overline{\nu}}- r^{\overline{\nu}}) \eeq
However, a different causal relation, $\prec_p$ will be introduced in addition to already existing one $\prec$. While $\prec$ matches flat space expectations only if the pair of points in interest includes $p$, $\prec_p$ does so for arbitrary pairs of points:
\beq a \prec_p b \Leftrightarrow \eta_{\overline{\mu} \overline{\nu}}(b^{\overline{\mu}}-a^{\overline{\mu}})(b^{\overline{\nu}}- a^{\overline{\nu}}) \geq 0 \eeq
At the same time, we will retain an original causal relation, $\prec$ for which the above is not true:
\beq \prec \neq \prec_p \eeq
Let define the following notation:
\beq J^+ (a) = \{r \succ a \} \; , \; J_p^+ (a) = \{r \succ_p a \}  \eeq
\beq J^- (a) = \{r \prec a \} \; , \; J_p^- (a) = \{r \prec_p a \} \eeq
\beq \alpha (a, b) = J^+ (a) \cap J^- (b) \; , \; \alpha_p (a, b) = J_p^+ (a) \cap J_p^- (b) \eeq
Also, lets define $V_p (a, b)$, which is not to be confused with $V (\alpha_p (a,b))$, as follows
\beq V_p (a, b) = k_d (\eta_{\mu \nu} (b^{\mu} - a^{\mu})(b^{\nu}-a^{\nu}))^{d/2} \eeq 
It is important to notice that 
\beq Vol (\alpha_p (a,b)) \neq V_p (a,b) \eeq
because the definition of $V_p$ neglects $\sqrt{(-1)^{d+1} g}$ factor in the volume element. But, for our purposes, $V_p$ as defined above is the simplest to use. On the other hand, $V(a,b)$ is defined as the actual volume of Alexandrov set:
\beq V(a, b) = Vol (\alpha (a, b)) \eeq
Then, the integral of the volume can be expanded as follows
\beq \int_{\alpha (p,q)} d^d r \sqrt{(-1)^{d-1} \det g} V(\alpha(p,r)) = \int_{\alpha_p (p,q)} d^d r V_p(p,q) + \nonumber \eeq
\beq + \Delta_{1d} (p,q) + \Delta_{2d} (p,q) + \Delta_{3d} (p,q) + 0 (\tau^{2d + 4}) \eeq
where $\Delta_{1d} (p,q)$ is an error due to the mismatch between $\alpha (p,q)$ and $\alpha_p (p,q)$, $\Delta_{2d} (p,q)$ is a correction due to the error in $V(\alpha (p,r))$  and $\Delta_3$ is an error due to the use of $d^d r$ instead of $d^d r \sqrt{(-1)^{d-1} det g}$ for a volume element. They are formally defined as follows: 
\beq \Delta_{1d}(p,q) = \int_{(\alpha (p,q) \setminus \alpha_p (p,q))} d^d r V_p (p,r) - \int_{\alpha_p (p,q) \setminus \alpha (p,q))} d^d r V_p (p,r) \eeq
\beq \Delta_{2d}(p,q)= \int_{\alpha_p (p,q)} d^d r (V(p,r) - V_p (p,r)) \eeq
\beq \Delta_{3d}(p,q) = \int_{\alpha_p (p,q)} d^d r (\sqrt{(-1)^{d-1} det g} -1) \eeq
Whenever any of these three correction terms are computted, the other two are neglected since the ``correction of the correction" is of the order $0(\tau^{2d+4})$ while the calculation is performed to order $0(\tau^{2d+2})$. Thus, the shape of Alexandrov set is assumed to be unchanged in the calculation of $\Delta_{2d}$ and $\Delta_{3d}$, $\sqrt{(-1)^{d-1} \det g}$ is dropped in calculation of $\Delta_{1d}$ and $\Delta_{2d}$ and the correction to $V(\alpha (p,x))$ is neglected in calculation of $\Delta_{1d}$ and $\Delta_{3d}$

Let's start with computting $\Delta_{1d}$.

In normal coordinates, the lightcone of $p$ is not deformed by curvature, while lightcone of $q$ still is: 
\beq J^+ (p) = J_p^+ (p) \; , \; J^- (q) \neq J_p^- (q) \eeq
Substitution of above into simple set theory algebra gives
\beq \alpha_p (p,q) \setminus \alpha (p,q) = J_p^+(p) \cap (J_p^- (q) \setminus J^-(q)) \subset J_p^- (q) \setminus J^-(q) \eeq
\beq \alpha (p,q) \setminus \alpha_p (p,q) = J_p^+(p) \cap (J^- (q) \setminus J_p^-(q)) \subset J^- (q) \setminus J_p^-(q) \eeq
Assuming that $J+p^-(q)$ and $J^-(q)$ are very close to each other, the above implies that most of the contribution to $\Delta_{1d}$ comes from the vicinity of $J_p^- (q)$. So, lets evaluate $V_p (p,r)$ for $r \in J_p^- (q)$. In other words, assume that 
\beq \sum (r^{\overline{k}})^2 = (\tau - r^{\overline{0}})^2 \eeq
This implies that 
\beq \eta_{\overline{\mu} \overline{\nu}} r^{\overline{\mu}} r^{\overline{\nu}} = (r^{\overline{0}})^2 - (\tau - r^{\overline{0}})^2 = \tau (2r^{\overline{0}} - \tau) \eeq
By using
\beq 2r^{\overline{0}} - \tau = r^{\overline{0}} - (\tau - r^{\overline{0}}) = r^{\overline{0}}- \sqrt{\sum (r^{\overline{k}})^2} \eeq
this becomes
\beq \eta_{\overline{\mu} \overline{\nu}} r^{\overline{\mu}} r^{\overline{\nu}} = \tau_0 \Big( r^{\overline{0}} - \sqrt{\sum (r^{\overline{k}})^2} \Big) \eeq
Let's define two functions $\chi_1$ and $\chi_2$ as follows:
\beq \chi_1 (r) = \int_{\alpha (r,q)} k_d (\eta_{\overline{\mu} \overline{\nu}} r^{\overline{\mu}} r^{\overline{\nu}})^{d/2} - \int_{\alpha_p (r,q)} k_d (\eta_{\overline{\mu} \overline{\nu}} r^{\overline{\mu}} r^{\overline{\nu}})^{d/2} \eeq
\beq \chi_2 (r) = \int_{\alpha (r,q)} k_d \tau_0^{d/2} \Big(r^{\overline{0}} - \sqrt{\sum (r^{\overline{k}})^2}\Big)^{d/2} - \nonumber \eeq
\beq -\int_{\alpha_p (r,q)} k_d \tau_0^{d/2} \Big(r^{\overline{0}} - \sqrt{\sum (r^{\overline{k}})^2} \Big)^{d/2} \eeq
As a consequence of the fact that $J^+ (p) = J_p^+ (p)$, 
\beq \chi_1 (p) = \chi_2 (p) \eeq
Furthermore, it is easy to see that 
\beq \chi_1 (q) = \chi_2 (q) = 0 \eeq
This means that 
\beq \Delta_{1d} = \chi_1 (p) = - (\chi_2 (q) - \chi_2 (p)) = -\int_0^{\tau (p,q)} d \tau \frac{d \chi_2 (r(\tau))}{d \tau} \eeq
where $r(\tau)$ is defined as a point on $\gamma_{pq}$ whose distance to $r$ is $\tau$:
\beq \gamma_{pr(\tau)} \subset \gamma_{pq} \; , \; l(\gamma_{pr(\tau)}) = \tau \eeq
This immediately implies that 
\beq \delta \tau >0 \Rightarrow r(\tau) \prec r(\tau + \delta \tau) \eeq
Furthermore, in normal coordinates
\beq r^{\overline{\mu}} (\tau) = \tau V^{\overline{\mu}} \eeq
where $V^{\overline{\mu}}$ is a tangent vector to $\gamma_{pq}$ at $p$. The fact that the latter is timelike implies that
\beq \delta \tau > 0 \Rightarrow r(\tau) \prec_p r(\tau + \delta \tau) \eeq
In order to compute $d \chi_2(r(\tau)) / d \tau$, notice that, schematically, 
\beq \delta \chi_2 (r) = \int_{S_1} f - \int_{S_2} f - \int_{S_3} f + \int_{S_4} f = \nonumber \eeq
\beq =\int_{S_1 \setminus S_2} f - \int_{S_2 \setminus S_1} f - \int_{S_3 \setminus S_4} f + \int_{S_4 \setminus S_3}f = \nonumber \eeq
\beq = \int_{(S_1 \setminus S_2) \setminus (S_3 \setminus S_4)} f - \int_{(S_3 \setminus S_4) \setminus (S_1 \setminus S_2)} f - \eeq
\beq - \int_{(S_3 \setminus S_4) \setminus (S_4 \setminus S_3)} f + \int_{(S_4 \setminus S_3) \setminus (S_3 \setminus S_4)} f \nonumber \eeq
where
\beq S_1 = \alpha (r (\tau + \delta \tau ), q) \; , \; S_2= \alpha_p (r (\tau + \delta \tau  ), q) \; , \nonumber \eeq
\beq  \; S_3 = \alpha (r (\tau ), q) \; , \; S_4= \alpha_p (r (\tau  ), q) \eeq
Substituting both $r(\tau) \prec r (\tau + \delta \tau)$ and $r(\tau) \prec_p r (\tau + \delta \tau)$ into simple set theory algebra, one obtains
\beq (\alpha (r(\tau), q) \setminus \alpha_p (r (\tau ), q)) \setminus (\alpha (r(\tau + \delta \tau),q) \setminus \alpha_p (r(\tau + \delta \tau),q))) = \nonumber \eeq
\beq = (J^+ (r (\tau)) \setminus J^+ (r(\tau + \delta \tau))) \cap (J^- (q) \setminus \alpha_p (r(\tau , q))) \subset J^+ (r (\tau)) \setminus J^+ (r(\tau+ \delta \tau)) \nonumber \eeq
\beq (\alpha_p (r(\tau), q) \setminus \alpha (r (\tau ), q)) \setminus (\alpha_p (r(\tau + \delta \tau),q) \setminus \alpha (r(\tau + \delta \tau),q)) = \eeq
\beq = (J_p^+ (r (\tau)) \setminus J_p^+ (r(\tau + \delta \tau))) \cap (J_p^- (q) \setminus \alpha (r(\tau , q))) \subset J_p^+ (r (\tau)) \setminus J_p^+ (r(\tau + \delta \tau)) \nonumber \eeq
\beq (\alpha (r (\tau+ \delta \tau), q) \setminus \alpha_p (r (\tau + \delta \tau), q)) \setminus (\alpha (r (\tau ), q) \setminus \alpha_p (r (\tau ), q)) = \nonumber \eeq
\beq = \alpha (r (\tau + \delta \tau ), q) \cap (J_p^+ (r (\tau )) \setminus J_p^+ (r (\tau + \delta \tau))) \subset J_p^+ (r (\tau )) \setminus J_p^+ (r (\tau + \delta \tau )) \nonumber \eeq
\beq (\alpha_p (r (\tau + \delta \tau), q) \setminus \alpha (r (\tau + \delta \tau), q)) \setminus (\alpha_p (r (\tau ), q) \setminus \alpha (r (\tau ), q)) = \nonumber \eeq
\beq = \alpha_p (r (\tau + \delta \tau ), q) \cap (J^+ (r (\tau )) \setminus J^+ (r (\tau + \delta \tau ))) \subset J^+ (r (\tau )) \setminus J^+ (r (\tau + \delta \tau )) \nonumber \eeq
Thus, all four integrals are performed either over a subset of $J^+ (r (\tau )) \setminus J^+ (r (\tau + \delta \tau ))$ or over a subset of $J_p^+ (r (\tau )) \setminus J_p^+ (r (\tau + \delta \tau ))$ In either case, the range of integration is in a vicinity of the lightcone of $r(\tau)$. 

Now let $s$ be an arbitrary point in that region. Remembering that $\tau$ denotes a distance from $p$ to the point of intersection of that region with geodesic $\gamma_{pq}$, and is not to be confused with a distance from $p$ to a floating point, we have  
\beq s \in (J^+ (r (\tau )) \setminus J^+ (r (\tau + \delta \tau ))) \cup  (J_p^+ (r (\tau )) \setminus J_p^+ (r (\tau + \delta \tau ))) \Rightarrow \nonumber \eeq
\beq \Rightarrow \sqrt{\sum (s^{\overline{k}})^2} = s^{\overline{0}} - \tau + 0 (\tau^2) \Rightarrow \eeq
\beq \Rightarrow k_d \tau_0^{d/2} \Big(s^{\overline{0}} - \sqrt{\sum (s^{\overline{k}})^2} \Big)^{d/2} = k_d (\tau \tau_0)^{d/2} + 0 (\tau^{d+2}) \nonumber \eeq
Thus, all integrals reduce to the integration over constant:
\beq \int_{S_1} f d^d x- \int_{S_2} f d^d x - \int_{S_3} f d^d x + \int_{S_4} f d^d x = \nonumber \eeq
\beq=  \int_{(S_1 \setminus S_2) \setminus (S_3 \setminus S_4)} f - \int_{(S_3 \setminus S_4) \setminus (S_1 \setminus S_2)} f - \int_{(S_3 \setminus S_4) \setminus (S_4 \setminus S_3)} f + \int_{(S_4 \setminus S_3) \setminus (S_3 \setminus S_4)} f = \nonumber \eeq
\beq = (\tau \tau_0)^{d/2} \int_{(S_1 \setminus S_2) \setminus (S_3 \setminus S_4)} d^d x - (\tau \tau_0)^{d/2} \int_{(S_3 \setminus S_4) \setminus (S_1 \setminus S_2)} d^d x - \nonumber \eeq
\beq - (\tau \tau_0)^{d/2} \int_{(S_3 \setminus S_4) \setminus (S_4 \setminus S_3)} d^d x + (\tau \tau_0)^{d/2} \int_{(S_4 \setminus S_3 \setminus (S_3 \setminus S_4)} d^d x =  \eeq
\beq = (\tau \tau_0)^{d/2} \Big( \int_{S_1} d^d x- \int_{S_2} d^d x  - \int_{S_3} d^d x + \int_{S_4} d^d x \Big) = \nonumber \eeq
\beq = (\tau \tau_0)^{d/2} (V(S_1)- V(S_2) - V(S_3) + V(S_4)) \nonumber \eeq
Now, $V(S_1)$, $V(S_2)$, $V(S_3)$ and $V(S_4)$ can be read off from Ref \cite{GibSol} as follows:
\beq V(S_1) = (k_d + (A_d Rg_{\mu \nu} + B_d R_{\mu \nu})(q^{\mu} - r^{\mu} (\tau + \delta \tau)) (q^{\nu} - r^{\nu} (\tau + \delta \tau))) \tau^d (r(\tau + \delta \tau) , q) \eeq
\beq V(S_2) = k_d  \tau^d (r(\tau + \delta \tau) , q) \eeq
\beq V(S_3) = (k_d + (ARg_{\mu \nu} + BR_{\mu \nu})(q^{\mu} - r^{\mu} (\tau)) (q^{\nu} - r^{\nu} (\tau ))) \tau^d (r(\tau) , q) \eeq
\beq V(S_4) = k_d  \tau^d (r(\tau) , q) \eeq
Since coordinate system is defined in terms of geodesics comming out of $p$, all of these geodesics, including $\gamma_{pq}$ are, by definition, straight lines in the chosen coordinate system. Therefore, it can be assumed that $\gamma_{pq}$ coincides with $t$-axis, which simplifies the above equations:
\beq V(S_1) = k_d (\tau _1 - \tau - \delta \tau) ^d + (A_d R+B_d R_{00}) (\tau_1 - \tau - \delta \tau)^{d+2}\eeq
\beq V(S_2) = k_d  (\tau_1 - \tau - \delta \tau)^d \eeq
\beq V(S_3) = k_d (\tau _1 - \tau ) ^d + (A_d R+B_d R_{00}) (\tau_1 - \tau )^{d+2}\eeq
\beq V(S_4) = k_d  (\tau_1 - \tau )^d \eeq
This implies that 
\beq \chi_2 (r (\tau + \delta \tau)) - \chi_2 (r (\tau)) = (\tau_1 \tau)^{d/2}(V(S_1) - V(S_2) - V(S_3) + V(S_4)) = \nonumber \eeq
\beq =  (A_d R+B_d R_{00}) (\tau_1 \tau)^{d/2} \frac{d}{d \tau} (\tau_1 - \tau)^{d+2} \delta \tau + 0 (\tau^{d+2} (\delta \tau)^2) \eeq
which implies
\beq \frac{d \chi_2 (r(\tau))}{d \tau} = (A_d R+B_d R_{00}) (\tau_1 \tau)^{d/2} \frac{d}{d \tau} (\tau_1 - \tau)^{d+2} \eeq
Thus, 
\beq \Delta_{1d} = \chi_1 (p) - \chi_1 (q) = \chi_2 (p) - \chi_2 (q) = \eeq
\beq = (A_d R+B_d R_{00}) \int_0^{\tau_1} d \tau (\tau_1 \tau)^{d/2} \frac{d}{d \tau} (\tau_1 - \tau)^{d+2}  \nonumber \eeq
The binomial expansion of $(\tau_1 - \tau)^{d+2}$ gives
\beq \Delta_{1d} = - (A_d R+B_d R_{00}) \int_0^{\tau_1} d \tau (\tau_1 \tau)^{d/2} \frac{d}{d \tau} \Big( \sum_{k=0}^{d+2} (-1)^k {{d+2} \choose k} \tau_1^{d+2-k} \tau^k \Big) \eeq
Evaluating of the derivative and combining it with $\tau^{d/2}$ factor gives
\beq \Delta_{1d} = - (A_d R+B_d R_{00}) \int_0^{\tau_1} d \tau \tau_1 ^{d/2}  \Big( \sum_{k=0}^{d+2} (-1)^k {{d+2} \choose k} \tau_1^{d+2-k} k \tau^{\frac{d}{2}+k-1} \Big) \eeq
Finally, the integration of that expression gives
\beq \Delta_{1d} = (A_d R+B_d R_{00}) \tau_1^{2d+2} \sum_{k=0}^{d+2} (-1)^k  {{d+2} \choose k} \frac{k}{\frac{d}{2}+k} \eeq
Now lets compute $\Delta_{2d}$. We will now go back to usual notation, where $r$ (which will be denoted by $x$ in order to avoid conflict with $r$ being a radius of a ball)is an arbitrary element of $\alpha (p,q)$, and it is no longer assumed to lie on $\gamma_{pq}$. As was stated earlier, $\Delta_{2d}$ correction is due to the error of computing the volume of $\alpha (p,x)$ where $x$ is far away from the boundary of $\alpha (p,q)$. Since that correction is already of the order $0 (\tau^{d+2})$, its integral over $\alpha (p,q)$ is of the order $0 (\tau^{2d+2})$, while its integral over the ``corrections" to the shape of $\alpha (p,q)$ is of the order $0 (\tau^{2d+4})$. For this reason, the latter term will be neglected, and it will be assumed that the shape of $\alpha (p,q)$ has not been affected by curvature, while the shapes of $\alpha (p,x)$ has. Furthermore, in light of the geodesic coordinates, while the corrections to volume of $\alpha (p,x)$ are not neglected, the flat space equation for the distance will be used in computting them. Thus, the equation for $\Delta_{2d}$ becomes
\beq \Delta_{2d} = A_d R \int_{\alpha_p (p,q)} d^d x  (x^{\overline{\mu}} x_{\overline{\mu}})^{1+\frac{d}{2}} + B_d R_{\overline{\mu} \overline{\nu}} \int_{\alpha_p (p,q)} x^{\overline{\mu}} x^{\overline{\nu}}  (x^{\overline{\rho}} x_{\overline{\rho}})^{\frac{d}{2}} \eeq
By using the cylindrical symmetry, this becomes
\beq \Delta_{2d} = A_d R \int_{\alpha_p (p,q)} d^d \overline{x} (x^{\overline{\mu}} x_{\overline{\mu}} )^{1+ \frac{d}{2}} + \nonumber \eeq
\beq + B_d R_{\overline{0} \overline{0}} \int_{\alpha_p (p,q)} d^d \overline{x} (x^{\overline{0}})^2 (x^{\overline{\rho}} x_{\overline{\rho}})^{d/2} + \eeq
\beq + B_d (\sum R_{kk}) \int_{\alpha_p (p,q)} d^d \overline{x} (x^{\overline{1}})^2 (x^{\overline{\rho}} x_{\overline{\rho}})^{d/2} \nonumber \eeq
where, due to cylindrical symmetry, $(x^{\overline{k}})^2$ was replaced with$(x^{\overline{1}})^2$. 

From cylindrical symmetry, 
\beq \int_{\alpha_p (p,q)} d^d \overline{x} (x^{\overline{\rho}} x_{\overline{\rho}})^{1 + \frac{d}{2}} = \eeq
\beq =\int_{\alpha_p (p,q)} d^d \overline{x} (x^{\overline{0}})^2 (x^{\overline{\rho}} x_{\overline{\rho}})^{ \frac{d}{2}} - (d-1) \int d^d \overline{x} (x^{\overline{1}})^2 (x^{\overline{\rho}} x_{\overline{\rho}})^{ \frac{d}{2}} \nonumber \eeq
This can be used to get rid of integral involving $(x^{\overline{1}})^2$ to get
\beq \Delta_2 = \tau^{2d+2} \Big[R \Big( \Big(A_d + \frac{B_d}{d-1} \Big) H_{d,2,d/2} \Big) + \nonumber \eeq
\beq + \frac{R_{00}}{d-1} \Big(dB_d H_{d,2,d/2} - B_d H_{d,0,1+\frac{d}{2}} \Big) \Big] \eeq
where
\beq H_{d,i,j} = \frac{1}{\tau^d} \int_{\alpha_p (p,q)} d^d \overline{x} (x^{\overline{0}})^i (x^{\overline{\mu}} x_{\overline{\mu}})^j \eeq
and
\beq H_{d,i,j+ \frac{1}{2}} = \frac{1}{\tau^d} \int_{\alpha_p (p,q)} d^d \overline{x} (x^{\overline{0}})^i (x^{\overline{\mu}} x_{\overline{\mu}})^{j+ \frac{1}{2}} \eeq
here $j$ can be either an integer or half integer. 

Let us now compute these coefficients. 

In the calculations that follows we will treat the above integrands as simply functions, and ``forget" that their source is a curvature. 

By slicing $\alpha_p (p,q)$ into balls $\overline{t}= {\rm const}$ and then slicing each ball into spheres, it is easy to see that 
\beq \int_{\alpha_p (p,q)} d^d \overline{x} (x^{\overline{0}} - p^{\overline{0}})^i ((x^{\overline{\mu}} - p^{\overline{\mu}})(x_{\overline{\mu}} - p_{\overline{\mu}}))^{j/2} d^d x = \nonumber \eeq
\beq = \frac{2 \pi^{(d-1)/2}}{ \Gamma ((d-1)/2)}   \Big( \int_0^{\tau/2} d \overline{t} \Big(\overline{t}^i \int_0^t d \overline{r} \overline{r}^{d-2} (\overline{t}^2 - \overline{r}^2)^{j/2} \Big) + \eeq
\beq + \int_{\tau/2}^{\tau}   d\overline{t} \Big(\overline{t}^i \int_0^{\tau - \overline{t} } d \overline{r} \overline{r}^{d-2} (\overline{t}^2 - \overline{r}^2)^{j/2} \Big) \Big) \nonumber \eeq
By changing variables to 
\beq \overline{u} = \frac{\overline{t}}{\tau} \; , \; \overline{s} = \frac{\overline{r}}{\overline{t}} \eeq
the above expression becomes
\beq \int_{\alpha_p (p,q)} d^d \overline{x} (x^{\overline{0}} - p^{\overline{0}})^i ((x^{\overline{\mu}} - p^{\overline{\mu}})(x_{\overline{\mu}} - p_{\overline{\mu}}))^{j/2}  = \nonumber \eeq
\beq  = \frac{2 \pi^{(d-1)/2}}{ \Gamma ((d-1)/2)} \tau^{i+j+d} \Big[ \int_0^{1/2} d \overline{u} \Big( \overline{u}^{i+j+d-1} \int_0^1 d \overline{s} \; \overline{s}^{d-2} (1-\overline{s}^2)^{j/2} \Big) + \nonumber \eeq
\beq + \int_{1/2}^1 d \overline{u} \Big( \overline{u}^{i+j+d-1} \int_{0}^{\frac{1}{u}-1} d\overline{s} \; \overline{s}^{d-2} (1-\overline{s}^2)^{j/2} \Big) \Big] \eeq
Since the limits of integration in the first term are constants, that  term can be represented as product of two separate integrals. After evaluating the u-integral, the expression becomes 
\beq \int_{\alpha_p (p,q)} d^d \overline{x} (x^{\overline{0}} - p^{\overline{0}})^i ((x^{\overline{\mu}} - p^{\overline{\mu}})(x_{\overline{\mu}} - p_{\overline{\mu}}))^{j/2}  = \nonumber \eeq
\beq = \frac{2 \pi^{(d-1)/2}}{ \Gamma ((d-1)/2)}  \tau^{i+j+d} \Big( \frac{1}{(i+j+d) 2^{i+j+d}} \int_0^1 d \overline{s} \; \overline{s}^{d-2} (1-\overline{s}^2)^{j/2} + \eeq
\beq + \int_{1/2}^1 d \overline{u} \; ( \overline{u}^{i+j+d-1} \int_0^{\frac{1}{u} -1} d \overline{s} \; \overline{s}^{d-2} (1-\overline{s}^2)^{j/2} ) \Big) \nonumber \eeq
From now on the calculation splits into four cases: even and odd $d$ and even and odd $j$. From the original intentions of the calculation it is clear that whenever $d$ is odd, $j$ is also add and visa versa. So only these two cases need be considered.

CASE 1: $d$ and $j$ are both even

Since $j$ is even, denote it as
\beq j= 2h \eeq
Expanding $(1-\overline{s}^2)^h$ binomially, the integral becomes
\beq \int_{\alpha_p (p,q)} d^d \overline{x}(x^{\overline{0}} - p^{\overline{0}})^i ((x^{\overline{\mu}} - p^{\overline{\mu}})(x_{\overline{\mu}} - p_{\overline{\mu}}))^{j/2}  = \nonumber \eeq
\beq = \frac{2 \pi^{(d-1)/2}}{ \Gamma ((d-1)/2)} \tau^{i+j+d} \sum_{k=0}^h (-1)^k {h \choose k} \Big[ \frac{1}{(i+2h+d)2^{i+2h+d}} \int_0^1 \overline{s}^{d-2+2k} d\overline{s} + \eeq
\beq + \int_{1/2}^1 d \overline{u} \Big( \overline{u}^{i+2h+d-1} \int_0^{\frac{1}{\overline{u}}-1} d \overline{s} \overline{s}^{d-2+2k} \Big) \Big] \nonumber \eeq
After evaluating the $\overline{s}$ integrals this becomes
\beq \int_{\alpha_p (p,q)} d^d \overline{x}(x^{\overline{0}} - p^{\overline{0}})^i ((x^{\overline{\mu}} - p^{\overline{\mu}})(x_{\overline{\mu}} - p_{\overline{\mu}}))^{j/2}  = \nonumber \eeq
\beq = \frac{2 \pi^{(d-1)/2}}{ \Gamma ((d-1)/2)} \tau^{i+j+d} \sum_{k=0}^h (-1)^k {h \choose k} \Big[ \frac{1}{2^{i+2h+d} (i+2h+d) (d-1+2k)} + \eeq
\beq + \int_{1/2}^1 d \overline{u} \Big( \overline{u}^{i+2h+d-1} \frac{(\frac{1}{\overline{u}} -1)^{d-1+2k}}{d-1+2k} \Big) \Big] \nonumber \eeq
After pulling expanding out $d-1+2k$ and expanding out $(\frac{1}{\overline{u}}-1)^{d-1+2k}$, that becomes
 \beq \int_{\alpha_p (p,q)}  d^d \overline{x} (x^{\overline{0}} - p^{\overline{0}})^i ((x^{\overline{\mu}} - p^{\overline{\mu}})(x_{\overline{\mu}} - p_{\overline{\mu}}))^{j/2}= \nonumber \eeq
\beq = \frac{2 \pi^{(d-1)/2}}{ \Gamma ((d-1)/2)} \tau^{i+j+d} \sum_{k=0}^h \frac{(-1)^k}{d-1+2k} {h \choose k} \Big(\frac{1}{2^{i+2h+d} (i+2h+d)} + \eeq
\beq + \sum_{l=0}^{d-1+2k} (-1)^l {d-1+2k \choose l} \int_{1/2}^1 d \overline{u} \; \overline{u}^{i+2h-2k+l} \Big) \nonumber \eeq
Evaluating the integral that is left this becomes
\beq \int d^d \overline{x} (x^{\overline{0}} - p^{\overline{0}})^i ((x^{\overline{\mu}} - p^{\overline{\mu}})(x_{\overline{\mu}} - p_{\overline{\mu}}))^{j/2}  = H_{dih} \tau^{i+j+d}  \eeq
where
\beq H_{dih}= \frac{2 \pi^{(d-1)/2}} \sum_{k=0}^h \frac{(-1)^k}{d-1+2k} {h \choose k} \Big( \frac{1}{2^{i+2h+d} (i+2h+d)} + \eeq
\beq + \sum_{l=0}^{d-1+2k} (-1)^l {d-1+2k \choose l} \frac{1-(\frac{1}{2})^{1+2h-2k+l+1}}{i+2h-2k+l+1} \Big) \nonumber \eeq
CASE 2: $d$ and $j$ are both odd

Since $j$ is odd, it will be replaced with
\beq j=2h+1 \eeq
Then binomial expansion tells us
\beq (1-\overline{s}^2)^j = \sqrt{1-\overline{s}^2} \sum_{k=0}^j {h \choose k} (-1)^k \overline{s}^{2k} \eeq
which means that the original integral can be rewritten as
\beq \int_{\alpha_p (p,q)} d^d \overline{x} \; (x^{\overline{0}} - p^{\overline{0}})^i ((x^{\overline{\mu}} - p^{\overline{\mu}})(x_{\overline{\mu}} - p_{\overline{\mu}}))^{j/2}  = \nonumber \eeq
\beq =\frac{2 \pi^{(d-1)/2}}{ \Gamma ((d-1)/2)} \tau^{2h+1+i+d} \sum_{k=0}^h {h \choose k} (-1)^k  \times \nonumber \eeq
\beq \times \Big[ \frac{1}{(i+2h+1+d)2^{i+2h+1+d}} \int_0^1 d \overline{s} \; \overline{s}^{d-2+2k} \sqrt{1-\overline{s}^2} + \eeq
\beq + \int_{1/2}^1 d \overline{u} \Big( \overline{u}^{i+2h+d} \int_0^{\frac{1}{\overline{u}}-1} d \overline{s} \; \overline{s}^{d-2+2k} \sqrt{1-\overline{s}^2}  \Big) \Big] \nonumber \eeq
Since $d$ is odd, so is $d-2+2k$. Thus,
\beq d-2+2k = 2a+1\;, \eeq
where
\beq a= \frac{d-3+2k}{2}\;.\eeq
Thus, the integral of interest is $\int \overline{s}^{2a+1} \sqrt{1-\overline{s}^2} d \overline{s}$ .

By using
\beq \overline{s} = \sin \overline{\theta} \eeq
the integral becomes 
\beq \int \overline{s}^{2a+1} \sqrt{1-\overline{s}^2}\, \dd\overline{s} = \int \sin^{2a+1} \overline{\theta} \cos \overline{\theta}\, \dd\sin \overline{\theta} = \int \sin^{2a+1} \overline{\theta} \cos^2 \overline{\theta}\, \dd\overline{\theta}  \eeq 
Combining one of the $\sin \overline{\theta}$ factors with $d \overline{\theta}$ gives 
\beq \int \overline{s}^{2a+1} \sqrt{1-\overline{s}^2}\, \dd\overline{s} = - \int \sin^{2a} \overline{\theta} \cos^2 \overline{\theta}\, \dd\cos\overline{\theta} \eeq
Expanding $\sin^{2a} \overline{\theta}$ as
\beq \sin^{2a} \overline{\theta} = (1 - \cos^2 \overline{\theta})^a = \sum_{b=0}^a (-1)^b {a \choose b} \cos^{2b} \overline{\theta} \eeq
the above integral becomes 
\beq \int \overline{s}^{2a+1} \sqrt{1-\overline{s}^2}\, \dd\overline{s} = - \sum_{b=0}^a (-1)^b {a \choose b} \int\cos^{2b+2} \overline{\theta}\, \dd\cos \overline{\theta} = \nonumber \eeq
\beq = - \sum_{b=0}^a (-1)^b {a \choose b} \frac{ \cos^{2b+3} \overline{\theta}}{2b+3} \eeq
By substituting $s= \sin \theta$ this becomes 
\beq
\int \overline{s}^{2a+1} \sqrt{1-\overline{s}^2}\, \dd \overline{s} = - \sum_{b=0}^a \frac{(-1)^b}{2b+3} {a \choose b} (1-\overline{s}^2)^{b+ \frac{3}{2}}
\eeq
Substituting $2a+1=d-2+2k$ we obtain
\beq
\int \overline{s}^{d-2+2k} \sqrt{1-\overline{s}^2}\, \dd\overline{s} = - \sum_{b=0}^{k+\frac{d-3}{2}} \frac{(-1)^b}{2b+3} {k+ \frac{d-3}{2} \choose b} (1-\overline{s}^2)^{b+ \frac{3}{2}}\;.
\eeq
Thus, the original integral becomes
\beq
\int_{\alpha_p (p,q)} \dd^d \overline{x}\,(x^{\overline{0}} - p^{\overline{0}})^i ((x^{\overline{\mu}} - p^{\overline{\mu}})(x_{\overline{\mu}} - p_{\overline{\mu}}))^{h+ \frac{1}{2}}  =  \nonumber
\eeq
\beq = \frac{2 \pi^{(d-1)/2}}{ \Gamma ((d-1)/2)} \sum_{k=0}^h \Big( (-1)^k {h \choose k} \sum_{b=0}^{k+\frac{d-3}{2}} \frac{(-1)^b}{2b+3} {k+\frac{d-3}{2} \choose b} \times \nonumber \eeq
\beq \times \Big( \frac{1}{(i+2h+1+d)2^{i+2h+1+d}} + \int_{1/2}^1 d \overline{u} \; \overline{u}^{i+2h+d} \Big( \Big(\frac{2}{\overline{u}} - \frac{1}{\overline{u}^2} \Big)^{b+ \frac{3}{2}} -1 \Big) \Big)\Big) \eeq
By evaluating the $-1$-term  of the integral over $\overline{u}$, and also pulling $\overline{u}$ out of the denominators, the expression becomes 
\beq \int_{\alpha_p (p,q)} d^d \overline{x} (x^{\overline{0}} - p^{\overline{0}})^i ((x^{\overline{\mu}} - p^{\overline{\mu}})(x_{\overline{\mu}} - p_{\overline{\mu}}))^{h+ \frac{1}{2}}  =  \nonumber \eeq
\beq = \frac{2 \pi^{(d-1)/2}}{ \Gamma ((d-1)/2)} \sum_{k=0}^h \Big( (-1)^k {h \choose k} \sum_{b=0}^{k+\frac{d-3}{2}} \frac{(-1)^b}{2b+3} {k+\frac{d-3}{2} \choose b} \times \eeq
\beq \times \Big( \frac{1}{(i+2h+1+d)2^{i+2h+1+d}} - \frac{1-(\frac{1}{2})^{i+2h+d+1}}{i+2h+d+1}  + \nonumber \eeq
\beq + \int_{1/2}^1 d \overline{u} \; \overline{u}^{i+2h+d-2b-3} (2 \overline{u}-1)^{b + \frac{3}{2}} \Big) \Big) \nonumber \eeq
By expanding $(2\overline{u}-1)^{b+1}$ we get
\beq \int_{1/2}^1 d \overline{u} \; \overline{u}^{i+2h+d-2b-3} (2 \overline{u}-1)^{b + \frac{3}{2}} =  \eeq
\beq = \sum_{c=0}^b (-1)^c {b+1 \choose c} 2^{b-c+1} \int_{1/2}^1 d \overline{u} \; \overline{u}^{i+2h+d-b-c-2} \sqrt{2\overline{u}-1} \nonumber  \eeq
By setting
\beq \overline{v}= \sqrt{2 \overline{u}-1} \eeq
this becomes 
\beq \int_{1/2}^1 d \overline{u} \; \overline{u}^{i+2h+d-2b-3} (2\overline{u}-1)^{b + \frac{3}{2}} =  \eeq 
\beq = \sum_{c=0}^b (-1)^c {b+1 \choose c} 2^{b-c+1} \int_0^1 d \overline{v} \; \overline{v}^2 \Big( \frac{\overline{v}^2 +1}{2} \Big)^{i+2h+d-b-2-c} \nonumber \eeq
By expanding $(\overline{v}^2+1)^ {i+2h+d-b-2-c}$ this becomes
\beq \int_{1/2}^1 \dd\overline{u}\, \overline{u}^{i+2h+d-2b-3}\,
(2\,\overline{u}-1)^{b + 3/2} \nonumber \eeq
\beq = \sum_{c=0}^b \Big( (-1)^c {b+1 \choose c} 2^{2b-i-2h-d+3} \int \dd\overline{v} \sum_{e=0}^{i+2h+d-b-2-c} \overline{v}^{2e+2} \Big) \eeq
\beq = \sum_{c=0}^b \Big( (-1)^c {b+1 \choose c} 2^{2b-i-2h-d+3} \sum_{e=0}^{i+2h+d-b-2-c} {1+2h+d-b-2-c \choose e} \frac{1}{2e+3}\Big) \nonumber
\eeq
Substituting these into original integral we obtain
\beq \int_{\alpha_p (p,q)} d^d \overline{x}(x^{\overline{0}} - p^{\overline{0}})^i ((x^{\overline{\mu}} - p^{\overline{\mu}})(x_{\overline{\mu}} - p_{\overline{\mu}}))^{h+\frac{1}{2}}  = H_{d,i,h+\frac{1}{2}} \tau^{2h+i+d+1} \eeq
where
\beq H_{d,i,h+1/2} = \frac{2 \pi^{(d-1)/2}}{\Gamma((d-1)/2)} \sum_{k=0}^h \Big\{ (-1)^k  {h \choose k} \sum_{b=0}^{k+\frac{d-3}{2}} \frac{(-1)^b}{2b+3} {k + \frac{d-3}{2} \choose b} \times \nonumber \eeq
\beq \kern62pt\times\ \Big[ \frac{1}{(i+2h+d+1)\,2^{i+2h+d+1}} - \frac{1-(\frac12)^{i+2h+d+1}}{i+2h+d+1} + \eeq
\beq +\ \sum_{c=0}^b \Big( (-1)^c {b+1 \choose c} 2^{2b-i-2h-d+3} \sum_{e=0}^{i+2h+d-b-2-c} {1+2h+d-b-2-c \choose e} \frac{1}{2e+3} \nonumber \Big) \Big] \Big\} \eeq
Finally, let's compute $\Delta_3$. In normal coordinates, 
\beq g_{\overline{\mu} \overline{\nu}} = \eta_{\overline{\mu} \overline{\nu}} - \frac{1}{3} R_{\overline{\mu} \overline{\rho} \overline{\nu} \overline{\sigma}} x^{\overline{\rho}} x^{\overline{\sigma}} \eeq
This implies that 
\beq \sqrt{(-1)^{d-1} \det g} = 1 - \frac{1}{6} R_{\overline{\rho} \overline{\sigma}} x^{\overline{\rho}} x^{\overline{\sigma}} \eeq
Thus,
\beq \Delta_{3d} = - \frac{k_d}{6} R_{\overline{\rho} \overline{\sigma}} \int d^d x x^{\overline{\rho}} x^{\overline{\sigma}} ((x^{\overline{\alpha}} - p^{\overline{\alpha}})(x_{\overline{\alpha}} - p_{\overline{\alpha}}))^{d/2} \eeq
As was done with $\Delta_2$, the correction to the correction term will be neglected, which means that integration is performed over $\alpha_p (p,q)$ instead of $\alpha (p,q)$ and no correction term is introduced to the $V(\alpha (p,x))$ when the expression $((x^{\overline{\alpha}} - p^{\overline{\alpha}})(x_{\overline{\alpha}} - p_{\overline{\alpha}}))^d$ was used. 

Substituting the above into expression for $\Delta_{3d}$ gives
\beq \Delta_{3d} = \frac{k_d}{6} R_{\overline{\rho} \overline{\sigma}} \int_{\alpha_p (p,q)} d^d \overline{x} x^{\overline{\rho}} x^{\overline{\sigma}} ((x^{\overline{\alpha}} - p^{\overline{\alpha}})(x_{\overline{\alpha}} - p_{\overline{\alpha}}))^{d/2} \eeq
By cylindrical symmetry, $(x^k)^2$ in the above integral can be replaced with $(x^1)^2$ which means
\beq \Delta_{3d} = \frac{k_d}{6} \Big[ R_{\overline{0} \overline{0}} \int_{\alpha_p (p,q)} d^d \overline{x} (x^{\overline{0}})^2 ((x^{\overline{\alpha}} - p^{\overline{\alpha}})(x_{\overline{\alpha}} - p_{\overline{\alpha}}))^{d/2} - \nonumber \eeq
\beq - \Big(\sum R_{\overline{k} \overline{k}} \Big) \int_{\alpha_p (p,q)} d^d \overline{x} (x^{\overline{1}})^2 ((x^{\overline{\alpha}} - p^{\overline{\alpha}})(x_{\overline{\alpha}} - p_{\overline{\alpha}}))^{d/2} \Big] \eeq
Again, by cylindrical symmetry, 
\beq \int_{\alpha_p (p,q)} d^d \overline{x} ((x^{\overline{\alpha}} - p^{\overline{\alpha}})(x_{\overline{\alpha}} - p_{\overline{\alpha}}))^{1+ \frac{d}{2}} = \int_{\alpha_p (p,q)} d^d \overline{x} (x^{\overline{0}})^2((x^{\overline{\alpha}} - p^{\overline{\alpha}})(x_{\overline{\alpha}} - p_{\overline{\alpha}}))^{d/2} - \nonumber \eeq
\beq - (d-1)\int_{\alpha_p (p,q)} d^d \overline{x} (x^{\overline{1}})^2((x^{\overline{\alpha}} - p^{\overline{\alpha}})(x_{\overline{\alpha}} - p_{\overline{\alpha}}))^{d/2} \eeq
which allows to express integral involving $(x^{\overline{1}})^2$ in terms of integrals involving $(x^{\overline{0}})^2$. Substituting this expression into the expression for $\Delta_{3d}$ and doing some simple algebra gives
\beq \Delta_{3d} = - \frac{k_d}{6(d-1)} \tau^{2d+2} (R (H_{d, 0, 1 + \frac{d}{2}} - H_{d, 2, \frac{d}{2}} ) + R_{00} (d H_{d, 2, \frac{d}{2}} - H_{d, 0, 1 + \frac{d}{2}} )) \eeq
By adding $\Delta_{1d}$, $\Delta_{2d}$ and $\Delta_{3d}$ the total correction becomes
\beq \Delta_d = \Delta_{1d} + \Delta_{2d} + \Delta_{3d} = \tau^{2d+2} (C_d R + D_d R_{00}) \eeq
where
\beq C_d = A_d \sum_{k=0}^{d+2} \Big( (-1)^k {d+2 \choose k} \frac{k}{\frac{d}{2} + k} \Big) + \nonumber \eeq
\beq + H_{d,0, \frac{d}{2}+1} \Big( A_d + \frac{B_d}{d-1} - \frac{k_d}{6(d-1)} \Big) + \frac{H_{d,2,\frac{d}{2}}}{d-1} \Big( \frac{k_d}{6} -B_d \Big) \eeq
and
\beq D_d = B_d \sum_{k=0}^{d+2} \Big( (-1)^k {d+2 \choose k} \frac{k}{\frac{d}{2} + k} \Big) + \nonumber \eeq
\beq + \frac{1}{d-1} \Big( \Big( B_d - \frac{k_d}{6} \Big) dH_{d,2,\frac{d}{2}} + \Big( \frac{dk_d}{6} - B_d \Big) H_{d,0,1+ \frac{d}{2}} \Big) \eeq
Now, as we were computting the corrections to the integral, we almost forgot the main term! Here it is:
\beq \int_{\alpha_p (p,q)} d^d x k_d ((x^{\overline{\mu}} - p^{\overline{\mu}})(x_{\overline{\mu}} - p_{\overline{\mu}}))^{d/2} = k_d \tau^{2d} H_{d,0,\frac{d}{2}} \eeq
This means that the total integral is 
\beq \int_{\alpha (p,q)} d^d x V(\alpha (p,x)) = k_d H_{d,0, \frac{d}{2}} \tau^{2d} + (C_d R + D_d R_{00} ) \tau^{2d+2} \eeq
In the beginning of this section it was shown that the gravitational pre-Lagrangian is given by 
\beq {\cal L} (\prec, E, p,q) = \frac{1}{4 \pi G} \Big( V^2 (\alpha (p,q)) + \nonumber \eeq
\beq + E \int_{\alpha (p,q)} d^d r \sqrt{(-1)^{d-1} \det g} V(p,r) \Big) \eeq
By substituting the above expression for the integral as well as 
\beq V(\alpha (p,q)) = k_d \tau^d + \tau^{d+2} (A_d R + B_d R_{00}) \eeq
we obtain 
\beq {\cal L} (\prec, E, p, q) = \tau^{2d} k_d (k_d + E H_{d,0,d/2}) + \eeq
\beq + \tau^{2d+2} (R(2k_d A_d + EC_d ) + R_{00} (2k_d B_d + ED_d )) \nonumber \eeq
Thus, in order to minimize variations, we have to get rid of $0(\tau^{2d})$ contribution which comes form $R_{00}$ term. Thus, $E_d$ is selected in such a way that would set the coefficient of $R_{00}$ to $0$ :
\beq 2k_d B_d + E_d D_d = 0 \Rightarrow E_d = - \frac{2k_d B_d}{D_d} \eeq
Substituting this into the expression for the Lagrangian gives
\beq {\cal L} (\prec, p, q) = \frac{1}{4 \pi G} \Big[ \tau^{2d} k_d^2 \Big( 1- \frac{2B_d}{D_d} H_{d,0,d/2} \Big) + 2k_d R \tau^{2d+2} \Big( A_d - \frac{B_d C_d}{D_d} \Big) \Big] \eeq
As long as $d$ is fixed, the first term in the above Lagrangian is constant and, therefore, does not affect physics. This means that Lagrangian that we care about is proportional to $R$ as expected. 

The gravitational Lagrangian can be rewritten as 
\beq {\cal L} = \frac{v_0}{8 \pi G_d} R \eeq
where $v_0$ is a dimension taken up by a single point of a causal set and $G_d$ is given by 
\beq G_d = G \frac{v D_d}{2k_d \tau_2^{2d+2} (A_d D_d - B_d C_d)} \eeq

However, if we release the assumption that dimension is fixed, the first term can be interpretted as dimension-Lagrangian which, for all the practical purposes, is quite separate from the gravity-Lagrangian in the second term, even though their origin is the same. That dimension-Lagrangian tries to ``force" the local neighborhood of each point to have one dimension rather than the other, which in general are non-integer. 

While at this point no work has been done to analyze the behavior of dimension-Lagrangian, it is apparent that any Lagrangian-based theory assumes the possibility of arbitrary small variation of the Lagrangian. This means that Lagrangian should be defined for fractal dimensions. 

We have to adress the fact that calculations performed above were assuming integer dimension. While the final form of expressions for $H_{dij}$ and $H_{d,i,j+ \frac{1}{2}}$ is no longer defined for fractal dimensions, some of the intermediate expressions in terms of integrals are. After all, integrals of non-integer value can always be computted numerically, if not analytically. 

One obstacle, however, is that geometric constructions that were used to obtain the integrals were based on integer dimensionality, and it might be a good project for a future research to see if they can be generalized to fractal dimension. One question that needs to be asked in the future research is how relevent the above expressions for fractal dimensionalities. 

A related issue is whether fractal dimensions are defined in such a way that every single causal set has some dimension, or whether there is a ``qualifying criteria" for causal set to be a fractal on the first place, which means that most causal sets don't have any dimensionality what so ever. 

In the former case, less assumptions can be made about any particular fractal since its criteria is more loose and therefore a fractal-based ``copy" of earlier derivations is more likely to fail. On the other hand, in the latter case, it is likely that Lagrangian will force a non-fractal causal structure, thus viewing Lagrangian as a function of dimensionality is physically meaningless, although it might still have a mathematical interest. 

Since topology of causal set is beyond the scope of the thesis, these issues will be postponed for future research. For our purposes right now we can settle with one version of an answer. According to this version, we can define fractal dimension in such a way that every single causal set is a fractal of some dimension. 

Furthermore, since Lagrangian density is formally well defined for every single causal set, we can simply define curvature $R$ pointwise in such a way that the relation between Lagrangian density, fractal dimensionality and $R$ perfectly satisfied point by point. In this case, proving this for non-integer dimension becomes non-issue since it becomes a statement that is true by definition.  

A more practical problem is that if non-integer dimensions are allowed on the first place, then the chances are that the ``preferred" dimensionality of the universe will also be non-integer.  This, of course, does not help our cause in trying to explain the actual observed four-dimensional topology. 

As will be seen in the next chapter, the latter is done by introducing vierbeins, which are viewed as literal physical fields coupled to causal relations via a type-1 Lagrangian. That Lagrangian ``encourages" a correlation between the distances inferred by vierbeins and the ones inferred by causal structure on the local regions of a causal set. 

This increases the probability of manifoldlike causal sets over non-manifoldlike ones. However, this should be weighted against the fact that from the pure combinatoric perspective manifoldlike causal sets are extremely unlikely. Thus, it is up to further research to see the overall probability of a causal set being manifold-like, based on vierbein Lagrangian taken together with combinatorics. 

However, assuming that vierbein theory is successful, the fact that vierbein fields are type 1 implies that these can play a role of ``constraints" on maxima and minima. This can not be said about type-2 fields for which the only thing that matters is the average. This might ultimately be a reason why the type-1 vierbeins ``outcompete" type-2 Lagrangian which results in our universe being four dimensional as opposed to a fractal.

Nevertheless, it is possbible that type-2 dimension-Lagrangian does have some very small effect, and makes our dimension to be $4 - \epsilon$ instead of $4$, thus providing a literal meaning to the dimensional regularization. Of course, it is up to further research to see if this will, indeed, be $4 - \epsilon$ as opposed to $4 + \epsilon$. 

Furthermore, one should also adress the fact that $\epsilon$ in dimensional regularization theory is assumed to be constant. One way of doing this is to show that this, indeed, is the prediction of the theory, which is possible in light of the fact that both vierbein Lagrangian and dimension-Lagrangian are constant throughout causal set. 

This, however, might not work due to vierbeins playing a role in fermionic Lagrangian or dimension-Lagrangian having a common source with gravity one, then one has to show that the generalized version of dimensional regularization theory based on varying $\epsilon$ still works.

\section{Chapter 5: Fermions on a Causal Set}

\subsection{Grassmann Numbers as Vector Space Elements}

In order to apply the principle of Lagrangian generators to fermions, one has to compare the magnitudes of Grassmann numbers. This, of course, can not be defined within a framework of standard theory. In order to go around this obstacle, Grassmann numbers have to be defined as individual elements of a vector space, outside of integration. 

In this dissertation, it is proposed to view them as elements of vector space, S, equipped both with commutting dot product ( $\cdot$ ) , anticommuting wedge product ($\wedge$), and measure $\xi$. Consequently, integral is well defined for all functions $\vec{F}$ , not neceserely linear ones. This statement is independent of the possibility of expressing $\vec{F}$ in algebraic form. 

Generically, integration is defined for any function $\vec{F} \colon S \rightarrow S \oplus (S \wedge S) \oplus (S \wedge S \wedge S) \oplus \ldots$ where $S \wedge S$ consists of elements of the form $a \wedge b$ where $a \in S$ and $b \in S$, $S \wedge S \wedge S$ consists of elements of the form $a \wedge b \wedge c$ where $a$ , $b$, and $c$ are elements of $S$, etc.

Noticing the difference between dot and wedge products, generic form of integral is
\beq
\int (\vec{\dd}_{\xi} x_1 \wedge \vec{\dd}_{\xi} x_2 \ldots \wedge \vec{\dd}_\xi x_n) \cdot \vec{F}(x_1, \ldots, x_n)\;, \label{integral}
\eeq
Here $\vec{\dd}_{\xi} x_k = \xi(x_k) \hat{x}_k\, \dd x_k$, $\xi(x_k)$ is a measure with both positive and negative values, $\hat{x}_k$ is a unit vector in the $x_k$ direction; and $\vec{x}_k = x_k \hat{x}_k$.

Of course, in order for the above integration to be considered Grassmann, certain conditions need to be met: If  $\vec{\dd}_{\xi} x = \xi(x)\, \hat{x}\, \dd x$ and $\vec{x} = x\, \hat{x}$, where $\hat{x}$ is a unit vector in the $x$ direction, then
\bea
& &\int \vd_{\xi}x \cdot \vec{x} = \int (\vd_{\xi} x \wedge 
\vd_{\xi} y) \cdot (\vec{x} \wedge \vec{y}) =1
\\
& &\int \vd_{\xi}x = \int \vd_{\xi}x \wedge 1
= \int (\vd_{\xi}x \wedge \vd_{\xi} y) \cdot \vec{x} = 0
\\ 
& &\int (\vd_{\xi} x\, \wedge \vd_{\xi} y) \cdot \vec{f} (x,y) = \int 
\vd_{\xi} x \cdot \bigg(\int \vd_{\xi} y \cdot \vec{f}(x,y) \bigg)\,.
\eea
Each of the first two of the above equations is what is expected of Grassmann 
variables. On the other hand, the last equation doesn't make sense in terms of standard 
Grassmann theory, since the expression $\int \dd\theta_1\, \dd\theta_2\,
\theta_1\, \theta_2 = \int \dd\theta_1\, (\int \dd\theta_2\, \theta_1\,
\theta_2)$ assumes that $\int \dd\theta_2\, \theta_1\, \theta_2$ is well defined, and the latter is Grassmannian. The goal of the proposed approach, however, is that all three expressions are equally-well-defined, among many others. 

By first trying to evaluate each of the above integrals, one can obtain restrictions on dot and wedge products that assure that above equalities hold: The equation
\beq
0 = \int \vec{\dd}_{\xi}\,x = \int \dd x\, \xi(x)\, \hat{x}
= \hat{x} \int \xi(x)\, \dd x\;,
\eeq
requires that
\beq
\int \xi(x)\, \dd x = 0\;;
\eeq
in other words, the measure has both positive 
and negative values. 

Furthermore, the expression
\beq
1 = \int \vd_{\xi}x \cdot \vec{x} = \int (\dd x\, \xi(x) \hat{x}) \cdot 
(\hat{x} x) = \hat{x} \cdot \hat{x} \int x\, \xi(x)\, \dd x\;,
\eeq
can be satisfied by setting
\beq
\hat{x} \cdot \hat{x} = 1\;,\qquad
\int x\, \xi(x)\, \dd x = 1\;.
\eeq
Now consider the multiple-integral example:
\bea
& &1 = \int \vd_{\xi} x \cdot \bigg(\int \vd_{\xi} y \cdot
(\vec{x} \wedge \vec{y} )\bigg)
= \int \bigg[\dd x\, \xi(x)\, \hat{x} \cdot \bigg(\int \dd y\, \xi(y)\,
\hat{y} \cdot (xy\, \hat{x} \wedge \hat{y})\bigg)\bigg] \nonumber\\
& &\kern9pt = \hat{x} \cdot(\hat{y} \cdot (\hat{x} \wedge \hat{y}))
\bigg(\int x\, \xi(x)\, \dd x \bigg) \bigg(\int y\, \xi(y)\, \dd y\bigg).
\eea
Since it was already established that
\beq
\int x\, \xi(x)\, \dd x = \int y\, \xi(y)\, \dd y = 1\;,
\eeq
the above calculation implies that
\beq
\hat{x} \cdot (\hat{y} \cdot (\hat{x} \wedge \hat{y})) = 1\;,
\eeq
which can be accomplished by setting
\beq
\hat{y} \cdot (\hat{x} \wedge \hat{y}) = \hat{x}\;.
\eeq
A similar argument shows that
\beq
(\hat{y} \wedge \hat{z}) \cdot (\hat{x} \wedge \hat{y}
\wedge \hat{z}) = \hat{x}
\eeq
and
\beq
\hat{z} \cdot (\hat{x} \wedge \hat{y} \wedge \hat{z})
= \hat{y} \wedge \hat{z}\;.
\eeq
However, this relationship makes it a little more tricky to define the
dot product consistently, due to the anticommutativity of $\wedge$:
\beq
\hat{y} \cdot (\hat{y} \wedge \hat{x})
= -\hat{y} \cdot (\hat{x} \wedge \hat{y})=-\hat{x}\;.
\eeq
This can be addressed by associating unit vectors with elements 
of totally ordered set, thus making a default decision between 
$\hat{x} \wedge \hat{y}$ versus $\hat{y} \wedge \hat{x}$. Then 
the power of $-1$ can be used to extend the definition of wedge product to the 
reverse orders. More precisely, vectors are associated with 
functions on the totally ordered set $S= \lbrace s_1, s_2, \ldots, s_n 
\rbrace$.

For simplicity, define the ordering in such a way that 
$s_i < s_j$ if and only if $i < j$. 

$s_1 \wedge s_2$ is a function defined as follows:
\beq
(s_1 \wedge s_2) ( \{s_1, s_2 \}) = 1 \; \; (s_1 \wedge s_2)(T) =0 , T \neq \{s_1, s_2\} \eeq
Anticommutativity implies that 
\beq
(s_2 \wedge s_1) ( \{s_1, s_2 \}) = -1\;.
\eeq
Remembering that $\{s_1, s_2\}=\{s_2, s_1\}$,
\bea
& &(s_1 \wedge s_2) ( \{s_2, s_1 \}) = 1 \nonumber \\
& &(s_2 \wedge s_1) ( \{s_2, s_1 \}) = -1\;.
\eea
Furthermore, $s_1 \wedge s_2 \wedge s_3$ is defined as 
\beq
(s_1 \wedge s_2 \wedge s_3) ( \{s_1, s_2, s_3 \}) =1 \;,
\quad (s_1 \wedge s_2)(T) =0 , T \neq \{s_1, s_2, s_3\}\;.
\eeq
The rule of addition of function, together with anticommutativity of the wedge, implies that $4s_6 + 3 s_5 \wedge s_7 + 8 s_8 \wedge s_{11} \wedge s_9$ is defined as a function $f$ satisfying the following properties: 
\beq f(\{6\}) = 4 \;; \quad f(\{5,7\})=3 \;; \quad f(\{8,9,11\}) = -8 \nonumber\ \eeq
\beq f (\{5,6,7\}) = f(\{5,8,9,11\})= \eeq
\beq = f(\{5,6,8,9,11\}) = f(\{5,6,7,8,9,11\}) = 0 \nonumber \eeq
More formally, dot and wedge products are defined as follows:

\noindent{\em Definition:\/} Let $p_1$ and $p_2$ be two polynomials
over $S$. Then $p_1 \cdot p_2$ is another polynomial over $S$ such
that for every $T \subset S$,
\beq
(p_1 \cdot p_2)(T) = \sum_{(U \setminus V) \cup (V \setminus U) = T} 
p_1(U)\, p_2(V)\;.
\eeq

The wedge product is defined as 
\beq
(p_1 \wedge p_2)(T) = \sum_{U \cup V = T \; ; \; U \cap V = \emptyset} (-1)^{\sharp \{ (a,b) \vert a>b \; ; \; a \in U \; ; \; b \in V \} } p_1 (U) p_2 (V) \; . 
\eeq
Finally, the definition of the derivative requires a definition of ratio. The situation is analogous with the set of integers where ratio is not defined everywhere and yet the notion of the ratio is used where it is. This is expressed in the following definition:

\noindent{\em Definition:\/} Let $\vec{a}$ and $\vec{b}$ be two Grassmann polynomials. If there exists a Grassmann polynomial $\vec{c}$ such that
$\vec{a} \wedge \vec{c} = \vec{b}$ then it is said that $\vec{c}= 
\vec{b}/\vec{a}$. If such $\vec{c}$ doesn't exist, then 
$\vec{b}/\vec{a}$ is not well defined.

The important thing is that the fraction was defined in terms of the wedge 
product, as opposed to the dot product, and also that the wedge product was 
ordered in the way it was. 

The other important component of definition of derivative is the definition of  a limit, which requires the notion of absolute values. Clearly, since $\vec{v} \wedge \vec{v} =0$, the wedge product can not be used as a definition of absolute value. Furthermore, since 
\beq (1+\hat{v}) \cdot (1+ \hat{v}) = 1 + 2 \hat{v} + \hat{v} \cdot \hat{v} = 2 + 2 \hat{v} \eeq
is non-real, the dot product can not be used as a definition of absolute value either. 

Instead, the absolute value of $\vec{V} \subset S \oplus (S \wedge S) \oplus (S \wedge S \wedge S) \oplus \ldots$ will be defined as a largest possible coefficient in its polynomial expression. As was done earlier, the coefficient of the $\vec{s_{a_1}} \wedge . . . \wedge \vec{s_{a_n}}$ term of $\vec{V}$ is identified with $\vec{V} (\{a_1, . . . , a_n \})$ up to possible sign difference. Since the sign difference does not affect absolute value, this gives the following definition: 

\beq \vert \vec{V} \vert = \sup \big\{ \vert \vec{V} (T) \vert \big\vert T \subset S \big\} \eeq

THEOREM: The above definition of absolute value satisfies triangle inequality

PROOF: Let $\vec{U}$ and $\vec{V}$ be two elements of $S \oplus (S \wedge S) \oplus (S \wedge S \wedge S) \oplus \ldots$. Suppose $\vert \vec{U} + \vec{V} \vert = c$. Then for any $\epsilon >0$ there exist $T \subset S$ such that $\vert \vec{U} (T) + \vec{V} (T) \vert > c- \epsilon $. The triangle inequality on $\mathbb{R}$ implies that $\vert \vec{U} (T) \vert + \vert \vec{V} (T) \vert > c - \epsilon$. But, by definition, $\vert U(T) \vert \leq \vert U \vert$ and $\vert V(T) \vert \leq \vert V \vert$ Therefore, $\vert \vec{U}  \vert + \vert \vec{V}  \vert > c - \epsilon$ Since this is true for all $\epsilon >0$, this implies that $\vert \vec{U}  \vert + \vert \vec{V}  \vert \geq c$, as desired. QED.

However, it is NOT true that absolute value of a dot product is equal to a product of absolute values. This can be seen from the following example:
\beq \vert (1+ \hat{e_1}) \cdot (1-\hat{e_1}) \vert = \vert 1 - \hat{e_1} \cdot \hat{e_1} \vert = \vert 1 - 1 \vert  = 0 \eeq 
yet,
\beq \vert 1+ \hat{e_1} \vert \vert 1-\hat{e_1}\vert  = 1 \times 1 = 1 \eeq
This, however, does not compromise the possibility of defining limits since, for a general metric space, there is no definition of product altogether, yet limits are well defined. 

Based on the above definition of absolute value, the limits on Grassmann space are defined as follows:

Let $\vec{F}$ be a function of the form $S \rightarrow S \oplus (S \wedge S) \oplus (S \wedge S \wedge S) \oplus \ldots$, and suppose $a \in S$ and $A \in S \oplus (S \wedge S) \oplus (S \wedge S \wedge S) \oplus \ldots$ Then
\beq \lim_{\vec{v} \rightarrow \vec{a}} \vec{F} (\vec{v}) = \vec{A} \Leftrightarrow  \eeq
\beq \Leftrightarrow \forall \epsilon >0 \exists \delta>0 (\forall \vec{v} \in S ((\vec{v}-\vec{a})\cdot(\vec{v}-\vec{a})<\delta^2 \Rightarrow \forall T \in S \vert \vec{F} (\vec{v})(T) - \vec{A}(T) \vert < \epsilon ) )\nonumber \eeq
By substituting the definitions for limit, the definition of the derivative becomes
\beq \vec{F}' (\vec{a}) = \vec{A} \Leftrightarrow \nonumber \eeq
\beq \Leftrightarrow \forall \epsilon >0 \exists \delta>0 \Big(\forall \vec{v} \in S \Big((\vec{v}-\vec{a})\cdot(\vec{v}-\vec{a})<\delta^2 \Rightarrow \eeq
\beq \Rightarrow \forall T \in S \Big\vert \vec{F} \Big( \frac{F(\vec{v})-F(\vec{a})}{\vec{v}-\vec{a}}\Big)(T) - \vec{A}(T) \Big\vert < \epsilon \Big) \Big)\nonumber \eeq
In order to make things more clean, the ratio can be removed in favor of its original definition. Thus, 
\beq  \Big\vert \vec{F} \Big( \frac{F(\vec{v})-F(\vec{a})}{\vec{v}-\vec{a}}\Big)(T) - \vec{A}(T) \Big\vert < \epsilon \nonumber \eeq
can be replaced with 
\beq  \exists \vec{D} \Big( \big(\vec{D} \wedge (\vec{v} - \vec{a}) = \vec{F}(\vec{v}) - \vec{F}(\vec{a}) \big)\wedge \vert \vec{D}(T)-\vec{A}(T)\vert < \epsilon \Big) \eeq 
Thus, the definition of derivative reads as follows: 
\beq \vec{F}' (\vec{a}) = \vec{A} \Leftrightarrow  \nonumber \eeq
\beq \Leftrightarrow \forall \epsilon >0 \exists \delta>0 \Big(\forall \vec{v} \in S \Big((\vec{v}-\vec{a})\cdot(\vec{v}-\vec{a})<\delta^2 \Rightarrow \eeq
\beq \Rightarrow \forall T \in S \exists \vec{D} \Big( \big(\vec{D} \wedge (\vec{v} - \vec{a}) = \vec{F}(\vec{v}) - \vec{F}(\vec{a}) \big)\wedge \vert \vec{D}(T)-\vec{A}(T)\vert < \epsilon \Big)\Big) \Big)\nonumber \eeq
In light of the fact that the absolute value of a product is NOT equal to the product of absolute values, this can NOT be further rewritten  as either $\vert \vec{F}(\vec{v}) - \vec{F} (\vec{a}) - \vec{A} (T) \wedge (\vec{v} - \vec{a})< \epsilon$ or $\vert \vec{F}(\vec{v}) - \vec{F} (\vec{a}) - \vec{A} (T) \cdot (\vec{v} - \vec{a})< \epsilon$, which means that the above definition of derivative is the final one. 

\newpage
\subsection{Vierbeins as Independent Fields}

As is the case for curved space, in the case of a general causal set a necessary ingredient to introduce spinor fields is the notion of vierbeins. In this thesis it is proposed to view vierbeins as physical fields subject to Lagrangian generator distinct from either gravitational or fermionic one. That Lagrangian generator would ``classically" force vierbeins to be orthonormal to each other. This, together with their coupling to causal structure, will increase the likelihood of the causal structure being manifold-like.

These vierbeins, however, do not match what one is used to be thinking of as vierbeins. In order to be able to claim that Lagrangian generator ``forces" vierbeins to be orthonormal, it is important to also claim that such relations do not exist apriori. Thus, in case of four dimensional manifold, each vierbein has 4 (not 3) degrees of freedom and the total number of degrees of freedom associated with 4 vierbeins is 16 rather than 10. These will be viewed as four independent vector fields. These vector fields are viewed as distinct both from fermions and from gravity, despite being coupled to both.

In case of general causal set, as before, vector fields can not be viewed as having a certain fixed number of degrees of freedom. Thus, by copying what was done for other vector fields, vierbeins are defined as simply a real valued functions on the set of pairs of points. In order to remind ourselves that these are independent, we use separate letters, $a$, $b$, $c$ and $d$ to define them:  
\beq a(p,q) = \int_{\gamma (p,q)} e_0^{\mu}\, \dd x^{\mu} \eeq
\beq b(p,q) = \int_{\gamma (p,q)} e_1^{\mu}\, \dd x^{\mu} \eeq
\beq c(p,q) = \int_{\gamma (p,q)} e_2^{\mu}\, \dd x^{\mu} \eeq
\beq d(p,q) = \int_{\gamma (p,q)} e_3^{\mu}\, \dd x^{\mu} \eeq
Intuitively, they represent $t$, $x$, $y$, and $z$ coordinates of $q$ with respect to $p$, respectively. However, it should be realized that the above are fields as opposed to coordinates, and neither coordinate system, nor manifold structure, were introduced by hand. Since no orthogonality is postulated, it is possible to take clearly one dimensional causal set, such as $\mathbb{R}$, and postulate four vierbein fields on it; it is also possible to start out with $\mathbb{R}^4$ and only postulate two vierbein vields. And, finally, it is also possible to postulate different systems of vierbeins with their own internal couplings on one and the same causal set, each system having different number of vierbeins, thus implying that the same causal set has different dimensions from itself!

The only reason the actual causal set ends up being manifold-like is that the system of vierbeins is subject to Lagrangian that makes them orthonormal. But even then, it is possible to imagine two competting systems of vierbeins, one has three vectors and the other has four vectors, both subject to the Lagrangians that make them orthonormal. If such two systems act on a random causal set, the result would neither approximate three dimensional manifold nor four dimensional one; but, while Lagrangian densities will be a lot more complicated, there will be no mathematical contradiction, contrary to what one would expect if one did, indeed, postulate two competting dimensionalities.  

However, the Lagrangian that ``encourages" vierbeins to be orthonormal should be ``stronger" than usual. After all, in case of four dimensional manifold, despite the fact that there are 16 vierbein degrees of freedom, it should closely approximate the situation with only 10 degrees of freedom. While classically this is easily done through the principle of least action, the same is not true quantum mechanically. 

In order to get a feel of what to do in quantum case, consider a simple example of one-dimensional non-relativistic quantum mechanics. The space degree of freedom is viewed as an analogue of one of the field degrees of freedom of quantum field theory. The goal is to ``effectively" get rid of the one space degree of freedom by setting $x \approx a$ while staying loyal to principles of quantum mechanics.This can be done by setting a potential 
\beq
V(x) = \Big(\frac{x}{a} \Big) ^n + \Big( \frac{x}{a} \Big)^{-n}\;.
\eeq
If $n$ is very large, the above would approximate a situation of a particle in the box, in which case the particle stays inside the box with 100 percent certainty. The purpose of $\vert x-a \vert^n$ term is to keep the particle from flying to the right, and the purpose of $\vert x-a \vert^{-n}$ term is to keep the praticle from flying to the left.

Since the causal set theory is Lagrangian-based rather than Hamiltonian-based, to make analogy closer, the above can be rewritten as
\beq {\cal L} = {\cal L}_{\rm kin} - \Big(\frac{x}{a} \Big)^n - \Big( \frac{x}{a} \Big)^{-n} \eeq
Thus, it is possible to get rid of quantum mechanical degrees of freedom by introducing two terms, similar to the above, into the Lagrangian. 

By viewing quantum field as multi-dimensional quantum mechanical system, this approach can be applied to vierbeins as well. The causal set version of this, however, is that instead of postulating such terms in a Lagrangian itself, they are postulated in a Lagrangian generator (and a choice should be made as to whether it is type 1 or type 2): 
\bea
& &{\cal K} (a,b,c,d; r, s) = \Big( \frac{a^2(r,s) - b^2(r,s) - c^2(r,s) - d^2(r,s)}{\tau^2 (r,s)} \Big) ^n \nonumber \\
& &\kern95pt +\ \Big( \frac{a^2(r,s) - b^2(r,s) - c^2(r,s) - d^2(r,s)}{\tau^2 (r,s)} \Big) ^{-n}\;,
\eea
where $n$ is a very large number. 

If the above Lagrangian generator is type 1, then in order for it to be close to $1$ at any given point $p$, one should be able to find an Alexandrov set $\alpha (p,q)$ such that the above expression is close to $1$ for every single pair of points selected inside of this Alexandrov set. On the other hand, if it is type 2, any given pair of points can violate that relation, as long as on average it is $1$. For this reason, I personally think that type 1 is preferable for vierbeins. But it is up to future numerical work to see if type 2 would still reproduce the predictions of quantum field theory. If so, then type 2 might be better in the sense that vierbeins would seem less ``forced". 

Of course, either of the two Lagrangian generators select ``minimizing" Alexandrov set and only affect what happens in its interio. This, however, matches what is expected of vierbeins. Due to lightcone singularity, in any given coordinate system, most of the points of the $\epsilon$-neighborhood of a point $x$ are arbitrary close to lightcone and, at the same time, arbitrary far away from $x$ coordinate-wise. Thus, the infinitesimal behavior of vierbeins is not something that happens in every single Alexandrov set, but only at a specifically selected one. This, does not affect how closely manifold-like it might get inside that, relevent, Alexandrov set, as illustrated by the fact that the approximation would hold for every single pair of elements of the latter, without fail, in type-1 scenario. 

Strictly speaking, this does not imply that causal set is manifoldlike. After all, for any given causal relations it is possible to adjust $a$, $b$, $c$ and $d$ point-wise in such a way that the causal structure given by $a^2-b^2-c^2-d^2$ matches the actual one. However, the more manifoldlike a given structure is, the more there are such ways to do that. This means that there will be more identical copies of the same causal structure in a path integrals, since all possible choices of vierbeins corresponding to each copy has to be counted separately. This amounts to manifoldlike causal structures to have higher weight in path integral than non-manifoldlike ones.

However, the fact that manifoldlike structures are overcounted should be balanced against the fact that they are very rare. It is up to future numerical work to see whether or not the total contribution of all manifold-like structures is greater or smaller than the total contribution of non-manifoldlike ones. This, of course, is also related to exactly how the notion of manifoldlike is defined. For instance, while a very thin cylinder can be viewed as a manfifold, the causal set produced by sprinkling of points onto that cylinder is not manifold-like. Exploring these issues, however, is beyond the scope of this thesis. 

\subsection{Vierbeins and Symmetry Properties of Fermionic Fields}

The next step of the theory is to define symmetry relations between vierbeins and fermionic fields. In Ref \cite{paper2} (which, for the purposes of general interest, will be summarized in the appendix to this chapter) it was shown that, in four dimensional manifold, in a toy model where spinor fields commute, one can ``trade" $2*4=8$ fermionic degrees of freedom with the 8 out of 10 vierbein ones. Thus, at any given point one can select a coordinate system in such a way that a fermionic field is a superposition of spin-up particle and spin-up antiparticle, with real coefficients. These two coefficients can be viewed as two real scalar fields, while the choice of frame can be viewed as four orthonormal vector fields. A spinor field can be viewed as the combination of these scalar and vector fields, which would contain $2+6=8$ degrees of freedom.  According to the above model, there is no such thing as spin 1/2 fermion. Rather, there are two spin 0 fields and four spin 1 fields coupled to each other. 

However, this approach does not work in light of the fact that vector fields are commutting while fermionic fields are anticommutting, which means that they can not be identified with each other. True, in the previous chapter Grassmann numbers were defined as individual elements of real vector space. However the price for doing this was the inclusion of $\xi$-measure. In case of several Grassmann numbers, $\xi$-measure is defined as 
\beq \xi(\theta_1 . . . \theta_n) = \xi (\theta_1) . . . \xi (\theta_n) \eeq
This is not invariant under rotation of $\theta$-s. One can experiment by defining multi-dimensional $\xi$ function that can not be expressed as above product, but still satisfies
\beq \int d \theta_1 . . . d \theta_n \xi (\theta_1, . . . , \theta_n) \theta_1 . . . \theta_n =1 \eeq
with all other integrals being $0$. But it is easy to see that in order to satisfy 
\beq \int d \theta_1 . . . d \theta_n \xi(\theta_1, . . . \theta_n) =0 \eeq
$\xi$ has to have both positive and negative values, which means that one has to ``choose" exactly where in the vector space is a ``transition point" where $\xi=0$. This choice will again violate the rotational symmetry of that space. 

For thses reasons, the idea of \cite{paper2} has to be dismissed for the case of anticommutting fermions. Instead, all the vierbein degrees of freedom has to be viewed as completely separate from fermionic ones. There is no such thing as symmatry transformation between fermionic field and vierbeins.  The field configurations that we would normally say are related by symmetries, are now viewed as completely different from each other, and the fact that their Lagrangian densities are the same is viewed as completely coincidental. 

This, of course, means that there is no formal justification in attempting to introduce Fadeev-Popov ghosts or use other techniques designed specifically for symmetries. Instead, there will be a lot of overcounting resulting from the fact that Lagrangian densities ``happened" to be the same at too many points. Infinities will be avoided by restricting path integration to large but finite range. 

Quite independently from the above, there is a separate source of extra degrees of freedom: as explained in section 5.2 in a lot more details, vierbeins are viewed as four independent vector fields that, apriori, are neither orthogonal to each other nor have norm 1. Thus, in case of four dimensional manifold, they contribute 16, rather than 10, degrees of freedom. At the same time, Lagrangian generator is introduced that ``enoourages" them to be orthonormal and that generator is ``strong" enough to make it ``effectively" appear as if there are only 10 degrees of freedom.  

Strictly speaking this means that there is no such thing as spin 1/2. Instead, there are four independent spin 1 fields and two independent spin 0 fields, the former commute and the latter anticommute. They are coupled to each other through various Lagrangian generators, which results in an appearance of spin 1/2 field which is not really there. 

\subsection{Lagrangian Generator for Type-1 Fermions}

As a consequence of the fact that Grassmann numbers are well defined outside of integral, the theory about going from Lagrangian generator to Lagrangian is well defined for Grassmann numbers. The next step is to find out what fermionic Lagrangian generators actually are. The goal is to come up with a Lagrangian generator that would, in the continuum limit, generate
\beq
{\cal L} = \overline{\psi} \gamma^m e_a^{\mu} \partial_{\mu} \psi + \overline{\psi} \gamma^m \sigma^{ab} \psi (e_m^{\mu} e_a^{\nu} (\partial_{\mu} e_{b \nu} - \partial_{\nu} e_{b \mu}) + e_a^{\rho} e_b^{\sigma} \partial_{\sigma} e_{\rho}^m )\;.
\eeq
This can be rewritten as 
\beq
{\cal L} = \gamma^m_{ij} \overline{\psi_i}  e_a^{\mu} \partial_{\mu} \psi_j + \overline{\psi} \gamma^m \sigma^{ab} \psi (e_m^{\mu} e_a^{\nu} (\partial_{\mu} e_{b \nu} - \partial_{\nu} e_{b \mu}) + e_a^{\rho} e_b^{\sigma} \partial_{\sigma} e_{\rho}^m)\;.
\eeq
Vierbeins are viewed as four separate vector fields, whose orthonormality is a consequence of Lagrangian generators introduced in sections 5.4 and 5.5. Similarly to what was done for a gauge field, vector fields are associated with scalar fields of pairs of points:
\beq
\tilde{e_a}(r,s) = (s^\mu-r^\mu)\, \big(e_{a\mu}(r) + \half\,(r^\nu + s^\nu)\, 
\partial_\nu e_{a\mu} \big) + O(\tau^3)\;.
\eeq
The expressions that need to be redefined for causal set scenario are $A^{\mu} B^{\nu} \partial_\nu C_\mu$ as well as $A^\mu\,B_\mu$.

Noticing that 
\beq A^{\mu} B_{\mu} = (A+B)^{\mu} (A+B)_{\mu} - A^{\mu} A_{\mu} -  B^{\mu} B_{\mu} \eeq
the Lagrangian generator corresponding to this expression is 
\beq {\cal K}_1 = \{ {\cal K}_{11},{\cal K}_{12},-{\cal K}_{13},-{\cal K}_{14},-{\cal K}_{15},-{\cal K}_{16} \}\eeq 
where $({\cal K}_{11}, {\cal K}_{12})$ correspond to $(A+B)^{\mu} (A+B)_{\mu}$, $({\cal K}_{13}, {\cal K}_{14})$ correspond to $A^{\mu} A_{\mu}$ and $({\cal K}_{15},{\cal K}_{16})$ correspond to $B^{\mu} B_{\mu}$.

For any fixed point $p$, $a(p,r)$ and $b(p,r)$ can be thought of as scalar functions of $r$ alone. Therefore, from section 3.3, they are given as 

${\cal K}_{11} (a, b, p, q) = k (a(p,q) + b(p,q))^2$ ,  $p \prec q$.

${\cal K}_{12} (a, b, p, q) = -k (a(p,q) + b(p,q))^2$ , $p$ and $q$ are unrelated.

${\cal K}_{13} (a, b, p, q) = k a^2 (p,q)$,  $p \prec q$.

${\cal K}_{14} (a, b, p, q) = -k a^2 (p,q)$ , $p$ and $q$ are unrelated.

${\cal K}_{15} (a, b, p, q) = k b^2(p,q)$ , $p \prec q$.
 
${\cal K}_{16} (a, b, p, q) = -k b^2(p,q)$ , $p$ and $q$ are unrelated.

As was seen in the scalar field section, the same equation applies both to spacelike and timelike gradients of scalar field. Thus, the above question will be the same if $A^\mu$ is replaced with one of the spacelike vectors. Therefore,

$ {\cal K}_{11} (b, c, p, q) = k (b(p,q) + c(p,q))^2$ ,  $p \prec q$
 
$ {\cal K}_{12} (b, c, p, q) = -k (b(p,q) + c(p,q))^2$ , $p$ and $q$ are unrelated

$ {\cal K}_{13} (b, c, p, q) = k b^2 (p,q)$,  $p \prec q$ 

${\cal K}_{14} (b, c, p, q) = -k b^2 (p,q)$ , $p$ and $q$ are unrelated 

${\cal K}_{15} (b, c, p, q) = k c^2(p,q)$ , $p \prec q$
 
$ {\cal K}_{16} (b, c, p, q) = -k c^2(p,q)$ , $p$ and $q$ are unrelated

A slight modification of the above defines the value of $A^\mu\, \partial_\mu\phi$. Namely, $b(r,s)$ should be replaced with $\phi(s) - \phi(r)$, which means replacing $B_\mu$ with $\partial_\mu \phi$. These substitutions produce

${\cal K}_{11} (a, \phi , p, q) = c (a(p,q) + \phi(q)-\phi(p))^2$ , $ p \prec q $ 

${\cal K}_{12} (a, \phi, p, q) = -c (a(p,q) + \phi(q)-\phi(p))^2$ , $p$ and $q$ are unrelated 

${\cal K}_{13} (a, \phi, p, q) = c a^2 (p,q)$ , $p \prec q$

${\cal K}_{14} (a, \phi, p, q) = -c a^2 (p,q)$ , $p$ and $q$ are unrelated

${\cal K}_{15} (a, \phi, p, q) = c (\phi(q)-\phi(p))^2$ , $p \prec q$
 
${\cal K}_{16} (a, \phi, p, q) = -c (\phi(q)-\phi(p))^2$ , $p$ and $q$ are unrelated

Again, the same equation applies if $A^\mu$ is replaced with something spacelike:

${\cal K}_{11} (b, \phi , p, q) = c (b(p,q) + \phi(q)-\phi(p))^2$ , $ p \prec q $ 

${\cal K}_{12} (b, \phi, p, q) = -c (b(p,q) + \phi(q)-\phi(p))^2$ , $p$ and $q$ are unrelated 

${\cal K}_{13} (b, \phi, p, q) = c b^2 (p,q)$ , $p \prec q$

${\cal K}_{14} (b, \phi, p, q) = -c b^2 (p,q)$ , $p$ and $q$ are unrelated

${\cal K}_{15} (b, \phi, p, q) = c (\phi(q)-\phi(p))^2$ , $p \prec q$
 
${\cal K}_{16} (b, \phi, p, q) = -c (\phi(q)-\phi(p))^2$ , $p$ and $q$ are unrelated

Now let us try to get $A^\mu\, B^\nu\, \partial_\nu C_\mu$.

Inspired by the similarity in structure with electromagnetism, a good starting point is to attempt to maximize the fluctuation of $a(r,s)\, b(r,s)\, c(r,s)\, d(r,s)$. Luckily, it is possible to simultaneously minimize the fluctuation of each of the four individual multiples, which would automatically minimize the fluctuation of the product. Based on the results of the scalar part, if the gradient of $\phi$ is timelike, in order for the fluctuations of $\phi$ to be minimized its gradient has to be parallel to the axis of the Alexandrov set; on the other hand, if the gradient is spacelike then in order for the fluctuations of $\phi$ to be minimized, that gradient should lie along the equator of the Alexandrov set. This means that in order for fluctuations of $a(r,s)$ to be minimized, $A^\mu$ should point along the axis of the Alexandrov set, while in order for fluctuations of $b(r,s)$, $c(r,s)$ and $d(r,s)$ to be minimized, then $B^\mu$, $C^\mu$ and $D^\mu$ should lie on the equator of the Alexandrov set. But the orthogonality condition implies that, not only these four statements are compatible, but in fact if the condition about the timelike vector $A^\mu$ is met, it {\em forces\/} the conditions about the three spacelike vectors $B^\mu$, $C^\mu$ and $D^\mu$ to be met as well! This means that minimizing the fluctuations of a product $a(r,s)\,b(r,s)\,c(r,s)\,d(r,s)$ implies minimizing each of the four multiples individually, which completely specifies the paramenters of the Alexandrov set.

Now select points $r_1$, $r_2$, $r_3$ and $r_4$ in a way that minimizes the fluctuations of $a(r_1,r_3)\, b(r_2,r_4)$. Again, each of these two multiplets is maximized separately. It is easy to see that $a(r_1, r_3)$ is maximized under $r_1 = p = (-\half\,\tau,0,0,0)$ and $r_3 = q = (\half\,\tau,0,0,0)$ while $b(r_2,r_4)$ is maximized under $r_2 = (0,\half\,\tau,0,0)$ and $r_4 = (0,-\half\,\tau,0,0,0)$ These two conditions are comparable with each other. Now suppose that we instead decided to maximize the fluctuations of $b(r_1,r_3)\, c(r_2,r_4)$. Again, fluctuations of each of these fields can be maximized simultaneousle. This time, since both of these fields have spacelike gradient, each one will be maximized by selecting points on the equator. In particular, to maximize $b(r_1,r_3)$, set $r_1 = (0,-\half\,\tau(p,q),0,0)$ and $r_3 = (0,\half\,\tau(p,q),0,0)$ and in order to maximize $c(r_2, r_4)$ set $r_2 = (0,0,\half\,\tau(p,q),0)$ and $r_4 = (0,0,-\half\,\tau(p,q),0)$. Again, this results in the same square loop that was used in the gauge case. Noticing that the square loop for the case of one timelike and one spacelike field is the same as the square loop for two spacelike fields, except that $t$ axis was replaced by the $y$ axis, it can be summarized that $u(r_1,r_3)\, v(r_2,r_4)$ is maximized under $U^\mu\, r_{1\mu} = -\half\,\tau$, $U^\mu\, r_{3\mu} = \half\,\tau$, $V^\mu\, r_{2\mu} = \half\,\tau$, $V^\mu\, r_{4\mu} = -\half\,\tau$ regardless of whether the fields are spacelike or timelike. 

Now if $w$ is a third holonomy, then, remembering that all vectors are unit vectors, the results from the gauge part can be rewritten as
\beq
w(r_1, r_2) + w(r_2, r_3) + w(r_3, r_4) + w(r_4, r_1)
= \tau^2\, (\partial_\rho W_\sigma - \partial_\sigma W_\rho)\;.
\eeq 
where $\rho$ and $\sigma$ are directions corresponding to $U$ and $V$. Remembering that $U$ and $V$ are of unit length, this can be rewritten as 
\beq
w(r_1, r_2) + w(r_2, r_3) + w(r_3, r_4) + w(r_4, r_1)
= \tau^2\, U^\mu\, V^\nu\, (\partial_\mu W_\nu - \partial_\nu W_\mu). \eeq
Now $U^\mu\, W_\mu = 0$ implies that
\beq
U^\mu\, V^\nu\, \partial_\nu W_\mu = - W^\mu\, V^\nu \partial_\nu U_\mu\;.
\eeq
Thus the above expression becomes 
\beq
w(r_1, r_2) + w(r_2, r_3) + w(r_3, r_4) + w(r_4 , r_1) = \nonumber \eeq
\beq = \tau^2(p,q)\,
(U^\mu\, V^\nu\, \partial_\mu W_\nu + W^\mu\, V^\nu\, \partial_\nu U_\mu).
\eeq
As was stated in the previous paragraph, the above was done under the condition of maximization of $u(r_1, r_3) v(r_1, r_3)$. In fact, this is the only dependence on $u$ and $v$ which means that if it was taken away, then $U^{\mu}$ and $V^{\mu}$ would no longer make sense on the right-hand side of the equation. 

To make this formally consistent with the generic prescription of Lagrangian generators, it is necessary to write that maximization explicit, within the framework of the former. The problem with this is that the selection of points is chosen to maximize one expression, $u(r_1, r_3)v(r_1, r_3))$, while the expression being evaluated is different, $w(r_1, r_2) + w(r_2, r_3) + w(r_3, r_4) + w(r_4 , r_1) $. In order to be consistent with the Lagrangian prescription, these two expressions should be the same. This is accomplished this by replacing both of them with $\big(u(r_1, r_3)v(r_1,r_3)\big)^n \big(w(r_1, r_2) + w(r_2, r_3) + w(r_3, r_4) + w(r_4 , r_1)\big)$ where $n$ is a very large number. If $n$ is large enough, the maximizing of above expression would approximately mean maximizing $u(r_1, r_3) v(r_1, r_3)$. At the same time, if the above maximization occurs, then the approximation $u(r_1, r_3) \approx v(r_1, r_3) \approx \tau$ will hold. Thus, the only non-trivial expression left will be $w(r_1, r_2) + w(r_2, r_3) + w(r_3, r_4) + w(r_4 , r_1)$, as desired. 

Thus, the equation can be rewritten as 
\bea
& &\kern-20pt\min_{\tau (p,q) = \tau_0}\; \max_{r_1, r_2, r_3, r_4 \in \alpha (p,q)} 
\big(u(r_1, r_3)v(r_1,r_3)\big)^n ( w(r_1, r_2) + w(r_2, r_3) + w(r_3, r_4) + w(r_4 , r_1))\nonumber \\
& &\kern100pt= \tau^{2(1-n)}(p,q)\,
(U^\mu\, V^\nu\, \partial_\mu W_\nu + W^\mu\, V^\nu\, \partial_\nu U_\mu)\;.
\eea
We can now permute this equation to get another two equations,
\bea
& &\kern-20pt\min_{\tau (p,q) = \tau_0}\; \max_{r_1, r_2, r_3, r_4 \in \alpha (p,q)} \big(v(r_1, r_3)w(r_1,r_3)\big)^n ( u(r_1, r_2) + u(r_2, r_3) + u(r_3, r_4) + u(r_4 , r_1))\nonumber \\
& &\kern100pt= \tau^{2(1-n)}(p,q)\,
(V^\mu\, W^\nu\, \partial_\mu U_\nu + U^\mu\, W^\nu\, \partial_\nu V_\mu)\;,
\eea
and
\bea
& &\kern-20pt\min_{\tau (p,q) = \tau_0}\; \max_{r_1, r_2, r_3, r_4 \in \alpha (p,q)} \big(w(r_1, r_3)u(r_1,r_3)\big)^n ( v(r_1, r_2) + v(r_2, r_3) + v(r_3, r_4) + v(r_4 , r_1)) \nonumber \\
& &\kern100pt= \tau^{2(1-n)}(p,q)\,
(W^\mu\, U^\nu\, \partial_\mu V_\nu + V^\mu\, U^\nu\, \partial_\nu W_\mu).
\eea
Subtracting the second equation from the sum of first and third, and then dividing the whole thing by 2, leads to
\bea
& & U^\mu\, V^\nu\, \partial_\mu W_\nu = \nonumber \\
& & =\half\, \Big(\frac{k_d}{V_0}\Big)^{1/d} \Big(
\min_{\tau (p,q) = \tau_0} \max_{r_1, r_2, r_3, r_4 \in \alpha(p,q)}\big(u(r_1,r_3)v(r_1,r_3)\big)^n \times \\
& & \times (w(r_1, r_2) + w(r_2, r_3) + w(r_3, r_4) + w(r_4 , r_1) ) + \nonumber \\
& &  + \min_{\tau (p,q)= \tau_0} \max_{r_1, r_2, r_3, r_4 \in \alpha(p,q)} \big(v(r_1, r_3)w(r_1,r_3)\big)^n (u(r_1, r_2) + u(r_2, r_3) + u(r_3, r_4) + u(r_4 , r_1) ) - \nonumber \\
& & -\min_{\tau(p,q) = \tau_0} \max_{r_1, r_2, r_3, r_4 \in \alpha(p,q)} \big(u(r_1,r_3)w(r_1,r_3)\big)^n (v(r_1, r_2) + v(r_2, r_3) + v(r_3, r_4) + v(r_4 , r_1) )\Big),
\nonumber \eea
which is the desired term.

The Lagrangian generator for the above term is $\{ {\cal K}_1, {\cal K}_2, -{\cal K}_3 \}$, where
\beq {\cal K}_1 (u,v,w; r_1, r_2, r_3, r_4) = \eeq
\beq = \big(u(r_1, r_3) v(r_1, r_3) \big)^n
\big(w(r_1, r_2) + w(r_2, r_3) + w(r_3, r_4) + w (r_4, r_1) \big) \nonumber \eeq
\beq {\cal K}_2 (u,v,w; r_1, r_2, r_3, r_4) = \eeq 
\beq = \big(v(r_1, r_3) w(r_1, r_3) \big)^n
\big(u(r_1, r_2) + u(r_2, r_3) + u(r_3, r_4) + u (r_4, r_1) \big) \nonumber \eeq
\beq {\cal K}_3 (u,v,w;r_1 , r_2, r_3, r_4) = \eeq
\beq = \big(u(r_1, r_3) w(r_1, r_3) \big)^n
\big(v(r_1, r_2) + v(r_2, r_3) + v(r_3, r_4) + v (r_4, r_1) \big)\nonumber \eeq
We can now rewrite the fermionic Lagrangian as 
\beq
{\cal L} = \gamma^m_{ij} \overline{\psi_i}  e_m^{\mu} \partial_{\mu} \psi_j + (\overline{\psi} \gamma^m \sigma^{ab} \psi - \overline{\psi} \gamma^a \sigma^{mb} \psi+ \overline{\psi} \gamma^m \sigma^{ab} \psi)\, e_b^\mu e_a^\nu \partial_\mu e_{m\nu}\;.
\eeq
This can be expressed in terms of the following Lagrangian generators:
\beq
{\cal K}_1^{mij} (\psi, e_0, e_1, e_2, e_3; p, q) = \gamma^m_{ij} \overline{\psi}_i (p) {\cal K}_1 (e_m, \psi_j; p, q ) \eeq 
\beq {\cal K}_2^{abm} (\psi, e_0, e_1, e_2, e_3; r_1, r_2, r_3, r_4) = \eeq
\beq  = (\overline{\psi} (r_1) \gamma^m \sigma^{ab} \psi (r_1) - \overline{\psi} (r_1) \gamma^a \sigma^{mb} \psi (r_1) + \overline{\psi} (r_1) \gamma^m \sigma^{ab} \psi (r_1)) \times \nonumber \eeq
\beq \times\ \{ {\cal K}_1, {\cal K}_2, -{\cal K}_3 \} (e_b, e_a,  e_m; r_1, r_2, r_3, r_4) \eeq
\beq {\cal K}_{\rm fermionic} = \{ {\cal K}_1^{mij} \vert \{m,i,j \} \subset \{ 0,1,2,3 \} \} \cup \nonumber \eeq
\beq \cup \{  {\cal K}_2^{abm} \vert \{m,i,j \} \subset \{0,1,2,3\} \eeq

\subsection{Lagrangian Generator for Type-2 Fermions}

Lagrangian for spinor field is given by 
\beq
{\cal L} = \overline{\psi} \gamma^{\mu} \partial_\mu \psi + \overline{\psi} \gamma^a \sigma^{ab} \psi (e_m^{\mu} e_a^{\nu} (\partial_\mu e_{b\nu} - \partial_\nu e_{b \mu})+e_a^{\rho}e_b^{\sigma}\partial_{\sigma}e_\rho^m)\;.
\eeq
As before, $e_a$, $e_b$, $e_c$ and $e_d$ are viewed as four separate vector fields, represented by four separate sets of holonomies. However, the information about orthogonality of these fields will be used, which might either be enforced through the separate type-1 Lagrangian generator for vierbeins talked about earlier or else through some Lagrange multiplier terms. 

We would like to compute separate type-2 Lagrangian generators for each of the key terms of the Lagrangian.

Let's start with the Lagrange multiplier terms for vierbeins. Since these terms are structually similar to the scalar field Lagrangian, lets try to borrow our results from the latter in order to save ourselves some work. 

From the section on type-2 scalar fields, we know that their Lagrangian generator is $({\cal J}_s, f, g)$ where 
\beq  {\cal J}_s (\phi, r, s) =  (\phi (r) - \phi(s))^2) \eeq
and corresponding pre-Lagrangian is 
\beq {\cal L} (\phi, E, p, q) = k_d \tau^{2d} (p,q) (q^{\mu} - p^{\mu})(q^{\nu} -p^{\nu})\partial_{\mu} \phi \partial_{\nu} \phi (k_d + 2E_d (I_{d0} + I_{d1} (d-1))) - \nonumber \eeq
\beq - 2 k_d E_d I_{d1}  \tau^{2d+2}(d-1) \partial^{\mu} \phi \partial_{\mu} \phi  \eeq
Replacing $\phi (r) - \phi (s)$ with $(r^{\mu} - s^{\mu})V_{\mu}$ it is easy to see that pre-Lagrangian corresponding to 
\beq {\cal J} (r,s) = v^2 (r,s) \eeq
is given by 
\beq {\cal L} (\phi, E, p, q) = k_d \tau^{2d} (p,q) (q^{\mu} - p^{\mu})(q^{\nu} -p^{\nu})v_{\mu} v_{\nu} (k_d + 2E_d (I_{d0} + I_{d1} (d-1))) - \nonumber \eeq
\beq - 2 k_d E_d I_{d1}  \tau^{2d+2}(d-1) v^{\mu} v_{\mu}  \eeq
While this is not necesserely true about actual Lagrangian, it is easy to see that pre-Lagrangian depends linearly on type-2 Lagrangian generator. Thus, by using
\beq u(r,s) v(r,s) = \frac{1}{2} ((u(r,s) + v(r,s))^2 - u^2 (r,s) - v^2 (r,s)) \eeq
and 
\beq u^{\mu} v_{\mu} = \frac{1}{2} ((u^{\mu} + v^{\mu})(u_{\mu} + v_{\mu}) - u^{\mu} u_{\mu} - v^{\mu} v_{\mu} ) \eeq
the pre-Lagrangian corresponding to 
\beq {\cal J} (u,v,r,s) = u(r,s) v(r,s) \eeq
with the same $f$ and $g$ as earlier is
\beq {\cal L} (u,v,E,p,q) = k_d \tau^{2d} (p,q) (q^{\mu} - p^{\mu})(q^{\nu} -p^{\nu})u_{\mu} v_{\nu} (k_d + 2E_d (I_{d0} + I_{d1} (d-1))) - \nonumber \eeq
\beq - 2 k_d E_d I_{d1}  \tau^{2d+2}(d-1) u^{\mu} v_{\mu}  \eeq
It is also easy to see that all coefficients are the same as the scalar case, so the solution for $E$ gives us identical result as it did back then which, upon substitution, gives us the same answer as for scalar case, if $\partial^{\mu} \phi \partial_{\mu} \phi$ is replaced with $U^{\mu} V_{\mu}$ and mass term is dropped:
\beq {\cal L} = \frac{I_{d1}k_d^2 (d-1)}{I_{d0} + I_{d1} (d-1)} \tau^{2d+2} U^{\mu}  V_{\mu}\eeq
We would also like to produce $U^{\mu} U_{\mu} + 1$ and $U^{\mu} U_{\mu} -1$. By inspecting the mass term of the Lagrangian generator for scalar field, while dropping both $m$ and $\phi_0$, it is easy to see that if Lagrangian generator is set to constant,
\beq {\cal J} (r,s)=1 \eeq
with $f$ and $g$ defined in the same way as before, the corresponding pre-Lagrangian is  
\beq {\cal L}_{const} (E, p,q)(1+E) k_d^2  \tau_2^{2d} (p,q) \eeq
Again, since the value of $E$ is identical to the one in scalar case, the actual Lagrangian is also identical to the mass term of scalar Lagrangian, where $\phi_0$ and $m$ are dropped: 
\beq {\cal L} (v, p, q) = k_d^2 \tau_2^{2d} \Big(1-\frac{k_d}{2(I_{d0} + I_{d1} (d-1))} \Big)  \eeq
In order for the sum of the two Lagrangians to be proportional to $V_d^{\mu} V_{d \mu} +1$, $V_d$ has to be defined as 
\beq V^{\mu}_d = V^{\mu} \tau_2^d \sqrt{\frac{I_{d1} (d-1)}{I_{d0} + I_{d1} (d-1)-k_d} } \eeq
The $+$ and $-$ signs are taken into account by replacing ${\cal J}=1$ with ${\cal J}=-1$. 

As far as $\overline{\psi} \gamma^{\alpha} e_{\alpha}^{\mu} \partial_{\mu} \psi$ goes, we can factor out $\overline{\psi} \gamma^{\alpha}$ and write a Lagrangian generator for $e_{\alpha}^{\mu} \partial_{\mu} \phi$. We can do that by copying what we had for $u^{\mu} v_{\mu}$, substituting $u_{\alpha}^{\mu}$ for $u^{\mu}$ and $\partial_{\mu} \psi$ for $v_{\mu}$. This substitution amounts to replacing $u(r,s)$ with $e_{\alpha} (r,s)$ and replacing $v(r,s)$ with $\psi(s) - \psi(r)$. Thus, the Lagrangian generator for kinetic term with constant vierbeins is 
\beq
{\cal L}_{{\rm flat} \;,\; {\rm kin}} (\psi, r, s)
=  \overline{\psi} \gamma^{\alpha} e_a (r, s) (\psi(s) - \psi(r))\;.
\eeq
Now let us move to type-2 Lagrangian generator for  $e_a^{\mu} e_b^{\nu} \partial_\nu e_{c\mu}$ terms. 

The plan is the following: 

PART 1: By using orthonormality of $e_k$, show that $e_a{}^\mu\, e_b{}^\nu\, \partial_{\nu} e_{c\mu}$ can be expressed as linear combination of the terms of the form $E_{lmn} = e_l{}^\mu\, e_m{}^\nu\, (\partial_\mu e_{n\nu} - \partial_\nu e_{n \mu} )$. This would simplify the situation tremendously since the latter somewhat resembles gauge theory which was already done.

PART 2: Find type-2 Lagrangian generator for $E_{abc}$ in a coordinate-free setting of causal set. Even though, as remarked above, the resemblence to guage theory should make it easy, there are still difference with gauge theory, including the fact that there are 3 holonomies rather than 1, which makes it somewhat difficult. But orthonormality of these three holonomies will be used in passing through this. 

\noindent PART 1

\noindent If the above expression is expanded and the dummy indices $\mu$ and $\nu$ are switched on the second term, this gives 
\beq
E_{abc} = e_a{}^\mu\, e_b{}^\nu\, \partial_\mu e_{c\nu}
- e_a{}^\nu\, e_b{}^\mu\, \partial_\mu e_{c\nu}\;.
\eeq
Apply $\partial_\mu (e_a{}^\mu\, e_{c\mu}) = \partial_\mu \eta_{ab} = 0$ to the second term of above equation gives
\beq
E_{abc} = e_b{}^\nu\, e_a{}^\mu\, \partial_\mu e_{c\nu}
+ e_c{}^\nu\, e_b{}^\mu \partial_\mu e_{a \nu}\;.
\eeq
Permuting the indices gives 
\bea
& &E_{bca} = e_c{}^\nu\, e_b{}^\mu\, \partial_{\mu} e_{a\nu}
+ e_a{}^\nu\, e_c{}^\mu \partial_\mu e_{b \nu}
\\
& &E_{cab} = e_a{}^\nu\, e_c{}^\mu\, \partial_{\mu} e_{b\nu}
+ e_b{}^\nu\, e_a{}^\mu \partial_\mu e_{c\nu}\;.
\eea
From these expressions it is easy to see that
\beq
E_{abc} + E_{bca} - E_{cab} = 2\, e_c{}^\nu\, e_b{}^\mu\, \partial_\mu e_a{}^\nu\;.
\eeq
Switching $a$ and $c$ and dividing the expression by 2 gives
\beq
e_a{}^\nu\, e_b{}^\mu\, \partial_\mu e_c^\nu
= \frac{1}{2}\, (E_{cba} + E_{bac} - E_{acb})\;.
\eeq
Thus, the problem of computing $e_a{}^\nu\, e_b{}^\mu\, \partial_\mu e_c{}^\nu$ reduces to the problem of computing $E_{cba}$ for causal set, as desired.

\noindent PART 2

\noindent 
Try Lagrangian generator $({\cal J}, f, g)$ where
\beq {\cal J} (u,v,w) = u(t,r) v(t,s) (w(r,s) + w (s,t) + w (t,r)) \eeq  
\beq f(r_1, r_2, r_3, r_4, r_5) = (r_1, r_3, r_4) \; , \; g(r_1, r_2, r_3, r_4, r_5) =  (r_3, r_4, r_t) \eeq
The pre-Lagrangian corresponding to this generator is given by 
\beq {\cal L} (u,v,w,E,p,q) = \int d^d r d^d s d^d t (J(p,r,s)+J(q,r,s)+2EJ(r,s,t)) = \nonumber \eeq
\beq = \Big( \int d^d t \Big) \int d^d r d^d s (J(p,r,s)+J(q,r,s)) + 2E \int d^d r d^d s d^d t J(r,s,t) = \nonumber \eeq
\beq = k_d \tau^d \int d^d r d^d s (J(p,r,s)+J(q,r,s)) + 2E \int d^d r d^d s d^d t J(r,s,t)  \eeq 
In the linear case that satisfies time reversal symmetry, the pre-Lagrangian becomes
\beq {\cal L} (u,v,w,E,p,q) = \eeq
\beq = 2 \Big( k_d \tau^d \int d^d r d^d s J(p,r,s) + E \int d^d r d^d s d^d t J(r,s,t) \Big) \nonumber \eeq 
Let  $U^{\mu}$, $V^{\mu}$ and $W^{\mu}$ be vector fields corresponding to $u$, $v$ and $w$ . As usual, set a coordinate system in which $t$ axis passes through $p$ and $q$, which are the end points of Alexandrov set $\alpha (p,q)$. In this coordinate system,  
\beq p= (- \tau/2, 0, 0, 0) \; , \; q= (\tau/2, 0, 0, 0) \eeq

From intuition we have from electrodynamics, 
\beq w(r,s)+w(s,t)+w(t,r) = (r^{\mu} - t^{\mu})(s^{\nu} - t^{\nu})(\partial_{\mu} W_{\nu} - \partial_{\nu} W_{\mu}) \eeq
Thus, in the above coordinate system, the first integral is given by 
\beq \int_{\alpha (p,q)} d^d r  d^d s u(p,r) v(p,s) (w(p,r)+w(r,s)+w(s,p)) = \nonumber \eeq
\beq = U_{\rho} V_{\sigma} (\partial_{\mu} W_{\nu} - \partial_{\nu} W_{\mu} ) \int d^d r d^d s (r^{\rho} r^{\mu} s^{\sigma} s^{\nu} - \frac{\tau^2}{4} \delta^{\sigma}_0 \delta^{\nu}_0 r^{\rho} r^{\mu} + \nonumber \eeq
\beq + \frac{\tau^2}{4} \delta^{\rho}_0 \delta^{\mu}_0 s^{\sigma} s^{\nu} + \frac{\tau^4}{16} \delta^{\mu}_0 \delta^{\nu}_0 \delta^{\rho}_0 \delta^{\sigma}_0 ) \eeq
By antisymmetry of $\partial_{\mu} W_{\nu} - \partial_{\nu} W_{\mu}$, $\mu$ and $\nu$ can never be equal in terms that survive the cancellation. Thus, in the second term inside the integral, the fact that $\mu=0$ (as inferred by $\delta^{\nu}_0$) implies that $\nu=k$ for some $k \geq 1$. In order for that term to survive integration, the indices of two $r$-s have to match which means that $\rho = \mu = k$. 

As far as the third term is concerned, the situation is the opposite: since $\mu =0$, we set $\nu=k$. In order for indices of $s$-s to match, it means that $\nu=k$ as well. Finally, as far as the last term is concerned, due to $\delta^{\mu}_0 \delta^{\nu}_0$, the coefficient outside the integral becomes $\partial_0 W_0 - \partial_0 W_0 =0$ which means that the last term is dropped. out. 

Thus, the integral becomes
 \beq \int_{\alpha (p,q)} d^d r  d^d s u(p,r) v(p,s) (w(p,r)+w(r,s)+w(s,p)) = \eeq
 \beq = \tau^{2d+4} (U_0 V_k (\partial_0 W_k - \partial_k W_0) I_{d0} I_{d1} + U_k V_0 (\partial_k W_0 - \partial_0 W_k) I_{d1} I_d0) + \nonumber \eeq
 \beq + \frac{k_d I_{d1}}{4} U_k V_0 (\partial_k W_0 - \partial_0 W_k) + \frac{k_d I_{d1}}{4} U_0 V_k (\partial_0 W_k - \partial_k W_0 ) ) \nonumber \eeq
 After some simple algebra this becomes
  \beq \int_{\alpha (p,q)} d^d r  d^d s u(p,r) v(p,s) (w(p,r)+w(r,s)+w(s,p)) = \nonumber \eeq
  \beq = \tau^{2d+4} I_{d1} (I_{d0} + \frac{k_d}{4}) (\partial_0 W_k - \partial_k W_0) (U_0 V_k - U_k V_0) \eeq
  Now lets move to the other integral,
 \beq \int d^d r d^d s d^d t u(r,s)v(r,t)(w(r,s)+w(r,t)+w(t,r)) =  \eeq
 \beq = U_{\rho} V_{\sigma} (\partial_{\mu} W_{\nu} - \partial_{\nu} W_{\mu}) \int d^d r d^d s d^d t (s^{\rho} - r^{\rho})(t^{\sigma} - r^{\sigma})(s^{\mu}-r^{\mu})(t^{\nu}-r^{\nu}) \nonumber \eeq
Let's expand out $(s^{\mu}-r^{\mu})(t^{\nu}-r^{\nu})$. It is easy to see that 
\beq r^{\mu} r^{\nu} (\partial_{\mu} W_{\nu} - \partial_{\nu} W_{\mu}) \eeq
which means that $r^{\mu} r^{\nu}$ term can be dropped. Thus, the integral becomes 
\beq \int d^d r d^d s d^d t u(r,s)v(r,t)(w(r,s)+w(r,t)+w(t,r)) =  \eeq
 \beq = U_{\rho} V_{\sigma} (\partial_{\mu} W_{\nu} - \partial_{\nu} W_{\mu}) \int d^d r d^d s d^d t (s^{\rho} - r^{\rho})(t^{\sigma} - r^{\sigma})(s^{\mu}t^{\nu} - s^{\mu} r^{\nu} - r^{\mu} t^{\nu}) \nonumber \eeq
 Upon further expanding all of the parentheses, only the even terms survive integration. Thus, the integral becomes 
 \beq \int d^d r d^d s d^d t u(r,s)v(r,t)(w(r,s)+w(r,t)+w(t,r)) =  \nonumber \eeq
 \beq = U_{\rho} V_{\sigma} (\partial_{\mu} W_{\nu} - \partial_{\nu} W_{\mu}) \Big[ \Big( \int d^d r \Big) \Big( \int d^d s s^{\rho} s^{\mu} \Big) \Big( \int d^d t t^{\sigma} t^{\nu} \Big) + \eeq
 \beq + \Big( \int d^d r r^{\sigma} r^{\nu} \Big) \Big( \int d^d s s^{\rho} s^{\mu} \Big) \Big( \int d^d t \Big) + \Big( \int d^d r r^{\rho} r^{\mu} \Big) \Big( \int d^d s \Big) \Big( \int d^d t t^{\sigma} t^{\nu} \Big) \Big] \nonumber \eeq
 Upon inspection of the above expression, in order for the terms to be even, we need $\rho = \mu$ and $\sigma = \nu$. Upon substitution of these and doing some simple algebra, the integral becomes
 \beq \int d^d r d^d s d^d t u(r,s)v(r,t)(w(r,s)+w(r,t)+w(t,r)) =  \nonumber \eeq
 \beq = 3k_d I_{1d} \tau^{3d+4} \Big( I_{0d} \sum_k (U_0 V_k - U_k V_0) (\partial_0 W_k - \partial_k W_0 ) + \eeq
\beq + I_{1d} \sum_{i,j} U_i V_j (\partial_i W_j - \partial_j W_i) \Big) \nonumber \eeq
Combining the two integrals tells us that the pre-Lagrangian is given by 
\beq {\cal L} (u,v,w,E,p,q) = \frac{k_d I_{d1} \tau^{3d+4}}{2} \Big[ 2 \Big( (1+3E) I_{d0} + \frac{k_d}{4} \Big) (\partial_0 W_k - \partial_k W_0)(U_0 V_k -U_k V_0) + \nonumber \eeq
\beq
+ 3E I_{d1} \sum_{i,j} (U_i V_j - U_j V_i) (\partial_i W_j - \partial_j W_i) \Big]
\eeq
In order to minimize variation of pre-Lagrangian we have to, as usual, adjust $E$ in such a way that non-covariant contribution cancels. This means 
\beq
I_{d0} + \frac{k_d}{4} + 3E_d I_{d0} = -3E_d I_{d1}\;,
\eeq
which gives
\beq
E_d = - \frac{I_{d0} + \frac{k_d}{4}}{3(I_{d0} + I_{d1})}\;.
\eeq
Substituting the above $E_d$ into pre-Lagrangian gives the Lagrangian
\beq
{\cal L} = - \frac{k_d I_{d1}^2 (I_{d0} + \frac{k_d}{4})}{I_{d0}+I_{d1}} (U^{\mu} V^{\nu} - U^{\nu} V^{\mu})(\partial_{\mu} W_{\nu} - \partial_{\nu} W_{\mu})\;.
\eeq
Unlike the bosonic cases, we can not scale $U$, $V$ and $W$ to adjust the coefficient because these fields, being a vierbein, need to have norm 1. Instead, we take advantage of the fact that the above terms are all multiplied by $\overline{\psi} \gamma^{\mu} \psi$ and scale $\psi$. 

In light of the fact that $\psi$ is Grassmannian, while scaling $\psi$, we have to take into account the $\xi$ measure that $\psi$ is subject to. From the case of one Grassmann variable we know that, after the re-scaling
\beq \theta_d = \frac{\theta}{C_d} \eeq
\beq \hat{\theta_d} = \hat{\theta} \eeq
\beq \xi_d (\theta) = C_d^2 \xi (C_d \theta) \eeq
 we still have 
 \beq \int d \theta_d \xi_d (\theta) =1 \eeq
 and 
 \beq \hat{\theta_d} \cdot \hat{\theta_d}=1 \eeq
 This means that we can proceed with re-scaling $\psi$ in the same way as we did for bosonic case. 

However, the situation is complicated by the fact that there is also $\overline{\psi} \gamma^{\alpha} e_{\alpha}^{\mu} \partial_{\mu} \psi$ term, for which re-scaling is very different. This is adressed by using similar idea to type-1 case where total Lagrangian generator is represented as ${\cal J} = ({\cal J}_1, {\cal J}_2, . . . )$ as opposed to ${\cal J} = {\cal J}_1 + {\cal J}_2 + . . . $ and separate these two terms as separate terms in Lagrangian generator. 

In this case, $m_d$ for the fermion will depend on whether we view mass term as independent one or whether we combine it with one of these other terms. Whatever is the case, if $\psi$ is replaced with $\psi_d = \psi / C_d$, then $m^2 \overline{\psi} \psi$ is replaced with $C_d^2 m^2 \overline{\psi_d} \psi_d$. Since, when it comes to mass term, both in scalar case and in fermionic case the field is considered to be a constant, we can copy the answer from fermionic case with appropriate adjustments, which gives us  
 \beq m^2 \overline{\psi} \psi k_d^2 \tau^{2d} \Big(1-\frac{k_d}{2(I_{d0} + I_{d1} (d-1))} \Big) = \eeq
 \beq = C_d^2 m^2 \overline{\psi_d} \psi_d k_d^2 \tau^{2d} \Big(1-\frac{k_d}{2(I_{d0} + I_{d1} (d-1))} \Big)  \eeq
 which implies
 \beq m_d = m k_d \tau^d C_d \sqrt{1-\frac{k_d}{2(I_{d0} + I_{d1} (d-1))}} \eeq
 where $C_d$ can have different values depending on whether mass term is combinded with the derivative terms of $\psi$ or with derivative terms of vierbeins or whether it is viewed as independent from both.  
 
\subsection{Appendix: Geometrical Interpretation of Non-Grassmanian Fermions}
 
In section 5.3 it was mentioned that if fermions were commutting, it would have been possible to replace $6$ of the $8$ fermionic real degrees of freedom with the degrees of freedom associated with vierbeins by pointwise selecting reference frames in such a way that spin-down particle as well as spin-down antiparticle components of fermionic field are $0$ point by point. 
 
What prevented us from doing it was the fact that fermionic field anticommutes while vierbein field commutes. From a different angle, the Grassmannian fermionic field is subject to $\xi$ measure, while vierbeins are not. However, it would be of mathematical interest to explore a toy model in which fermions commute and thus the above mentioned concerns do not apply. This will be the subject of this chapter. 
 Since this toy model does not apply in real life, this chapter is for mathematical interest only and can be skipped as far as the rest of the thesis is concerned. 
 
 If we have an arbitrary spinor at a point, we can always rotate it into a state of the form $\chi_p u_1 + \chi_a v_1$ (here ``p" stands for particle, and ``a" stands for antiparticle). This can be seen by counting degrees of freedom. The rotation group in 4 dimensions has 6 degrees of freedom, while multiplication by an arbitrary complex scalar adds 2 degrees of freedom. This means that if the actions of these two groups were independent, we would obtain a total of 8 real degrees of freedom, which matches the number of real degrees of freedom in a 4-spinor. 

In light of the above, I would like to get rid of the notion of fermion in favor of more ``geometrical" quantities, which are:

\noindent(1) 6 orthonormal vierbeins which define local frame in which spinor has a form $\chi_p u_1 + \chi_a v_1 $, where both $\chi_p$ and $\chi_a$ are real. 

\noindent(2) $\chi_p$ and $\chi_a$ (see above).

\noindent Thus, I will rotate the reference frame from point to point in such a way that the fermionic field is always in the desired form.
 
In order to stress the fact that vierbeins are now viewed as fields, I will replace $e_0^{\mu}$, $e_1^{\mu}$, $e_2^{\mu}$ and $e_3^{\mu}$ by $A^{\mu}$, $B^{\mu}$, $C^{\mu}$ and $D^{\mu}$ respecively, and introduce Lagrange multipliers to assure that 
\bea
& &A^{\mu}A_{\mu}=1\;,\quad B^{\mu}B_{\mu} = C^{\mu}C_{\mu}
= D^{\mu}D_{\mu} = -1\;, \\
& &A^{\mu}B_{\mu} = A^{\mu}C_{\mu} = A^{\mu}D_{\mu}
= B^{\mu}C_{\mu} = B^{\mu}D_{\mu} = C^{\mu}D_{\mu} = 0
\eea
(I am using a metric of signature $(+,-,-,-)$). We will then relax the assumption about rotation of reference frames and go back to the flat Minkowski case. Thus, the final form for my notation for a spinor will be $(A^\mu, B^\mu, C^\mu, D^\mu, \phi, \chi)$.

The geometrical model I propose has also an intuitive appeal: if we take the word ``spin" literally and imagine a particle spinning,  we would need to know the plane in which the particle spins. This gives us two axes, which are described by two vectors, $B^{\mu}$ and $C^{\mu}$. Now, since spin is subject to Lorentz transformations, we also need to know the rest frame of the particle, and this is determined by a timelike vector $A^{\mu}$. As far as $D^{\mu}$ is concerned, due to the orthogonality and unit-norm conditions, it is completely determined by the above 3 vectors. As you will see from the results of the ``Lagrangian" section, things can indeed be visualized in terms of spinning.

The biggest objection one can have is that vector fields and fermions have different transformation properties. However, if one realizes that the transformation properties are completely determined by the Lagrangians and inner products, we can cure the problem by drawing attention towards the latter two, and away from the transformation properties. For example, the implication of spin-$\frac12$ is that a $360^\circ$ rotation in vector space is the same as a $180^\circ$ rotation in spinor space. This problem can be cured by redefining what we mean by a rotation: Instead of simply using $U \mapsto M(\theta)\, U$, where $M(\theta)$ is the usual rotation matrix, we use $U \mapsto \ee^{\ii\theta/2}\, M(\theta)\, U$; by adding a phase, the complex amplitude switches sign upon a rotation by $\theta = 360^\circ$, despite the fact that the vectors are rotated back to their original positions. The reason for this feature is that SU(2) is not the full symmetry group; rather, the full symmetry group is ${\rm SU}(2) \times {\rm U}(1)$. This gives us the freedom of selecting a subgroup $R$ of ${\rm SU}(2) \times {\rm U}(1)$ such that $R \times {\rm U}(1) = {\rm SU}(2) \times {\rm U}(1)$. Any such $R$ can be used as a definition of rotation group, and the freedom of choosing this $R$ corresponds to a freedom in defining the value of the spin: spin-$\frac12\ \times$ spin-0 = spin-1 $\times$ spin-0. Another example: suppose we perform a $180^\circ$ rotation in the space of vectors. In this case, the fact that vectors determine a coordinate system doesn't stop us from {\em defining\/} the inner product between two flipped coordinate systems to be 0 instead of $-1$. After all, we can define the inner product any way we like, so we chose to do it this way. These two features will be implemented in the remainder of the section.

Throughout this section I will use the following representation:
$$\gamma^0 = \left( \begin{matrix}
{\bf1} & \hfill0\, \\ 0 & -{\bf1}\\ \end{matrix} \right),
\qquad
\gamma^k = \left( \begin{matrix}
0 & \sigma^k \\ -\sigma^k & 0 \end{matrix} \right).
$$
where $\bf 1$ is the $2\times2$ unit matrix, $\sigma^k$ for $k=1,2,3$ are the Pauli matrices, and the basis column state vectors will be defined as follows:
$$u_1 = \left( \begin{matrix}
1 \\ 
0 \\ 
0 \\ 
0 \end{matrix} \right)\;,\qquad
u_2 = \left( \begin{matrix}
0 \\ 
1 \\ 
0 \\ 
0 \end{matrix} \right)\;,\qquad
v_1 = \left( \begin{matrix} 
0 \\ 
0 \\ 
1 \\ 
0 \end{matrix} \right)\;,\qquad
v_2 = \left( \begin{matrix}
0 \\ 
0 \\ 
0 \\ 
1 \end{matrix} \right)\;.$$
Even though in this section we are only dealing with a toy model in which there are no Grassmann numbers, we are still free to get rid of $\chi_p^2$ and $\chi_a^2$ terms of the Lagrangian. This means that as far as spin connection terms are concerned, we are looking only at $\chi_p \chi_a$ terms. Based on the fact that spinors take the above form, it is apparent that the only term of $\omega_{mab} \overline{\psi} \gamma^m \sigma^{ab} \psi = \omega_{mab} \psi^{\dagger} \gamma^0 \gamma^m \sigma^{ab} \psi$ that survives is the one where $\gamma^0 \gamma^m \sigma^{ab}$ is off-diagonal matrix in the $2 \times 2$ block representation. This will happen only if m, a and b are all non-zero, which identifies them as $1$, $2$ and $3$ up to permutations, which means they are all proportioanl to $\psi^{\dagger} \gamma^5 \psi =2 \chi_p \chi_a$. 
As far as derivative terms, we do have to keep $\psi_p \partial \psi_p$ and $\psi_a \partial \psi_a$ terms as well as we still have to keep the ``mixed" ones. This means that we are looking both at the diagonal and off diagonal matrices in block diagram. However, since there are no spin down components of either particle or antiparticle, each block needs to be diagonal. The matrices that satisfy these constraints are $\gamma^0$ and $\gamma^3$ . $\gamma^0$ will give us $\overline{\psi} \gamma^0 e^{0 \mu} \partial_{\mu} \psi = e^{0 \mu} (\chi_p \partial_{\mu} \chi_p + \chi_a \partial_{mu} \chi_a)$ and $\gamma^3$ will give us $\overline{\psi} \gamma^3 e^{3 \mu} \partial_{\mu} \psi = e^{3 \mu} (\chi_p \partial_{\mu} \chi_p - \chi_a \partial_{mu} \chi_a)$ Thus, the Lagrangian becomes 
\bea & & {\cal L}_{free}= k \psi^{\dagger} \gamma^5 \psi (\omega^1_{23}-\omega^2_{13}+\omega^3_{12}) + \overline{\psi} e^{0 \mu} \gamma_0 \partial_{\mu}  \psi +  \overline{\psi} e^{3 \mu }\gamma_3 \partial_{\mu}  \psi \; \\ \nonumber
& &  = 2k \chi_p \chi_a (\omega^1_{23}-\omega^2_{13}+\omega^3_{12}) + e^{0 \mu} (\chi_p \partial_{\mu} \chi_p + \chi_a \partial_{\mu} \chi_a)+ e^{3 \mu} (\chi_p \partial_{\mu} \chi_a - \chi_a \partial_{\mu} \chi_p) \eea
Finally, in order to stress the fact that vierbeins are viewed as fields, we will replace $e^{0\mu}$ through $e^{3 \mu}$ with $A^{\mu}$ through $D^{\mu}$ respectively, and introduce Lagrange multipliers to enforce orthonormality. We will also replace $\omega^1_{23}$ with $\omega^B_{CD}$ and do similarly with all the other indeces. These $\omega$-s are now functions of our vector fields that are defined based on formal substitution of these in place of Vierbeins without making an assumption of orthonormality, since the latter is only a consequence of Lagrange multipliers.  Thus, Lagrangian becomes 
\bea & & {\cal L}_{free}=2k \chi_p \chi_a (\omega^B_{CD}-\omega^C_{BD}+\omega^D_{BC}) + A^{\mu} (\chi_p \partial_{\mu} \chi_p + \chi_a \partial_{\mu} \chi_a)+ D^{ \mu} (\chi_p \partial_{\mu} \chi_a - \chi_a \partial_{\mu} \chi_p) \; \nonumber \\
&  & +\ \lambda_1\,(A^\mu A_\mu-1) + \lambda_2\,(B^\mu B_\mu+1)
+ \lambda_3\,(C^\mu C_\mu+1) + \lambda_4\, (D^\mu D_\mu+1) \nonumber\\
& &\kern25pt+\ \lambda_5\,A^\mu B_\mu + \lambda_6\,A^\mu C_\mu
+ \lambda_7\, A^\mu D_\mu + \lambda_8\, B^\mu C_\mu
+ \lambda_9\, B^\mu D_\mu + \lambda_{10}\, C^\mu D_\mu \;,
\eea
where
\bea \omega^U_{VW} = U^{\mu} V^{\nu} (\partial_{\mu} W_{\nu} - \partial_{\nu} W_{\mu}) + V^{\rho} W^{\sigma} \partial_{\sigma} U_{\rho} \eea
Now I would like to introduce interaction terms into Lagrangian. Since it is possible that we have interaction of more than one fermion, I would like to be able to define $\overline{\xi} \psi$ and $A_{\mu} \overline{\xi} \gamma^{\mu} \psi$.  In general, this means I would like to define $\overline{\xi} \Lambda \xi$ . Suppose vierbeins that are needed to put $\xi$ in the form $\chi_p u_1 + \chi_a v_1$ are $e_0^{\mu}=A^{\mu}$ through $e_3^{\mu}=D^{\mu}$ while vierbeins that are needed to put $\psi$ in the form $\eta_p u_1 + \eta_a  v_1$ are $f_0^{\mu}=E^{\mu}$ through $f_3^{\mu}=H^{\mu}$

Now, suppose the transformation from the $e$-basis to $f$-basis,  $e^{-1} f$ , lies in the connected component of identity matrix. In other words, they are either both forward-moving or both backward-moving. In either case, they are both forward-moving relative to each other. This means that we can write $e^{-1} f = \exp(\ln (e^{-1} f))$.  Thus, $\ln (e^{-1} f)$ can be viewed as generated by infinitesimal transformations. The infinitesimal spinor transformation that corresponds to $\ln (e^{-1} f)$ is $-\frac{\ii}{4} (\ln (e^{-1}f))_{\mu \nu}\, \sigma^{\mu \nu}$, where $\sigma^{\mu \nu} = \frac{\ii}{2} [ \gamma^{\mu}, \gamma^{\nu} ]$. Now, by exponentiating it back, we will get the finite spinor transformation corresponding to the transformation between these two coordinate systems:  $\exp\{-\frac{\ii}{4} (\ln (e^{-1} f))_{\mu \nu}\, \sigma^{\mu\nu}\}$. Thus, 
\beq \overline{\xi} \Lambda \psi =  (\chi_p \langle u_1 \vert + \chi_a \langle v_1 \vert)  \Lambda \exp\{-\frac{\ii}{4} (\ln (e^{-1} f))_{\mu \nu}\, \sigma^{\mu\nu}\} (\eta_p \vert u_1 \rangle + \eta_a \vert v_1 \rangle ) \eeq
Now in the case where one reference frame is forward-moving and the other one is backward-moving, all we have to do is insert a time-reversal operator inside the log, namely $\exp\{-\frac{\ii}{4} (\ln(T L^{-1}M))_{\mu\nu} \sigma^{\mu\nu}\}$, where $T$ is time reversal. This gives us
\beq \overline{\xi} \Lambda \psi =  (\chi_p \langle u_1 \vert + \chi_a \langle v_1 \vert)  \Lambda T \exp\{-\frac{\ii}{4} (\ln (e^{-1} f))_{\mu \nu}\, \sigma^{\mu\nu}\} (\eta_p \vert u_1 \rangle + \eta_a \vert v_1 \rangle ) \eeq
From the point of view of mathematical rigor, the equation we just got is as far as we can get. However, it would be fruitful to note that there is an intuitive correlation between that equation and the concept of ``spinning", which the word ``spin" represents. This discussion is not rigorous, and can be skipped by mathematically minded readers.

In order to visualize the ``spinning" that goes on, one can replace infinitesimal points in spacetime by small arrows. However, unlike the way it is normally done, these arrows will {\em not\/} be aligned with a spin axis. Instead, these arrows will be, themselves, spinning around some other axis. Thus, spin around the $z$ axis can be visualized as a vector pointing in the $x$ direction whose end is moving in the $y$ direction. Now, due to the fact that spin is subject to Lorentz transformations, we also need to know the reference frame, and it is given by $A^{\mu}$. And finally $D^{\mu}$ is a cross product of $B^{\mu}$ with $C^{\mu}$ in a reference frame in which the particle is at rest. Thus $D^{\mu}$ is what is usually thought of as the direction of spin.

Now let's look at each term in the Lagrangian to see what it represents. First, consider the $C^{\mu} A^{\nu} \partial_{\nu} B_{\mu}$ term. In the reference frame of the point around which the arrow spins, $A^{\nu}=\delta^{\nu}{}_0$ is just a vector pointing along the $t$ axis. Thus, in this reference frame, the term becomes $C^{\mu} \partial_0 B_{\mu}$. Now, if we visualize $B^{\mu}$ as pointing along the $x$ axis and $C^{\mu}$ along the $y$ axis, this expression reads off as ``how fast does the end of the $x$ axis move in the $y$ direction"? Note that the end of the $x$ axis can also move in the $z$ direction, but this speed would simply have no contribution to the Lagrangian. In other words, the way to think of it is this: $A^{\mu}$, $B^{\mu}$ and $C^{\mu}$ are vector fields with a weird coupling between them. They are coupled to each other in such a way that the end of vector $B^{\mu}$ is ``forced" to move in the direction of the vector $C^{\mu}$ by the Lagrangian, which would then be interpreted by the observer as a rotation around $D^{\mu}$.

Now let's look at the $A^{\mu} B^{\nu} \partial_{\nu} C_{\mu}$ and $A^{\mu} B^{\nu} \partial_{\nu} C_{\mu}$ terms. One difference between the first term and these two terms is that in the ``directional derivative" part of the equation (which in the first term is $A^{\nu} \partial_{\nu}$) $A^{\nu}$ is being replaced by $B^{\nu}$ and $C^{\nu}$, respectively.  This means that the differentiation is no longer in the time direction, but rather in a spatial direction. The interpretation of this might be that, as opposed to speaking of the rotation of one arrow (in which case it travels along the $y$ direction) we are comparing the motions of different arrows (and the spatial direction differentiates the arrows we are looking at). The other difference is that while the first term refers to angular motion, both of the other two terms refer to linear motion: $A^{\mu} B^{\nu} \partial_\nu C_\mu$ refers to ``boosting" of the spatial axis $C_{\mu}$ in the time direction $A^\mu$ (which would be proportional to the {\em negative\/} velocity in the $C_\mu$ direction), while $B^\mu C^{\nu} \partial_\nu A_\mu$ refers to ``boosting" of the time axis $A_\mu$ in the spatial direction $B^\mu$ (which can be interpreted as a {\em positive\/} velocity in the $B_\mu$ direction). Thus, by identifying $B^{\mu}$ with the $x$ axis and $C^\mu$ with the $y$ axis, the Lagrangian tells us that we ``want" points that are located away along the $x$ direction to move in the negative $y$ direction, and we want points located away along the $y$ direction to move in the positive $x$ direction. This is equivalent to saying that we want to have a ``planetary system" and we want the orbits of the planets to have spin in the $-z$ direction.

This can be summarized as follows: we can envision space to be constructed of mini-atoms.  The first term in our Lagrangian tells us about the spin of each electron in an atom, while the last two terms tell us about the orbital rotation of electrons. However, these atoms are ``glued together" so that the orbital rotation of one atom gets ``passed" onto neighboring ones, which is why this actually looks like a derivative globally. At the same time, while there is ``cohesion" between atoms in the second two terms, the first term has no such thing: the arrow is infinitesimal and doesn't extend to a neighboring point. Similarly, when we talk about linear motion in the last two terms, this linear motion is really a similar arrow pointing in the $t$ direction, which is also infinitesimal. Thus, while there {\em is\/} spatial cohesion in the second two terms, there is no time cohesion.

\newpage                                              
\section{Chapter 6: Conclusion}

A causal set is a locally finite set of points, the only reliable structure for which are causal relations. The assumptions normally made for ordinary physics, such as manifoldlike-ness and symmetries, no longer apply to causal sets. This allows the causal set to have more general, non-manifoldlike, structure, while the manifoldlike structure arising on larger scale is a possible result of more general, non-manifold-based, theory.

Causal set theory appeals to observation made by Hawking and Malevent that a manifold structure can be completely identified based on the scaling together with the causal relations between its events. In discrete case, the scaling is identified as simple count of points, thus causal relations are viewed as the only defining factor for gravitational field. Strictly speaking, causal relations are defined as a partial order.

Scalar field on a causal set are defined as one point real or complex valued functions; Vector fields on a causal set are defined as a two-point real valued function. Grassmann numbers are defined as an individual elements of a vector space equipped with commuting dot-product and anticommuting wedge-product, and spinor field is defined as a combination of vector field, corresponding to vierbeins, and Grassmann-valued one point functions. 

Two classes of fields are postulated: type 1 and type 2. For every type-1 field, a Lagrangain generator is defined as a function of the behavior of that field together with a choice of few ``representative" points. It turns out that if Alexandrov set is selected in such a way that would minimize the fluctuations of Lagrangian generator for the choices of points varying within its interior, this would result in Lorentz covariant expression. 

For type-2 fields, on the other hand, Lagrangian corresponding to a given Lagrangian generator is a linear combination of two different integrals of that generator; the difference in the integrals is due to the difference in the choice of parameters integration is performed. Arbitrary constant is introduced as a multiplying factor of one of these integrals, and minimization is performed mainly with respect to that constant; the minimization with respect to the choices of Alexandrov set is unimportant in linear cases, but it is retained to handle non-linear situations and specifically to adress lightcone singularity that might arise. 

Whether the field is type 1 or type 2, the final expression can be thought of as a relativistic covariant generalization of non-covariant expression used for Lagrangain generator. If Lagrangian generator is properly chosen, this will typically produce Lagrangians we are used to. Since in standard physics Lagrangians, too, are introduced as relativistic generalizations of non-covariant expressions, what was done for causal sets is simply a non-manifold equivalent of what we already used to. 

Relavitistic covariance is obvious from the construction: the only topoogy used is causal relations which is inherently covariant. Thus, it is shown that if any field is viewed as type 1, we do end up with covariant expression. The purpose of introducing type-2 fields is mainly to handle unwanted terms that might arise in type-1 cases, although these terms are still very much covariant. 

For example, in the case of a type-1 scalar field, $\partial^{\mu} \phi \partial_{\mu} \phi$ possibly appears with different coefficients, depending whether it is positive or negative at a given point. For a type-1 gauge field, an unwanted term $\epsilon_{\alpha \beta \gamma \delta} F^{\alpha \beta} F^{\gamma \delta}$ came along, and for type-1 gravity, unwanted contractions, such as $R^{\alpha \beta} R_{\alpha \beta}$ appear. 
 
 However, in each of the fields except for gravity, this is dealt with by a simple addition or subtractions of different terms, which makes gravity a prime reason for introducing the notion of type-2 fields. However, once the notion of type-2 fields exists anyway, we might as well apply it to other, non-gravitational, fields. The prime motivation for doing the latter is to be able to postulate SU($n$) symmetry for gauge fields.
 
For type-1 electroweak field, we are forced to view electromagnetic and weak interactions as completely separate, and the appearance of symmetry is only a result, rather than a cause, of similarity of Lagrangians. To make it worse, symmetry is only approximate rather than exact in discrete cases. On the other hand, if electroweak field is viewed as type 2, the symmetry can be formally postulated on a fundamental level and, consequently, it is exact. 

Finally, in an Appendix that follows, a model of relativistically-covariant collapse of the behaviors of the fields, including gravity, was introduced. It is assumed that the existing, non-fluctuating, fields co-exist with the other versions of these same fields fluctuating within a ``corridor" around the non-fluctuation configurations. Thus, the probability of each non-fluctuating configuration is simply a path integral of the fields fluctuating inside of the corresponding corridor. While this model has an advantage compared to decoherence in a sense that it is manifestly covariant, it was argued that in non-relativistic case it reproduces the key predictions of decoherence model, including entropy being a defining feature of measuring apparatus. 

That model allows one to ``collapse" causal relations into a specific configuration, despite the fact that they, being identified with gravitational field, undergo quantum fluctuations. This allows for the topological background necessary to be able to introduce propagators. However, while this adresses the issue of topological background, it does not adress the issue of it being manifold-like. It turns out that a separate Lagrangian generator needs to be introduced for that purpose alone (see sec 5.8). There is a qualitative argument that causal set, subject to Lagrangian generator of sec. 5.8 is more manifold-like than it would have been otherwise, but it is not clear whether or not it is sufficient to make it manifold-like. This is a subject of future research.

As we have seen in this dissertation, a fundamentally different way of viewing causal sets was presented. Instead of statistically generating it as is traditionally done, it was claimed that a set of Lagrangians should be a starting point of generating a causal set. This amounts to saying that geometry is no longer a background of Lagrangians but rather the set of Lagrangians is a cause of geometry. This is in line with a traditional picture of general relativity that metric is a gravitational field which is just as much of a field as any other field. 

However, the generating Lagrangians that were used, especially the fermionic one, might look ``manufactured" and one might have a question as to why not use some other Lagrangians? There are two answers to this argument. The first answer is that the same question can be asked for standard theory as well. For example, in case of Einstein-Hilbert action, we can ask why do we have $R$ instead of $R^2$ in the Lagrangian. 

The fermionic Lagrangian serves even better example to illustrate the point. While it is true that the Dirac equation was originally invented as a square root of Klein-Gordon equation, this is not true for the Lagrangians. So, in a desperate attempt for something to ``look" like a square root on a classical limit, they had to cleverly ``manufacture" a Lagrangian that would produce that. So, if they are allowed to do that, why can't we? In particular, we might as well be even more clever and manufacture a Lagrangian generator for a fermionic field, so that, after a few-step process of first getting actual Lagrangian, then applying it to a manifold, and finally taking classical limit, we will get ``square root" at the end of the day. 

Of course, however, the argument ``if everyone can do it badly so can I" is not a good one. Especially if I am trying to come up with a fundamental theory. Thus, this is not an official standpoint of the theory, and I consider the issue to still be unresolved. The only reason for the above argument is to explain why my theory ``has a right to live". But the fact that the theory has a right to live does not imply that all of its problems are resolved. In fact, to this day, none of the theories can make that claim. 

Another argument in defense of that theory is that, in my opinion, the above weakness is not as serious as the foundational problems of physics the theory proposes to solve. After all, if we were offered an explanation of all of our experiments in terms of Newton-based classical physics alone, few would object, even if it meant that we were to introduce some complicated-looking forces that were not originally present in Newtonian physics.

Of course, the response to this argument is that this theory, as it is, does not claim to ``explain all experiments on modern physics". On the contrary, it's only hope is to reproduce some of the most basic things that are otherwise taken for granted: Lagrangians as well as basic geometry. Even then, the manifold-like structure that is being predicted in section 5.2 is approximate at best. However, in light of the fact that this area of physics is very young, it is hoped that situation will change in future when more research is done. In this light, this dissertation should be viewed as a necessary starting point for that future work.

It should also be understood that there are technical obstacles in developping this theory to the point of making predictions. In case of square lattice, it is possible to systematically perform path integration over arbitrary many degrees of freedom since the entire information can be captured in the definition of a lattice. In case of causal set, in light of the randomness of causal structure, this is no longer true. Thus, if we were to perform path integral, we would have to do so numerically rather than analytically and we have to separately count each degree of freedom. 

In my visit to Perimeter Institute, Sorkin told me that if my causal set consist of 100 -- 1000 points, it would take few years for the modern computer to finish calculations with the Lagrangains that I proposed. However, my hope is to come closer to be able to replace the numeric methods with analytic ones. Of course, actually doing that is simply unrealistic. But it is possible to do something in the spirit of perturbation theory: assume that somehow we know the value of $Z(\prec_0, {\cal F}_0)$ and try to calculate $Z(\prec, {\cal F})$ provided that ${\cal F}$ is sufficiently close to ${\cal F}_0$ and $\prec$ is sufficiently close to $\prec_0$.

On the other extreme we can do something else in parallel: instead of using small-number-simplifications characteristic of perturbation theory, I can use large-number-simplifications characteristic of classical limit. This would predict behavior of classical objects which would serve as a potential test of the theory. Furthermore, by applying it to black holes, we can have a conceptual framework to try to tackle questions such as information paradox.

It should also be remembered that even classical Einstein's equation is only solvable analytically for the most simple situations, such as spherical mass distribution. This, however, does not disqualify classical general relativity from being a valid theory. So, even if I would not be able to make any predictions beyond black holes, I can claim that black holes to causal set theory is the same as spherical mass distribution to general relativity. This might be the ultimate defense of causal set theory against such objections.

\newpage
\appendix
\section{Appendix: Interpretation of Quantum Mechanics on Causal Sets}

\subsection{Quantum Gravity and the Need for an Interpretation of Quantum Mechanics}

By viewing causal relations as a quantum field, one is forced to admit that causal relations undergo quantum fluctuations. If the concept of quantum fluctuations is left unaltered, causal relations would be forced to undergo all possible structures, which means that there is no appriori causal relation. This leads to absurd situation: without any apriori topology, any pair of points is apriori just as close to or just as far from each other as any other pair of points, which means that the propagators between any pair of points should be identical!

The solution to this problem can be accomplished as a special case of a solution to a more general problem, the one of interpretation of quantum mechanics.  Essentially, a successful solution to the latter is a theory that ``localizes" fluctuating quantities. Thus, if successfully done, its application to gravitational field will provide the necessary background topology to introduce propagators.

Before proceeding, it is important to ask a question whether or not it is possible to ``temporary ignore" the problem of interpritation of quantum mechanics the way it is done in standard non-gravitational quantum field theory. In flat space quantum field theory, ``sources" and ``sinks" designate the presence of classical objects that cause the collapse of wave function. These classical objects are localized in spacetime, which means that there is some unknown mechanism to allow such localization to occur. 

It is possible to deliberately overlook the question of localization of sources/sinks by simply saying, for now, that all the sources/sinks are color blue while all particles are color red, and only color red is fluctuating. Similar thing can be tried for spacetime by picturing that spacetime, being color blue, is non-fluctuating. But then it is no longer possible to derive Einstein's equation from variation of the action, since the latter is nothing but a classical limit of path integration, although one can attempt to go around this issue by simply imposing approximate validity of Einstein's equation as a constraint (this is done in the last section of the Appendix) 

This, however, in itself will be ``more" than the ``formal" prescription would give. Thus, since we have to do that anyway, nothing stops us from saying that we ``don't like" imposing approximate validity of Einstein's equation by hand and, instead, we ``like" a lot more the idea of interpretation of quantum mechanics way of ``localizing" the gravitational field. This will be the mindset of most of the chapter.

\subsection{Quantum Corridors}

The standard way of interpritation of quantum mechanics is the one of decoherence. However, this model is not acceptable for two different reasons. First of all, it viewes parallel universes as non-interfering components of a single wave function of one, large, universe. As such, they will interact with each other gravitationally, which means that they will not be trully parallel once gravity is introduced. Secondly, the definition of wave function of a universe is  based on the concept of ``simultaneous events" which violates relativity. 

For this reason, in this section a new model of interpritation of quantum mechanics will be introduced that avoids the above difficulties. Its comparison with decoherence model will be discussed in more detail in the next section. The idea is to replace the cross graining in spacelike hypersurface which violates relativity with coarse graining in the set of spacetime histories which is manifestly covariant. 

The idea of doing that have been around for a while. For example, it was mentioned by Mensky in Chapters 5 and 6 of Ref \cite{decoherence} as well as by hartle in Chapter 8 of Ref \cite{hartle}. Since I didn't have time to study these references in detail, I am not sure regarding the extend to which my approach differs from theirs and whether these differences are good or bad. Nevertheless, I chose to follow more closely Mensky than Hartle just because I understand it better, and I will borrow his terminology of ``quantum corridors" and ``quantum tunnels". 

Since my reading of Mensky was relatively superficial, it is not clear whether he was implying the same procedure as I was carrying out here. If his corridor was referring to a size of a lab or some other variable that depends on specific circumstances, I would disagree; on the other hand, if his corridor has a fixed size, given by a constant of nature, it would be exactly what I am trying to do here. 

It is also unclear just how far from this approach Hartle stands. The ``words" part of his Chapter 8 seem to match closely with what I propose in this chapter and also the way I would paraphrase Mensky, although mathematics part is rather different. On the surface reading, his mathematics is a lot more complicated than mine. Potentially, this might mean that my approach is too simplistic and in need of further elaboration. Once again, this is something I am leaving for further research. 

Let's get down to business. One can argue that when quantum system is being measured, it is not literally being seen; rather, an observer sees an arrow on the measuring apparatus pointing in a certain direction. That arrow is a classical object. Thus, the ultimate goal of the theory is simply a prediction of behavior of classical objects, and nothing more. 

It is proposed that the entire classical history of the universe -- both distant past, distant future, and everything in between -- is a single outcome of a single measurement performed by a single observer living outside the space-time "system" (or a causal set if you will). This is a "quantum field theory" version of measurement as opposed to "quantum mechanics" one. That is, fields are measured point by point. 

The degree of precision of the measurement is identical at each point of the universe, and has nothing to do with any labs that might and might not be present there. After all, lab itself is still part of a quantum system, which means that it is an outcome of the measurement as opposed to a setup. 

As far as setup is concerned, no spacetime event is apriori different from any other one, thus it will only make sense that the degree of measurement of the fields at each point is identical. That degree of precision can be viewed as fundamental constant of nature, just as fundamental as speed of light or charge of electron. 

A classical behavior of a system can be viewed as a collection of all possible quantum mechanical behaviors that are ``compatible" with it, which will be denoted as ${\cal C}$. The definition of "compatible" is equivalent to the degree of the above described universal measurement. The probability amplitude associated with ${\cal C}$ is given by 
\beq
Z \big( {\cal F} \in {\cal C} \big) = \int_{\cal C} {\cal DF}\,
\ee^{\ii\, S({\cal F} )}\;.
\eeq
Now, the only ``acceptable" choices of ${\cal C}$ are the ones for which any two of its elements are ``approximately equal" to each other, with respect to approximations done up to smallest ``classical" scales. This, of course, required definition of approximation. 

In case of a scalar field, $\phi_1 \approx \phi_2$ means that $\phi_1 \approx \phi_2$ as long as $\vert \phi_1 (x) - \phi_2 (x) \vert < \epsilon$ for every single value of $x$. It is then easy to see that there is one to one correspondence that associates  every choice of ${\cal C}$ with the average value of $\phi \in {\cal C}$. 

Thus, a range of fields $\phi$ can be replaced with a single, localized, field $\phi$, and the probability density associated with each of the versions of such field is given by  
\beq
Z \big( \phi (x) = \phi_0 (x) \big) = \int_{\vert \phi(x) - \phi_0 (x) \vert
< \epsilon} \ee^{\ii\,S(\phi_0)}\;,
\eeq
where the integration will be replaced by a sum in a causal set. 

In the above expression, in light of the fact that the range is constrained to a certain width, there is one to one correspondence between a field $\phi_0$ and the range of variations of $\phi$. Thus, the above will define the probability of fixed values of $\phi_0$ as opposed to the one of a range of values of $\phi$. The issues of consistency with decoherence theory will be adressed in the next section. 

There is another issue: different nearby fields $\phi_0$ might resemble each other so well that the same probability can be counted many times which would skew the results. This can be adressed by replacing the definition of probability with  rigid yes-or-no criteria of whether a given history is ``allowed". That criteria is coupled with a principle that there exist exactly one parallel universe to ``realize" each of the ``allowed" histories. The goal is for that to imply that the probability of a system falling into a history that looks a certain way is consistent with the probability predicted by the above equation. This can be done by postulating the following constraint:

CONSTRAINT 1: A field $\phi_0$ is allowed if and only if $\vert Z(\phi(x)= \phi_0 (x)) \vert^2 > f(\vert Z(\phi(x)= \phi_0 (x)) \vert^2)$ where $f(x)$ is a decimal expression of $x$ starting from the 100-th digit. 

In above the constraint, the function $f$ was used as random number generator, and the probability of $x > f(x)$ is proportional to $x$ as long as $x$ is between $0$ and $1$

In case of causal set, the value of $\phi_0$ is discretized in order to remove potential conceptual difficulties by making the number of parallel universes finite. This amounts to imposing another constraint:

\noindent CONSTRAINT 2: for any point $p$, $\phi_0(p)$ is a factor of some small number $\delta$

For vector fields similar trick is done:
\beq
Z \big( v(p,q) = v_0 (p,q) \big) = \int_{\vert v(p,q) - v_0 (p,q) \vert
< \epsilon v_0 (p,q)} \ee^{\ii\, S(v(p,q))}\;.
\eeq
\noindent CONSTRAINT 1: A field $v(p,q)$ is allowed if and only if
$$
\vert Z(v(p,q)=v_0(p,q)) \vert^2 > f(\vert Z(v(p,q)= v_0(p,q)) \vert^2)\;,
$$
where $f(x)$ is the decimal expression of $x$ starting from the 100-th digit.

\noindent CONSTRAINT 2: for any point $p$ and $q$, $v(p,q)$ is a factor of some small number $\delta$.

Now the key to localizing geometry is to do similar trick to gravitational field. This requires a definition of a corridor on a space of causal relations (i.e. gravitational fields).

A corridor around causal relation $\prec$ is a set of all causal relations $\prec^*$ such that the respective gravitational fields defined in terps of $\prec$ and $\prec^*$ approximate each other. This is defined as ``neighborhood" of $\prec$, and denoted as $n (\prec, \epsilon, N)$  (where $N$ is the a low bound on a distance scale imposed in order to avoid unwanted discrete effects) and $G(T, \prec)$ is replaced with $H(T, \prec, \epsilon, N)$ where $H(T, \prec, \epsilon, N) = G(T, \prec) \cap n(\prec, \epsilon, N)$ 

In light of the fact that classical Einstein equation can be derived by means of variation of an action with respect to $g_{\mu \nu}$, ideally one would like to define $n(\prec)$ in terms of small variation of the same. However, imposing separate restrictions for each of the choice of $\mu$ and $\nu$ is not a relativistically covariant procedure, which means it is not well defined for causal set. However, due to the fact that length of geodesic depends linearly on $g_{\mu \nu}$ this becomes an easy replacement. Timelike geodesic is defined much simpler than spacelike geodesic is, so the former will be used. It has to be taken into account that, due to the fact that since $\prec$ and $\prec^*$ are distinct, some pairs of points are related by one causal relation and not by the other, which means that only one of the two geodesic segments in question is present. The way to deal with this case is to notice that if $g_{\mu \nu} \approx g^*_{\mu \nu}$ then whenever the Lorentzian distances according to these two metrics differ in sign, they should both be close to $0$. Thus, a constraint to impose is that if $p \prec q$ AND their distance is large according to $\prec$, then $p \prec^* q$. The constraint regarding comparable geodesic length should also be imposed. This, too, requires the scenario where two points have large enough Lorentzian distance according to relevent partial orders, as this would reduce stochastic fluctuations in manifold-like scenario. These two constraints can be combined into one: if $p \prec q$ and they are far enough from each other according to $\prec$, then $p \prec^* q$ and their distance according to $\prec^*$ will be comparable to the one according to $\prec$ :

\noindent DEFINITION: Let $\prec$ be a partial order on a causal set $S$. Let $\epsilon$ be some small real number and let $N$ be an integer. Let $\prec^*$ be some other partial ordering on $S$. Then $\prec^*$ is an element of $n_{\tau}( \prec, \epsilon, N)$ if for any points $p \prec r_1 \prec . . . \prec r_N \prec q$ there are points $s_1, . . . , s_M$ satisfying $p \prec s_1 \prec . . . \prec s_M \prec q$ where $M \geq (1- \epsilon) N$. Likewise, for any points $p \prec^* r_1 \prec^* . . . \prec^* r_N \prec q$ there are points $s_1, . . . , s_M$ satisfying $p \prec s_1 \prec . . . \prec s_M \prec q$, where $M \geq (1- \epsilon) N$.

The meaning of the subscript $\tau$ in $n_{\tau}$ is that $n$ is defined in terms of Lorentzian distances, $\tau$. It is also possible to define $n$ in terms of volumes of Alexandrov set: 

\noindent DEFINITION: Let $\prec$ be a partial order on a causal set $S$. Let $\epsilon$ be some small real number and let $N$ be an integer. Let $\prec^*$ be some other partial ordering on $S$. Then $\prec^*$ is an element of $n_{V}( \prec, \epsilon, N)$ if whenever there are more than $N$ choices of $r$ satisfying $p \prec r \prec q$, there are also more than $(1-\frac{\epsilon}{4})N$ choices of $s$ satisfying $p \prec^* s \prec^* q$. Likewise, if there are more than $N$ choices of $r$ satisfying $p \prec^* r \prec q$, there are also have more than $(1-\frac{\epsilon}{4})N$ choices of $s$ satisfying $p \prec s \prec q$.  

The reason $\frac{\epsilon}{4}$ was used in the above definition instead of $\epsilon$ is to make sure that geodesic length, as opposed to the volume, deviates by order of $\epsilon$ since the former rather than the latter linearly depends on the metric. Similar conversion was not done for $N$ since the latter is just intended to put some low bound on the number of points to avoid discreteness effects, which means that it doesn't make a physical difference whether it refers to lengths or volumes.  

Of course, this still doesn't adress the renormalization issue. But since causal set theory can be done numerically, the theory is still well defined.

\subsection{Quantum Corridors vs Decoherence: The Non-Gravitational Case}

In order for previous section to stand, it has to be argued that interpritation of quantum mechanics proposed in the previous section is consistent with more standard decoherence model. Of course, there is no exact match between the models -- for one thing, decoherence model is non-relativistic while tunnel one is. However, it is still possible to show that the model at hand reproduces key predictions of decoherence model -- in particular the fact that the defining difference between ``measuring apparatus" and other systems is large entropy of the former. Since decoherence theory is non-relativistic, it is suffice to show that the two theories agree in non-relativistic case, which is what most of the argument will be devoted to. Meanwhile, it will be apparent that tunnel theory is written in relativistic form, hence it is Lorentz invariant. 

According to the decoherence theory, the defining feature of quantum measurement lies in a very large number of degrees of freedom of measuring device. Take, for example, localization of a particle on the screen. The complete system, consisting with incident particle as well as all the particles that compose a screen, can be represented as a point in a phase space. If there are total of $n$ particles, the phase space has $3n$ dimensions, which correspond to space dimensions of each of the particles. This system evolves in non-relativistic time $t$ according to Schr\"odinger's equation, with evolution equation $v(t)$.

The screen can be viewed as a collection of many different small pieces glued together, numbered $1$ through $n$. After the interaction of a particle with a screen occurs, the resulting state becomes
\beq
\vert v(t) \rangle = \sum a_i \vert v_i (t) \rangle\;.
\eeq
where a state $\vert v_i (t) \rangle$ is a component of a wave function corresponding to the possibility of interaction of a particle with a piece number $i$. 

In case of absolutely smooth screen,  $\langle v_i (t) \vert v_j (t) \rangle \neq 0$ as long as $t \neq t_0$, where $t_0$ is the time that a particle collapses on the screen. However, in real situation the screen is NOT totally smooth; rather, it is a complex system involving a lot of different particles. Thus, in the real life
\beq
\vert v_i (t) \rangle = \sum_j c_{ij} \vert v_{ij} (t) \rangle\;.
\eeq
On the one hand, it is still true that 
\beq
\langle v_{ij} \vert v_{kl} \rangle \neq 0\;.
\eeq
However, due to the averaging up of all the different degrees of freedom, 
\beq
\langle v_i (t) \vert v_j (t) \rangle = \sum_{k,l} c^*_{ik} c_{jl}\,
\langle v_{ik} (t) \vert v_{jl}  (t) \rangle \approx 0\;.
\eeq  
Physically this means that there is no well defined phase shift between the component of wave function reflected from pieces $i$ and $j$ due to the fact that there is too much variation of phases of Fourier components of signal emitted from each of these two pieces. Consequently, all interference terms cancel, which means that the law of addition of probability amplitudes become equivalent to the law of addition of probabilities themselves. Mathematically, this is expressed as 
\beq
\langle v_i (t) \vert v_j (t) =0 \Longleftrightarrow \langle v_i(t) + v_j (t) \vert v_i (t) + v_j (t) \rangle = \langle v_i (t) \vert v_i (t) \rangle  +\langle v_j (t) \vert v_j (t) \rangle\;.
\eeq
Since both $\vert v \rangle$ and  $\vert v_i \rangle$ correspond to functions in a phase space rather than a position space, they take into account the behavior of all existing particles put together, which means that there is nothing ``outside" of these functions to interact with them. What is normally viewed as interaction now becomes an evolution of free state. In light of this, the fact that $\vert v_i (t) \rangle$ and $\vert v_j (t) \rangle$ stay non-overlapping means that they represent imaginary quantum systems evolving completely independently of each other, although sharing the same space. In other words, they act like ``parallel universes".

On the first glance it might seem counterintuitive that the two peaks would never overlap with each other in the future. After all, two delta functions evolving according to Schrodinger's equation are predicted to overlap. In order to answer this, we have to remember that above argument was not made in real space, but in imaginary phase-space. In order to get a physical intuition of what actually happens, we have to "translate" that into a real space language. 

If we have a classical object, such as billiard ball, it has a basic trajectory plus the internal distribution of its molecules. Its representation in position space takes into account the former and ignores the latter, while its represenation in phase space takes both into account to the same extend. On the one hand, a billiard ball can eventually reach any space location, regardless of the hole it moved through. On the other hand, however, its complex internal structure allows it to "remember" which hole it passed. Thus, if it passed through a hole A, it will never be able to reach a state of "remembering" that it passed through a hole B and vise versa. 

Roughly speaking, a phase space of a system of particles is its location on the actual space plus its memory. All systems, whether simple or complex, are very likely to reach the same space locations at some future point after very different behaviors in the past. Thus, the lack of overlap in phase space is blamed entirely on memory. 

 Parallel universes, being defined in phase space, take memory into account just as much as the do position. Thus, in order for a system of particles to move into a different universe and then come back to the earlier one, it has to be able to "erase" all of its memory. Thus, its ability to do that is a key to whether it is classical or quantum mechanical. 

Electrons, being one-particle systems, have no memory. Thus, the inevitable overlap in position space will imply overlap in phase space. This corresponds to earlier mentioned observation that the two delta functions evolving according to Schrodinger's equation are predicted to overlap. Speaking of "erasing the memory" comment, electrons "erase" their memories every second. Thus, they jump between parallel universes all the time, so, effectively, each of these parallel universes has all possible locations of each electron. 

Few particle systems do have some memory, but that memory is not very detailed. Thus, they can remember that they were in graduate school, but they won't remember whether it was in Montana State or University of North Texas. This means that they can still go through holes far away from each other and end up having identical memory, resulting in possible overlap in phase space. 

However, once the system reaches sufficient complexity, its memory is detailed enough to remember exact numbers of leafs on the trees and stones on the ground. As a result of the entropy of the environment, it is extremely unlikely for two different cities to match in all these details. Consequently, the probability for the system to pass through different holes and end up having the same memory is very small. This is equivalent to the earlier statement that the components of wave function on phase space are non-overlapping. 

Now let's see whether this can be reproduced by quantum tunnels. According to the tunnel model proposed in this paper, the tunnel is defined in terms of pointwise values of quantum field as opposed to the locations of each particle. 

This is done for a good reason: the reason quantum tunnel was introduced on the first place is that, unlike decoherence model, it is relativistically covariant. Implication of being relativistically covariant is the use of second quantization instead of the first. This means that the position degrees of freedom corresponding to each particle are replaced with field degrees of freedom of each point. 

However, this leads to the following question: a billiard ball can not "shrink" to one point. Thus, no matter how large it is, each point will only have at most one of its electrons. The field of each electron is much smaller than the width of the tunnel, as evidenced by the fact that electrons are quantum mechanical and not classical. So, then, how can pointwise-defined tunnels distingush the behavior of billiard balls, or their memories, no matter how large they are? 

I propose to answer this question in the following way. While electrons, being fermions, can not occupy the same location, the photons, being bosons, can. Thus, in order to be able to claim that billiard ball, being "large" has field value greater than the tunnel width, we have to claim that the photons emitted from the different particles that comprise it all met at the same location. 

Now, on average, this does not happen. If the photons are evenly distributted throughout the billiard ball, then every point of the billiard ball will on average receive as many photons as it would have if it was one particle system. However, due to entropy, there are some random high peaks. 

While, of course, this is subject to further research, I hypothesize that the size of some of these peaks is larger than the width of the corridor. Thus, the tunnel can detect the magnitudes and location of these peaks, and nothing more. However, if the billiard ball is sufficiently complex, the distribution of these peaks will be complex as well. 

I claim that these peaks alone serve as a "code" for the memory of the billiard ball.  If that ball passes through different locations in space, it might still have similar peaks, but these peaks might be slightly shifted as a result of interaction with an environment. Since, again, the measurement is based on second quantization rather than first, that tiny shift of the peaks is detected no matter how small it is. 

After all, no one measures position of each peak. Instead, the field strength is measured point-wise. Thus, an infinitesimal shift of a peak is equivalent to two totally independent events: its annihilation at point $r$ and its creation at point $r+ \epsilon$. Both events are on a scale larger than the width of a tunnel, hence both are detectable. 

\subsection{Including Fermions in the Picture}

As was mentioned in the previous section, our ability to measure the position of billiard ball is due to the photons emitted by its different particles meeting at the same location. This is a consequence of photons being bosons since fermions are not allowed to occupy the same state. Thus, the implication of the theory is that bosonic fields are the only ones being measured, and we simply infer the behavior of fermions based on the bosons they emit. 

However, in this section, I will ignore the above argument and try to restrict fermions into a tunnel anyway just to see what happens. It will be shown that this would only lead to absurd results, which would further justify the proposed idea of measuring only bosons. 

In light of the fact that Grassmann variables are defined in a literal sense, one might ask the following question: is it possible to confine fermionic field to a corridor the way it was done with a bosonic one? Unfortunately, the answer to this is no. After all, if a corridor was narrower than the width of the $\xi$ measure imposed on Grassmann space, then the integral of $\xi$ would no longer give $0$ , which means that the predicted behavior would no longer be fermionic. 

It would  not necesserely be bosonic either, since  the corridor might be wide enough for SOME variation of $\xi$ to occur, which would contradict the constant measure for bosonic case. Thus, the prediction of the theory would be the presence of bosons, fermions, as well as in-between particles. Bosons would correspond to a situation of corridor being much narrower than the width of the $\xi$ function, fermions would correspond to the corridor being much wider than the width of the $\xi$ function, and in-between particles would correspond to the width of a corridor being a fraction of a width of $\xi$ function, but that fraction is not too small. 

The advantage of this approach is that if there was a way to somehow exclude ``in-between" particles, it would be possible to claim that both bosons and fermions have the same measure, namely $\xi$ function, which might help with a unification arguments. It would also be possible to hypothesize that, in the similar way as the locations of centers of corridors are being excluded based on low probability, the same applies to their widths. Thus, spin statistics theorem implies that both narrow corridors for fermionic Lagrangian generators as well as wide corridors for bosonic Lagrangian generators give near-zero probability density. This would ``select out" the corridors where spin statistics theorem works. 

However, in light of the fact that weight function is unnatural, it is questionable whether imposing it on bosons for consistency sake is worth it. Besides, total consistency would not be achieved anyway since for bosons dot and wedge products characteristic of Grassmann integration will not be implimented. Finally, there is no natural way of excluding ``mixed particles" that are neither bosonic nor fermionic. 

Of course, one can still impose quantum corridors on fermions in order to be consistent with the setup of the theory. But, in order to avoid the above difficulties, one would have to restrict himselves only to the corridors that are much wider than non-zero region of $\xi$ function. Thus, they would either contain all of $\xi$ or none of it. In the former case, one would have regular fermionic behavior, but one would not be able to specify exactly into what part of the peak the system is localized. In the latter case, the integral will be $0$ which means that the whole scenario is ruled out by the constraint $\vert Z \vert^2 \geq f(\vert Z \vert^2)$.

However, it can be argued that being able to localize bosons is enough. Consider, for example, a double-slit experiment. When an electron hits the screen, the observer doesn't literally see the electron. Rather, he sees light coming into his eyes that is emitted from the location where electron supposedly hit the screen. That light is electromagnetic radiation. Since the electromagnetic field is bosonic, it can be localized. Thus, the statement ``it is not likely for electron to fly from point $A$ to point $B$" can be replaced with the statement ``it is not likely for light to first be emitted from point $A$ and then be emitted from point $B$. After all, the main way this can happen is by electron to travel from point $A$ to point $B$ and the latter is unlikely". 

In the language of Feynman diagrams this means that all of the external lines are only photons, while internal ones can be both photons and electrons. As a result of that, electrons have impact on photon-alone scattering process. Since measurement predicts only external lines rather than internal ones, quantum corridors are only applied to photons.  However, since Lagrangian includes internal lines as well, the electron part of the Lagrangian affects the probabilities of photon-based corridors. 

The localization of electron in space can be treated in the similar way. In the hypothetical situation, if fermions were localized directly, it would have been argued that the ``initial configuration" of fermions had large enough entropy for decoherence to occur.  On the other hand, in the real situation electrons are being replaced with the photons in above argument, which means that the claim of the theory is that the initial configuration of photons forces decoherence. Indeed, if initial configuration of photons indicates a specific structure of electrons, then it would be highly unprobable that it happens without such structure of electrons actually being there. This means that the main contribution to path integral arises from electrons actually forming that pattern. 

This means that integral will predict a decoherence picture similar to the one that would occur if there were indeed electrons spaced in that fashion. That decoherence implies that certain corridors of photons are much more probable than others. In light of the fact that due to decoherence different space locations of electron will imply very different behaviors of a photon, these ``more probable" corridors correspond to localization of electron in space.

\subsection{Manifold Structure: Revisited}

In section 5.2 it was discussed how introducing vierbein fields might increase the likelihood of manifold-like structure of a causal set. However, at the very beginning of the section, an important disclaimer was made: due to the fact that path integral is taken over all possible geometries, there is no such thing as making a prediction of geometry to approximate anything in particular, regardless of degrees of approximation. Rather, the only thing that was done in section 5.2 was to make an argument that manifold-like geometries have higher weight in a path integral. But, whether we are adding up all possible geometries, or only all possible manifold-like ones, the fact remains that there is no overall geometrical background. 

But now in this Appendix, the model of ``quantum collapse" of geometry (or, equivalently, collapse of causal relations) was introduced. This means that, regardless of presence or absence of Lagrangian generator of sec 4.4, there will indeed be the fixed geometry. On the other hand, if the Lagrangian generator of sec 4.4 is not introduced, there is no reason to expect that fixed geometry to be manifold-like. The causal structure, which is identified as geometry, can collapse to virtually anything, including, for example, tree-like causal relations.

Thus, introducing Lagrangian generator that would encourage geometry to be manifold-like, and introducing collapse mechanism to make geometry fixed are two very different parts of the same puzzle. If these two parts are both successfully done, then putting them together would lead to a manifold-like background for propagators to be introduced. 

As far as this thesis is concerned, the proposed solutions to either of these pieces of puzzle.

Since a manifold-like structure can not be introduced by means of 0 curvature, vierbeins are used to fulfill that purpose, and they are viewed as additional fields. Lagrangian generator is given by
\bea
& &{\cal K} (a,b,c,d; p, q) = \Big( \frac{a^2(p,q) - b^2(p,q) - c^2(p,q) - d^2(p,q)}{\tau^2 (p,q)} \Big) ^n \nonumber \\
& &\kern95pt +\ \Big( \frac{a^2(p,q) - b^2(p,q) - c^2(p,q) - d^2(p,q)}{\tau^2 (p,q)} \Big) ^{-n}\;,
\eea
where $n$ is a very large number. It should be understood that $a$, $b$, $c$ and $d$ are viewed as fields rather than coordinates. 

Due to the fact that $n$ is very large, it is clear that if that fraction is outside of a very small neighborhood of $1$, one of these two terms will be very large, depending on whether fraction is greater than $1$ or smaller than $1$, which would cause the sum to be very large in either case. If rapid increase of Lagrangian generator will ``leap" into rapid increase of Lagrangian itself, interference between Lagrangians of nearby configurations will lead to probability amplitude being very close to $0$. 

However, it is still okay for the Lagrangian generator to increase very fast, as long as it has no impact on Lagrangian. Since ${\cal L} = \min\, \max\, {\cal K}$, there has to be at least one Alexandrov set inside of which $\max \, {\cal K}$ is small. Thus, around each point there has to be at least one Alexandrov set in which the Lorentzian equation for the distance nearly holds, but it is not true for arbitrary Alexandrov set.

This, in fact, matches the observations for Lorentzian manifold. For example, suppose an electron is sent into a black hole with near-lightcone velocity. Then the proper time between emission of electron and electron reaching the center of black hole is very small. But it is not true that the flat space geometry is approximately valid inside of the Alexandrov set defined in terms of these two events. However, it is still true that if electron was sent somewhere else, its velocity can, indeed, be adjusted in such a way that metric is, indeed, flat in its frame. This means that the statement of approximate local validity of Lorentzian geometry is true ``for at least one frame" rather than ``for every single frame". This is exactly what Lagrangian generator is telling us.

It should be noticed, however, that even if  $\tau^2 \approx a^2(p,q) - b^2(p,q) - c^2(p,q) - d^2(p,q)$ is enforced, it does not imply a manifoldlike structure. In fact, for every single causal set, including the one with a lot of posts, it is possible to choose $a$, $b$, $c$ and $d$ pair-wise in such a way that the approximation holds for every single pair of points. What distinguishes manifoldlike causal set is that the number of such choices of $a$, $b$, $c$ and $d$ for manifoldlike causal sets is considerably larger than it is for non-manifoldlike ones. Thus, the key element of the theory is that larger number of similar choices of $a$, $b$, $c$ and $d$ implies larger probability. 

This, indeed, was done in constraint 1 of section 2 of this appendix 
\beq
\vert Z \vert ^2 > f (\vert Z \vert ^2)\;,
\eeq
where $f(x)<1$ is the decimal expression for $x$ starting from the 100-th digit.

That constraint, effectively, replaces the notion of probability with a notion of ``allowing" or ``forbidding" histories. Every ``allowed" history is represented in exactly one parallel universe, regardless of its probability; however, the ``density" of allowed histories approximately correlates with probability. 

Once all the fields are discretized according to constraint 2, there will appear a close correlation between the number of closely matching histories and the so-called probability. Consequently, if there is some other reason for the number of histories to be larger, it will imply larger probability. In particular, if causal relation $\prec$ happens to be manifoldlike, there will be larger number of choices of $a$, $b$, $c$ and $d$ that are NOT rulled out by the above constraint. Since every ``allowed" choice is represented in exactly one parallel universe, this means that there will be a lot more parallel universes to represent manifold-like causal relation then there is to represent non-manifoldlike one, which is why randomly selected causal set is manifold-like with large enough probability. 

\subsection{Appendix: Dyson-Based Model of Gravity for Continuum Manifold}

In the first section of this Appendix it was stated that in order to have a topological background for propagators, geometry, being identified as gravitational field, were to undergo quantum fluctuations, there has to be a mechanism to ``collapse" fluctuating geometry into some fixed one in order to have a topological background needed to propagate other fields. 

However, it was also mentioned that there is an alternative: a possibility suggested by Dyson that gravitational field does not exist altogether, which means that the geometry is not subject to quantum fluctuations to start with. This possibility was put aside on the basis that Einstein's equation is a consequence of variational principle which, in turn, is a classical limit of path integration. 

Nevertheless, it was acknowledged that it is possible to avoid any reference to path integration by viewing Einstein's equation as a postulate as opposed to a consequence of variational principle. In this section, this alternative path will be taken up. 

One of the attractive features of this path is that by avoiding quantizing gravitation one can avoid dealing with non-renormalizeable theory. Furthermore, this might allow one to postulate gravity in a usual continuum. The only reason to discritize space-time at all is renormalization of non-gravitational fields which is something we are already used to from flat space quantum field theory. 

It should be understood that this is a competing model with the one proposed in the rest of the thesis which means that this section can be skipped without compromising the understanding of other parts. 

In Section 2 of this Appendix it was found that laws of physics can be replaced with a set of all possible universes which are ``constrained" in some way. This means that an additional constraint can be imposed while staying perfectly consistent with philosophy of the theory, namely the one that Einstein's equation is approximately satisfied. Of course, it can not be exactly satisfied since Bianchi identity would then demand a conservation of energy momentum tensor. But an approximate constraint can still be imposed. Thus, there are two constraints: 

CONSTRAINT: Let $f(x)$ be a value between $0$ and $1$ corresponding to the decimal expression of $x$ starting from 100-th digit. A history $(g_{\mu \nu}(x), \phi_0 (x))$ is allowed if and only if the following is true

1) $\vert Z(\phi=\phi_0; g_{\mu \nu}) \vert ^2 > f(\vert Z(\phi=\phi_0; g_{\mu \nu}) \vert ^2)$, where $Z(\phi=\phi_0; g_{\mu \nu})$ is given by 
\beq Z \Big(\phi (x)= \phi_0 (x) \Big) = \int \big[{\cal D} \phi \big] \exp \Big( \int d^4 x \sqrt{-g} \big(i{\cal L}(\phi; x) - k(\phi (x)-\phi_0 (x))^2 \big) \Big) \eeq
2) Einstein's equation approximately holds, where by ``approximately" it is meant
\beq
R_{\mu\nu} - \frac12\, R\,g_{\mu\nu} = T_{\mu\nu} + t_{\mu\nu}\;,
\eeq
where
\beq
T_{\mu\nu}\, t^{\mu\nu} < \epsilon_1\;, \quad g_{\mu\nu}\,t^{\mu\nu} < \epsilon_2\;.
\eeq
It should be noted that the above assumes that the values of $T_{\mu \nu}$ has been localized throughout spacetime. This, of course, requires quantum corridors. However, while non-gravitational fields are still subject to quantum corridors, gravity itself no longer is. 

Since the former is renormalizeable while the latter isn't, this allows to compute the behavior of non-gravitational fields by standard methods and then apply the above constraints to estimate gravity without having to compute any graviton propagators. Then, based on estimated gravity, adjust the estimation of  non-gravitational propagators, and based on that adjust the estimation of gravity, etc.   In every case, the propagators are computted assuming aforegiven gravitational background. This allows to make predictions without dealing with non-renormalizeable theories.

\end{document}